\documentclass[reprint, superscriptaddress, bibliography]{revtex4-1}
\usepackage{latexsym,amssymb,amsfonts,mathrsfs,amsmath,bm}
\usepackage{graphicx}
\usepackage[usenames,dvipsnames]{xcolor}
\usepackage{soul}
\usepackage[linktocpage,colorlinks=true,linkcolor=blue,citecolor=blue,breaklinks=true,urlcolor=blue]{hyperref}

\renewcommand{\vec}[1]{\bm{#1}}
\newcommand{\mummy}{\mu\text{m}}
\newcommand{\mummys}{\mu\text{m}/\text{s}}

\newcommand{\br}{ {\bf r} }
\newcommand{\bhu}{ \hat{\bf u} }

\newcommand{\swh}{h}%

\begin{document}

\title{Nutrient transport driven by microbial active carpets}  

\author{Arnold J. T. M. Mathijssen}
\email[Correspondence: ]{amath@stanford.edu}
\affiliation{Department of Bioengineering, Stanford University, 443 Via Ortega, Stanford, CA 94305, USA}

\author{Francisca Guzm{\'a}n-Lastra}
\affiliation{Facultad de Ciencias, Universidad Mayor, Av. Manuel Montt 367, Providencia, Santiago, Chile}
\affiliation{Departamento de F\'isica, FCFM Universidad de Chile, Beauchef 850, Santiago, Chile}

\author{Andreas Kaiser}
\affiliation{Department of Biomedical Engineering, Pennsylvania State University, University Park, 16802, USA}

\author{Hartmut L{\"o}wen}
\affiliation{Institut f{\"u}r Theoretische Physik II: Weiche Materie, Heinrich-Heine-Universit{\"a}t, D-40225 D{\"u}sseldorf, Germany}

\date{\today}

\begin{abstract}
We demonstrate that active carpets of bacteria or self-propelled colloids generate coherent flows towards the substrate, and propose that these currents provide efficient pathways to replenish nutrients that feed back into activity. A full theory is developed in terms of gradients in the active matter density and velocity, and applied to bacterial turbulence, topological defects and clustering. Currents with complex spatiotemporal patterns are obtained, which are tuneable through confinement. Our findings show that diversity in carpet architecture is essential to maintain biofunctionality.
\end{abstract}

\maketitle



The collective motion of microorganisms and active colloids has sparked great interest, as biological functions can emerge from self-organisation of local power injection \cite{vicsek2012collective, marchetti2013hydrodynamics, lauga2009hydrodynamics, koch2011collective, elgeti2015physics, cates15a,  bechinger2016active, zottl2016emergent, needleman2017active}. 
To sustain these processes, self-propelled particles increase nutrient uptake \cite{magar2003nutrient,short2006flows, michelin2011optimal, tam2011optimal} 
and redistribute oxygen \cite{tuval2005bacterial}
by hydrodynamically enhanced mixing \cite{wu2000particle, kim2004enhanced, thiffeault2010stirring},
bioconvection \cite{pedley1988growth, hill2005bioconvection, karimi2013gyrotactic},
and particle entrainment \cite{pushkin2012fluid, jeanneret2016entrainment, mathijssen2018universal, vaccari2018cargo}. 
The vast majority of these flow-driving swimmers accumulate at surfaces \cite{berke2008hydrodynamic, li2009accumulation, molaei2014failed, sipos2015hydro, elgeti2015run, mathijssen2015hotspots, figueroa2015living, jin2018chemotactic, daddi2018state, ohmura2018simple, mathijssen2018oscillatory}, at concentrations an order of magnitude larger than in the bulk \cite{berke2008hydrodynamic, molaei2014failed, mathijssen2015hotspots}, and thus form `active carpets'.
Instead of wall attachment, these freely roaming carpets are stabilised by mutual cell attraction or chemotaxis.
However, this crowding drains reserves rapidly, and renewal is restricted by the boundary \cite{mathijssen2015tracer}, so biofunctionality is curtailed.
Moreover, swimmer-generated flows cancel each other in the case of homogeneous coverage, by symmetry, so supply of nutrients is limited by diffusion.
A steady advection changes this situation radically; it opens effective pathways for resource replenishment and reinforce activity.

\begin{figure}[b!]
\includegraphics[width=1\columnwidth]{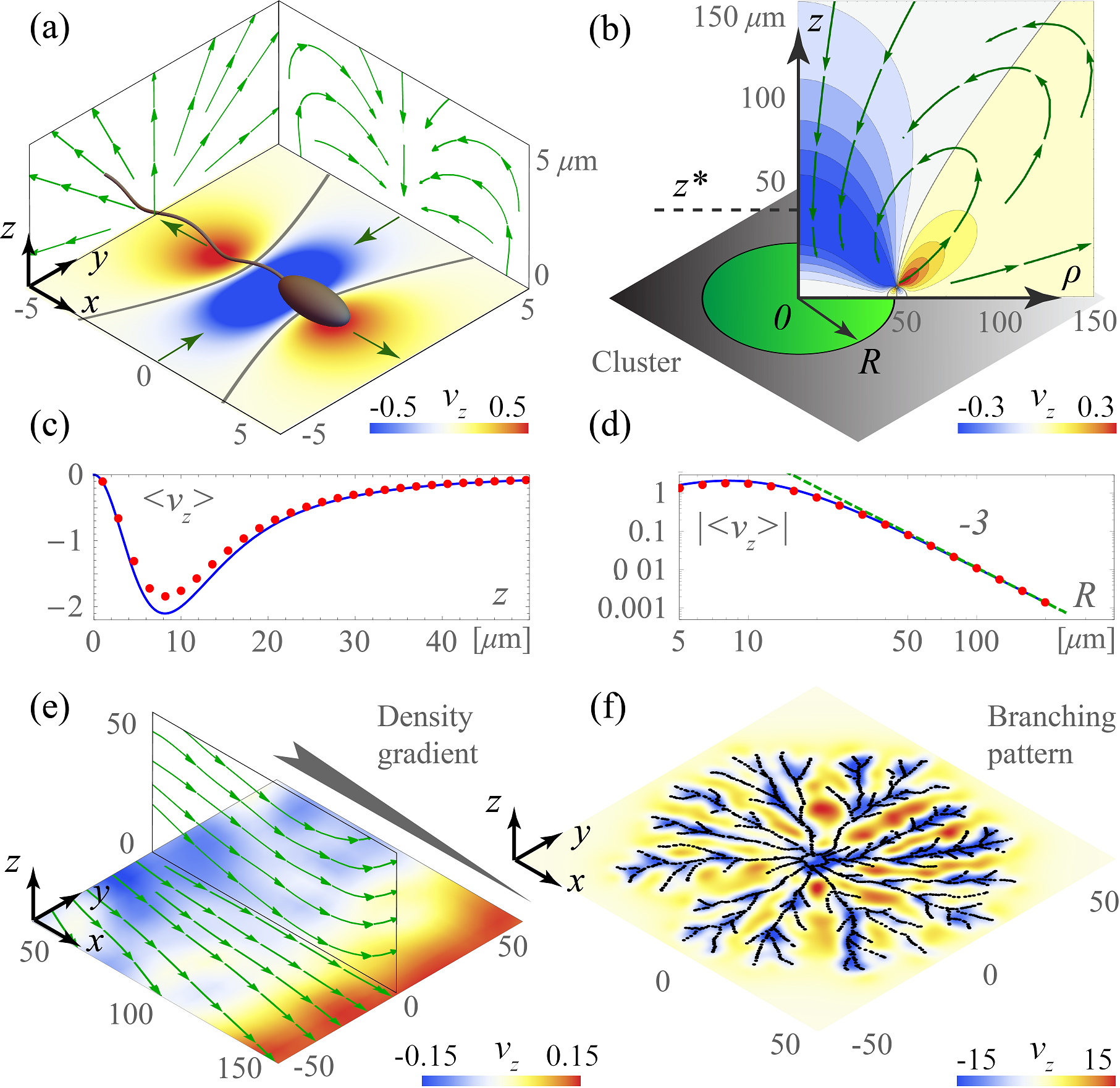}
\caption{\label{Fig1} 
Currents generated by clusters and density gradients, in $\mummys$. 
(a) Individual bacterial flow, shown for $z=5 \mummy$ (top view), $x=\pm5 \mummy$ (front view), $y=\pm5 \mummy$ (side view).
(b) Bacterial cluster flow, $\langle \vec{v}(\rho,z) \rangle$, with uniform density $n=0.1 / \mummy^2$ and size $R=50 \mummy$.
(c) Vertical flow as a function of $z$, evaluated at $\rho=0$ and $R=10 \mummy$, obtained numerically (red points) and analytically (blue lines, Eq.~\ref{Eq:AverageClusterFlow}). 
(d) Same; as a function of $R$, evaluated at $\rho=0$ and $z=10 \mummy$.
(e) Bacterial density gradient, simulated with $N=12,500$ and $R=200 \mummy$, shown for the planes $z=20 \mummy$ (top view) and $y=0$.
(f, enlarged in SI Fig.~1) Bacteria arranged in a branching pattern, simulated with $N=1800$ cells (black points), shown for $z=5 \mummy$ (top view).
}
\end{figure}

In this Letter, we demonstrate that such coherent transport arises from gradients in density, activity or orientation, which emerge naturally from the long-ranged order in collective behaviour \cite{tonertu1995long}, such as in
bacterial vortex arrays \cite{riedel2005self, lushi2014fluid, ingham2008swarming, wioland2013confinement, wioland2016ferromagnetic},
bacterial turbulence \cite{dombrowski2004self, sokolov2007concentration, zhang2010collective, cisneros2010fluid,  sokolov2012physical, dunkel2013fluid, wensink2012meso},
and giant density fluctuations \cite{narayan2007long, yaouen2012athermal, buttinoni2013dynamical, cates15a, sepulveda2017wetting, reichhardt2018clogging}.
Topology and geometry play a crucial role in these living fluids \cite{toth2002hydrodynamics, elgeti2011defect, sanchez2012spontaneous, marchetti2013hydrodynamics, keber2014topology, giomi2015geometry}, providing a bridge with material sciences and cell biology \cite{needleman2017active, saw2017topological}. We focus on bacteria as a concrete example, but this theory applies to the broader class of active carpets to which no external forces and torques are applied.

First, we show that a bacterial cluster, despite random orientations, creates a net nutrient transport towards the surface.
Second, in uniform-density carpets, gradients in swimmer orientation produce flows instead.
We derive and implement these to topological defects commonly found in living fluids.
Combining these fundamental ingredients, the nutrient transport by vortex arrays and bacterial turbulence are evaluated, and the spatiotemporal correlations of the flows compared to the collective dynamics.

\subsection*{Individual swimmer flows}

\begin{figure}[t!]
\includegraphics[width= \linewidth]{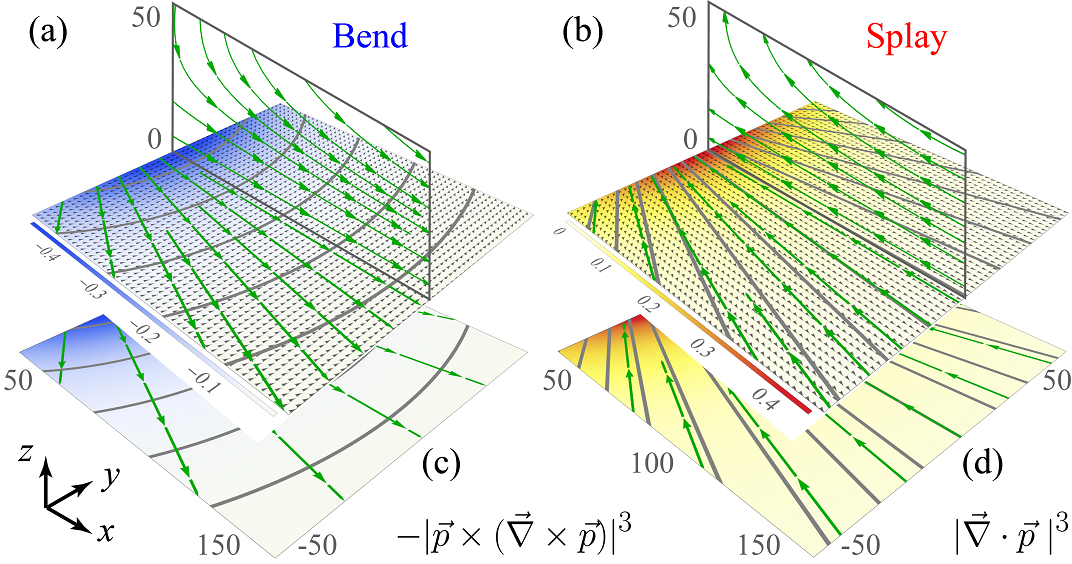} \\
\caption{\label{Fig2} 
Flows generated by gradients in the director field.
(a) `Bend-type' gradients: bacteria are oriented along circles centred at the origin. 
(b) `Splay-type' gradients: orientations along the corresponding radial lines. 
We simulate $N\sim200,000$ swimmers (grey arrows) arranged on a large uniform lattice with $R=500 \mummy$ and $n=0.25 / \mummy^2$, so that density gradients and edge effects are negligible. 
Colours indicate flows in the $z$ direction, in $\mummys$, evaluated for the plane $z=20 \mummy$, and green arrows show stream lines, also for $y=0$. 
(c,d) Corresponding theoretical estimates (\ref{Eq:BendFlows},\ref{Eq:BendFlowVertical}).
}
\end{figure}

We consider a colony of microswimmers with balanced propulsion and drag forces.
These are located at $\vec{r}_s$ and oriented along $\vec{p}$ parallel to a solid surface, which is fixed at $z=0$ in Cartesian coordinates.
Each swimmer generates a flow $\vec{u}(\vec{r})$ that can displace nutrients, represented by a tracer particle located at $\vec{r}$.
At low Reynolds numbers, and for distances $d = |\vec{r} - \vec{r}_s|$  larger than a few body lengths, this individual flow is well described by a Stokes dipole aligned with the swimming direction \cite{drescher2011fluid, lauga2009hydrodynamics}, given by 
\begin{align}
\label{Eq:IndividualDipoleFlow}
\vec{u} (\vec{r}, \vec{r}_s, \vec{p}) = \kappa [ ( \vec{p} \cdot \vec{\nabla}_s) \mathcal{B}(\vec{r}, \vec{r}_s) ] \cdot \vec{p},
\end{align}
where the dipole strength is $|\kappa| \sim 3 v_s a_s^2/4$ in terms of the swimmer's speed $v_s$ and size $a_s$ \cite{mathijssen2018universal}.
The no-slip condition at the wall is accounted for using the Blake tensor $\mathcal{B}(\vec{r}, \vec{r}_s)$ formalism \cite{blake1971note,mathijssen2015hydrodynamics} [see Supplementary Information (SI) \S1].
Throughout this paper, as an example, we use swimmer height $h = z_s= 1 \mummy$ and dipole moment $\kappa = 30 \mu\text{m}^3/\text{s}$ for the pusher \textit{E. coli} \cite{drescher2011fluid}.

Figure~\ref{Fig1}(a) shows the resulting flow driven by a single bacterium.
Nutrients are attracted towards the surface directly above the swimmer (blue regions), but pushed upwards in front of and behind the cell (red regions). 
The net flux across any plane in $z$ vanishes due to the incompressibility of the liquid, $\int \vec{u} dx dy=0$, but across a plane recirculating vortices can emerge (green stream lines).
For pullers, $\kappa <0$, the flow direction is inverted.
Taken together, the average flow velocity due to all swimmers on the surface combined is
\begin{align}
\label{Eq:AverageTotalFlow}
\langle \vec{v} (\vec{r}) \rangle
&= \int \vec{u} (\vec{r}, \vec{r}_s, \vec{p}) f(\vec{r}_s, \vec{p}) d\vec{r}_s d\vec{p},
\end{align}
where $f$ is the probability density of finding a swimmer at position $\vec{r}_s$ and orientation $\vec{p}$.

\subsection*{Clusters \& density gradients}

We examine a cluster of $N$ bacteria that assemble around a chemoattractant [Movie S1]. 
Remarkably, this active carpet generates a steady current that brings nutrients down towards the surface.
To analyse this, we first imagine a circular cluster of radius $R$ centred at the origin with constant density, $n=N/(\pi R^2)$, and uniformly distributed swimmer orientations in the plane.
The total flow, derived in SI \S2A and shown in Fig.~\ref{Fig1}(b), is found by inserting this profile, $f \propto \frac{n}{2\pi}$, into Eq.~(\ref{Eq:AverageTotalFlow}).
As in the movie, this yields a downwelling region for all lateral distances $\rho < R$ and all heights $z > h$, where $\rho = \sqrt{x^2 + y^2}$, despite the random swimmer orientations and thermal particle diffusion.
Subsequently, the nutrients move from the centre to the edge of the cluster, to $\rho > R$, where incompressibility demands that liquid be transported back up, causing a large toroidal recirculation.
Directly above the cluster, along the $z$ axis and in the limit $z \gg h$, the result simplifies to the mean drift velocity
\begin{equation}
\label{Eq:AverageClusterFlow}
\langle \vec{v} (z,R) \rangle=-12\pi n h \kappa\frac{z^2 R^2}{(z^2+R^2)^{5/2}} \hat{\vec{z}}.
\end{equation}
For a typical bacterial density, $n \sim 0.1/ \mummy^2$ \cite{figueroa2015living} and cluster size $R \sim 50 \mummy$ we expect significant nutrient transport up to $\langle \vec{v} (z^*) \rangle \sim 25 \mummy / \text{min}$ [Fig.~\ref{Fig1}(b,c)].
This can be orders of magnitude larger than sedimentation velocities for micron-sized particles.
Compared to diffusion the P\'eclet number is large, $\text{Pe} = \frac{v R}{D} \sim 48$, and the transport is additive over time.
Moreover, flows are $\sim 2$x stronger for more realistic Gaussian clusters [SI \S 2B].

\begin{figure*}[t!]
\includegraphics[width=\linewidth]{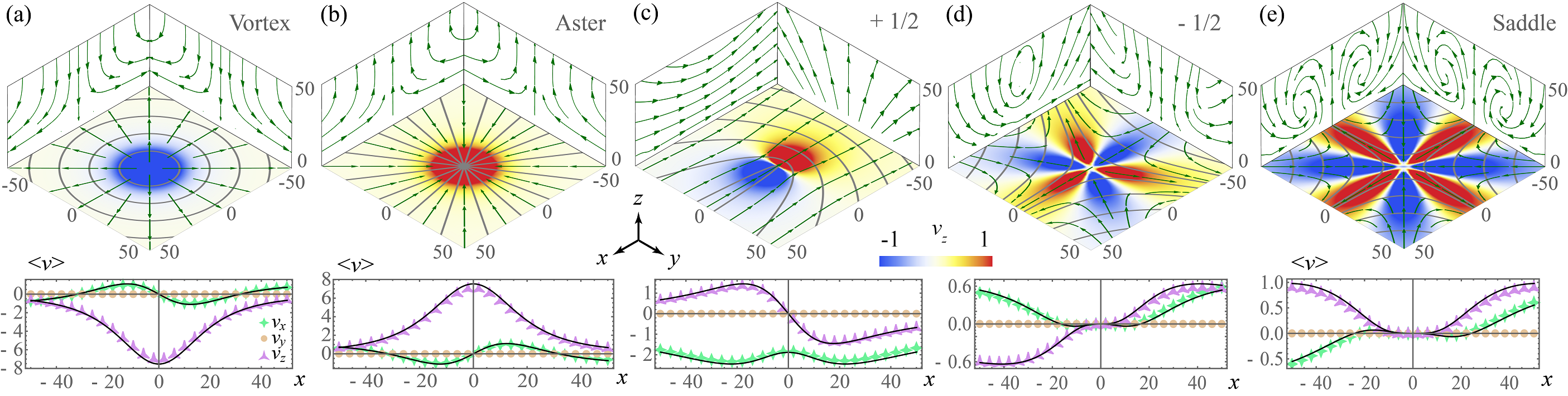} \\
\caption{\label{Fig3} 
Defects in the director field generate strong flows because of large orientation gradients. 
Swimmers are arranged in a dense uniform lattice with orientation $\phi_s = \phi_0 + m \theta$ (grey lines). 
Upper panels: Colours indicate vertical flows in $\mummys$, simulated for the plane $z=5 \mummy$, and green arrows are stream lines, also for the planes $x,y=-50 \mummy$.
(a) Vortex defect with $(m,\phi_0)=(1, \pi/2)$. 
(b) Aster defect with $(1,0)$. 
(c) Plus half defect with $(\frac 1 2 , 0)$. 
(d) Minus half defect with $(- \frac 1 2, 0)$. 
(e) Saddle defect with $(-1, 0)$.  
Lower panels: Flows in $\mummys$ for the plane $y=0$, obtained numerically (markers) and analytically (lines).
}
\end{figure*}

Counterintuitively, larger homogeneous clusters do not transport faster.
To be precise, in the thermodynamic limit where $R,N \to \infty$ with constant $n$, the individual swimmer flows cancel each other out, on average, so the surface attraction vanishes.
Indeed, the mean flow (Eq.~\ref{Eq:AverageClusterFlow}) decays as $1/R^3$ in this limit [Fig.~\ref{Fig1}(d)]. 
Maximising $\langle \vec{v} (z,R) \rangle$ with respect to $R$, for a given distance from the surface $z$, we obtain the optimal cluster size $R^* = \sqrt{2/3} z$.

More generally, all gradients in swimmer density or activity can drive currents.
To see this we simulate a cluster with a linearly decreasing density [SI \S 5C].
As before, this generates a horizontal flow along the gradient with downwelling at the high end [Fig.~\ref{Fig1}(e)].

Using this information, one can also predict transport driven by clusters of a more complex morphology. 
Figure~\ref{Fig1}(f) depicts flows generated by bacteria arranged in a branching pattern [SI \S 5D].
In agreement with the previous simplified cases, flows move downwards to the high-density regions, the branches.
This configuration is of course arbitrary, but serves to emphasize the robustness with respect to cluster shape.

An important prerequisite for steady flows is that gradients are sustained.
Stable gradients in metabolism can arise by e.g. local nutrient hotspots, and density gradients by chemotaxis or light control \cite{steager2007control, palacci2013living, arlt2018painting, frangipane2018dynamic}.
To quantify this, we analyse the stability of a cluster around chemoattractant [SI \S 6].
While we considered an instantaneous swimmer distribution above, we explicitly model their dynamics here, together with rotational (or run-tumble) fluctuations $D_r$.
We find that with increasing chemotactic strength, $\Omega_c$, a stable cluster forms and a net nutrient flux emerges, which saturates when $\Omega_c > D_r$.

\subsection*{Orientation gradients}

In the previous scenario with random orientations, the mean flows vanish in the absence of gradients in density. 
Furthermore, if all swimmers are oriented in the same direction, through collective motion or alignment interactions, then the currents also cancel in the thermodynamic limit [SI \S 3A].
However, gradients in swimmer orientation give rise to a second source of flow generation.

To classify the relevant orientation derivatives, it is important to note that the swimmer flow (Eq.~\ref{Eq:IndividualDipoleFlow}) is nematically symmetric [Fig.~\ref{Fig1}(a)], i.e. invariant under $\vec{p} \to -\vec{p}$. 
Hence, the only first-order derivatives that obey this symmetry in a 2D active carpet are, expressed in liquid crystal terminology \cite{degennes1993physics}, the `bend' and `splay' contributions, 
\begin{align}
\label{Eq:Bend}
B &= (\vec{p} \times (\vec{\nabla}_s \times \vec{p}))^2,
\\
\label{Eq:Splay}
S &=  (\vec{\nabla}_s \cdot \vec{p})^2.
\end{align}
The effect of these gradients is illustrated in Fig.~\ref{Fig2}.
We consider actives particles that swim collectively (a) in concentric circles, $\phi_s = \theta + \frac \pi 2$, or (b) towards a chemoattractant source, $\phi_s = \theta$, where $\phi_s = \text{arctan}(p_y/p_x)$ and $\theta = \text{arctan}(y/x)$, and they are spread out uniformly in space to minimise swimmer density gradients [SI \S 5E].
In both cases the orientation gradients decay with distance from the centre quadratically; for (a) we have $B(\rho) =1/\rho^2$  and $S(\rho) = 0$, and vice-versa for (b).
Then, a strong correlation is observed between bend gradients and liquid moving downwards and outwards.
Conversely, splay gradients drive flows inwards and upwards.

To make analytical progress, we realise that it is not always possible to find a general formula for the \textit{local} flow in terms of the gradients, $\langle \vec{v} \rangle (\vec{r}) = \chi(B,S)$, because the velocity is generated by a \textit{region} of swimmers in which the gradients vary.
These variations increase for larger $z$ values as the number of equidistant swimmers, i.e. this region of influence, grows.
However, the gradients are approximately constant far from the circle centre, when $z \ll \rho$, so we can couple the gradients and flows in that area [Fig.~\ref{Fig2}(a,b)].
Therefore, by expanding the mean current (Eq.~\ref{Eq:AverageTotalFlow}) in terms of $1/ \rho$ [SI \S 3B,3C], we find the first-order contributions to the horizontal and vertical flows due to bend and splay gradients,
\begin{align}
\label{Eq:BendFlows}
\langle v_\rho \rangle
&\approx 8\pi n h \kappa \left ( \sqrt{B(\rho)} - \sqrt{S(\rho)} \right ) ,
\\
\label{Eq:BendFlowVertical}
\langle v_z \rangle
&\approx - 8\pi n h \kappa z^2 \left ( [B(\rho)]^{3/2} -  [S(\rho)]^{3/2} \right).
\end{align}
This approximation, shown in Fig.~\ref{Fig2}(c,d), offers a good agreement with its numerical counterpart.
It also follows that for weak gradients, the horizontal flows are stronger than the vertical transport.

\subsection*{Topological defects}

\begin{figure}[t]
\includegraphics[width= \linewidth]{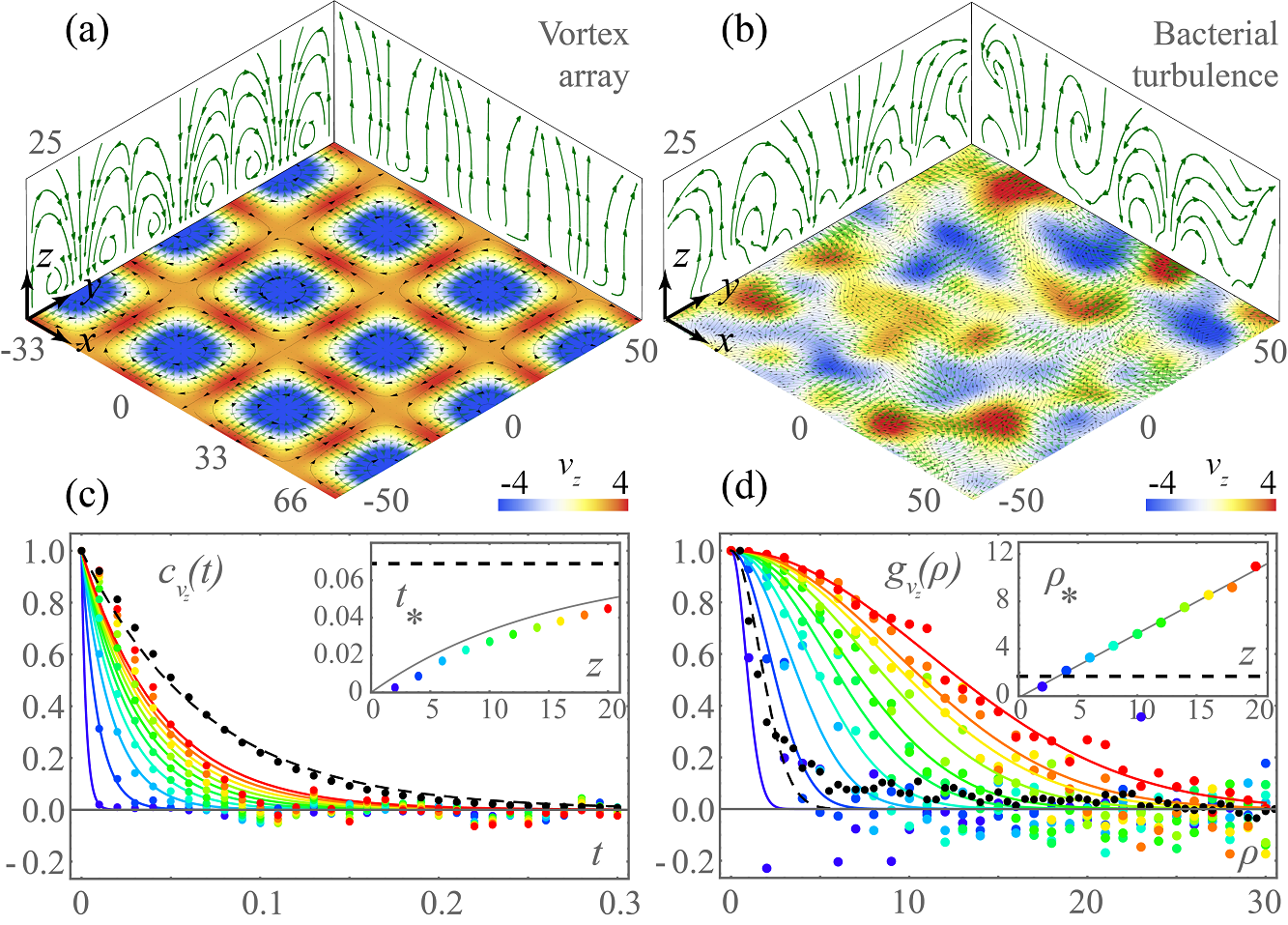} 
\caption{\label{Fig4} 
Flows created by collectively moving swimmers.
(a; enlarged in SI Fig.~2) Taylor-Green vortex pattern with unit cell size $\lambda = 33 \mummy$, and uniform swimmer density $n=0.25 /\mummy^2$. 
(b; enlarged in SI Fig.~4) Bacterial turbulence, simulated with the SPR model with aspect ratio $a=5$, packing fraction $\Phi  = 0.7$ and $n=0.25 /\mummy^2$.
(a,b) Colours indicate vertical flows in $\mummys$, simulated for $z=10 \mummy$. Green arrows are stream lines and black arrows the swimmer orientations.
(c,d) Bacterial turbulence. Temporal and spatial correlation functions of $v_z$, respectively, for $z\in [2,20]$ (blue-red), with corresponding correlations of swimmer orientation (dashed black).
Fits (solid lines) provide the correlation time ($t_*$) and length  ($\rho_*$).
Insets show these swimmer (dashed) and flow (blue-red, fits in grey) correlations against height.
}
\end{figure}

Like we saw for density gradients, it is now possible to interpret more complex carpet designs in terms of the fundamental ingredients, bend and splay. 
The first non-trivial orientation patterns with significant orientation gradients are the lowest-order topological defects [Fig.~\ref{Fig3}].
Their director fields are defined as $\phi_s = \phi_0 + m \theta$, where $\phi_0$ is a phase angle and $m=\pm \frac{1}{2}, \pm 1, \pm \frac{3}{2}, \dots$ is the topological charge \cite{degennes1993physics}.
Because these defect arrangements are well characterised mathematically, it is possible to find analytical solutions for the flows they generate [SI \S 4].

Swimmers with polar order feature integer-charge defects. 
For $m=1$  [Fig.~\ref{Fig3}(a,b)], there is a continuous transition from nutrient attraction near `vortex' defects ($\phi_0 = \frac \pi 2$), via no flow `spiral' defects ($\phi_0 = \frac \pi 4$),  to repulsion near `aster' defects ($\phi_0 = 0$), 
\begin{equation}
\label{Eq:VortexAsterFlow}
\langle v_z \rangle^{m=1}= 8 \pi n h \kappa \frac{z^2 \cos (2\phi_0)}{(\rho^2+z^2)^{3/2}}.
\end{equation}
Active particles with nematic order feature half-integer charges.
Near an $m=\frac 1 2$ defect [Fig.~\ref{Fig3}(c)], cooperation between bend and splay gradients drives horizontal currents, outwards from the bend curvature.
The flows in $z$ follow from recirculation, down towards the defect and back up again, with extrema at $\rho = z/\sqrt{2}$.
Also near $m=-\frac 1 2$ defects and near `saddle' defects, $m=-1$, the horizontal flows move in towards the convex side of the bends and out in the regions of converging splay [Fig.~\ref{Fig3}(d,e)]. 
In all cases, the calculated flows [SI \S 4] agree well with the simulated ones [Fig.~\ref{Fig3}, lower panels].

An important observation is that splay gradients (divergence of $\vec{p}$ in Eq.~\ref{Eq:Splay}) and density gradients are coupled in time, via motility. 
Specifically, bacteria can accumulate or deplete from defects, as observed in liquid crystals \cite{genkin2017topological}.
Therefore, vortex defects [Fig.~\ref{Fig3}a] remain stable over time, but steady states of aster defects [Fig.~\ref{Fig3}b] must feature more complex dynamics, such as defect ordering \cite{doostmohammadi2016stabilization} or ejection of swimmers from the carpet into the bulk.
Otherwise the defects can be motile, with time-dependent flows, as we discuss below for bacterial turbulence.


\subsection*{Vortex arrays}

The topological building blocks can be used to comprehend the currents created by active carpets featuring collective motion.
Particularly common in nature, and microfluidically controllable, are vortex patterns that bacteria or spermatozoa at high surface densities can self-organise into \cite{riedel2005self, lushi2014fluid, ingham2008swarming, wioland2013confinement, wioland2016ferromagnetic}.
Note, high surface densities go hand in hand with association and dissociation of swimmers in the bulk \cite{elgeti2015physics}.
Therefore, even if bulk swimmers are an order of magnitude more sparse \cite{berke2008hydrodynamic, molaei2014failed}, they will also generate diffusive flows \cite{wu2000particle, kim2004enhanced, thiffeault2010stirring}.

We first consider a Taylor-Green Vortex (TGV) carpet, which periodically features `vortex' and `saddle' defects ($m=\pm1$) at the centre and corners of the unit cell, respectively [Fig.~\ref{Fig4}a, SI \S 5F].  
Nutrients are attracted down to the vortex centres (locally described by Eq.~\ref{Eq:VortexAsterFlow}), and recirculated upwards with 4-fold symmetry at the face centres of the unit cell, in agreement with the individual defect flows [Fig.~\ref{Fig3}a,e].
Changing the vortex size with confinement can therefore tune the flows.

\subsection*{Bacterial turbulence}

Similarly, we consider the more complex patterns generated by bacterial turbulence \cite{dombrowski2004self, sokolov2007concentration, zhang2010collective, cisneros2010fluid,  sokolov2012physical, dunkel2013fluid, wensink2012meso}. 
Their collective dynamics are simulated using the Self-Propelled Rod (SPR) model \cite{wensink2012meso} to determine swimmer positions and orientation [Movie S2, SI \S 7A,B].
Because of the high volume fraction, density gradients remain negligible but orientation gradients are abundant.
Hence, recirculatory currents are generated, as shown in Fig.~\ref{Fig4}b.
Weak flows occur in the regions where swimmers are aligned with each other [SI \S 3A], but defects give rise to strong bend and splay gradients and thus nutrient transport.

Movies S3 - S5 show how these currents develop during the onset of turbulence, giving top views at $z=10, 25 \mummy$, respectively, and a side view for the cross section $y=0$. 
Interestingly, further from the active carpet the downwelling and upwelling regions are slower but larger.
We quantify this by computing the temporal and spatial correlation functions, $c_{v_z}(t)$ and $g_{v_z}(\rho)$, for different heights $z$ [SI \S 7C,D].
Hence, we obtain the correlation time $t_*(z)$ and correlation length $\rho_*(z)$ from their fits [Fig.~\ref{Fig4}c,d].
At short timescales the nutrient transport is ballistic but, of course, after this memory time it is diffusive.
Far from the carpet this memory is set by the decorrelation of swimmer orientations (dashed black), but nearby $t_*$ reduces to the mean free time between collisions with individual swimmers.
Conversely, the correlation length $\rho_*$ grows linearly with $z$, and it is not bound by the correlation length of swimmer orientations because the region of influence by more equidistant bacteria grows beyond the turbulent swirl radius.
Indeed, the renormalised correlations $g_{v_z}(\rho/z)$ collapse onto one another [SI Fig.~5], highlighting the scaling relation of the flow's long-rangedness.

Topological analysis of active carpets can be a powerful technique:
Knowing only the defect configuration in homogeneous carpets, one can interpolate the director field and thus predict the resulting flows.
We describe this for a monolayer of bacteria, but at higher cell densities the carpet could be thicker with multiple layers moving collectively.
Our analysis might still apply then, provided the carpet thickness is smaller than the correlation length, before transitioning to 3D turbulence \cite{ishikawa2011energy, shendruk2018twist}.

\subsection*{Conclusions}

We studied the emergence of large-scale recirculation by a carpet of force-free actuators.
Surprisingly, finite clusters of randomly oriented bacteria drive non-diffusive currents, in contracts with ciliary arrays \cite{elgeti2013emergence, ding2014mixing, uchida2010synchronization, uchida2010bsynchronization} and grafted cells \cite{darnton2004moving, kim2008microfluidic, hsiao2014collective, hsiao2016impurity} where alignment is essential for microbiological transport [SI \S2C].
Moreover, in the context of diversity in carpet architecture, it might be beneficial for an individual organism not to generate a flow to maximise the collective flux.
To consolidate this, a mathematical foundation is derived in terms of gradients in the carpet activity, density and orientation fields.
In nature, stable density gradients or clustering can arise by self-assembly \cite{cates15a, bechinger2016active, sepulveda2017wetting} and chemo-, thermo-, photo-, or rheotaxis \cite{marcos2012bacterial, mathijssen2018oscillatory}. 
Orientation gradients can form through individual actuation or collective instabilities \cite{tonertu1995long, simha2002instabilities}.
To stabilise these, topological constraints are key, through defect ordering \cite{doostmohammadi2016stabilization} or confinement by liquid drops \cite{vladescu2014filling} and spherical manifolds \cite{janssen2017aging}.
Experimental realisations may be achieved by chemoattractants, thermokinetic or light-controlled coordination \cite{steager2007control, palacci2013living, arlt2018painting, frangipane2018dynamic}.
Lithographic surface patterning and rectification \cite{galajda2007wall, wan2008rectification, koumakis2013targeted, simmchen2016topographical} could also make complex flux patterns, when correcting for disturbance flows due to cell-wall interactions \cite{ lauga2009hydrodynamics, koch2011collective, elgeti2015physics}.
Hence, these currents may be employed to drive active flow networks \cite{woodhouse2016stochastic} and provide understanding for transport by complex-shaped clusters, for bacterial turbulence \cite{dombrowski2004self, sokolov2007concentration, zhang2010collective, cisneros2010fluid,  sokolov2012physical, dunkel2013fluid, wensink2012meso}, and biofilm architecture \cite{vidakovic2018dynamic}.

\subsection*{Acknowledgements}

We would like to thank Manu Prakash and Deepak Krishnamurthy for helpful discussions.
AM acknowledges funding from the Human Frontier Science Program (Fellowship LT001670/2017).
FGL acknowledges Millennium Nucleus ``Physics of active matter'' of the Millennium Scientific Initiative of the Ministry of Economy, Development and Tourism, Chile.
HL acknowledges support from the Deutsche Forschungsgemeinschaft, DFG project SPP 1726.

\bibliography{bibliography}

\begin{thebibliography}{118}%
\makeatletter
\providecommand \@ifxundefined [1]{%
 \@ifx{#1\undefined}
}%
\providecommand \@ifnum [1]{%
 \ifnum #1\expandafter \@firstoftwo
 \else \expandafter \@secondoftwo
 \fi
}%
\providecommand \@ifx [1]{%
 \ifx #1\expandafter \@firstoftwo
 \else \expandafter \@secondoftwo
 \fi
}%
\providecommand \natexlab [1]{#1}%
\providecommand \enquote  [1]{``#1''}%
\providecommand \bibnamefont  [1]{#1}%
\providecommand \bibfnamefont [1]{#1}%
\providecommand \citenamefont [1]{#1}%
\providecommand \href@noop [0]{\@secondoftwo}%
\providecommand \href [0]{\begingroup \@sanitize@url \@href}%
\providecommand \@href[1]{\@@startlink{#1}\@@href}%
\providecommand \@@href[1]{\endgroup#1\@@endlink}%
\providecommand \@sanitize@url [0]{\catcode `\\12\catcode `\$12\catcode
  `\&12\catcode `\#12\catcode `\^12\catcode `\_12\catcode `\%12\relax}%
\providecommand \@@startlink[1]{}%
\providecommand \@@endlink[0]{}%
\providecommand \url  [0]{\begingroup\@sanitize@url \@url }%
\providecommand \@url [1]{\endgroup\@href {#1}{\urlprefix }}%
\providecommand \urlprefix  [0]{URL }%
\providecommand \Eprint [0]{\href }%
\providecommand \doibase [0]{http://dx.doi.org/}%
\providecommand \selectlanguage [0]{\@gobble}%
\providecommand \bibinfo  [0]{\@secondoftwo}%
\providecommand \bibfield  [0]{\@secondoftwo}%
\providecommand \translation [1]{[#1]}%
\providecommand \BibitemOpen [0]{}%
\providecommand \bibitemStop [0]{}%
\providecommand \bibitemNoStop [0]{.\EOS\space}%
\providecommand \EOS [0]{\spacefactor3000\relax}%
\providecommand \BibitemShut  [1]{\csname bibitem#1\endcsname}%
\let\auto@bib@innerbib\@empty
\bibitem [{\citenamefont {Vicsek}\ and\ \citenamefont
  {Zafeiris}(2012)}]{vicsek2012collective}%
  \BibitemOpen
  \bibfield  {author} {\bibinfo {author} {\bibfnamefont {T.}~\bibnamefont
  {Vicsek}}\ and\ \bibinfo {author} {\bibfnamefont {A.}~\bibnamefont
  {Zafeiris}},\ }\href {\doibase 10.1016/j.physrep.2012.03.004} {\bibfield
  {journal} {\bibinfo  {journal} {Phys. Rep.}\ }\textbf {\bibinfo {volume}
  {517}},\ \bibinfo {pages} {71} (\bibinfo {year} {2012})}\BibitemShut
  {NoStop}%
\bibitem [{\citenamefont {Marchetti}\ \emph {et~al.}(2013)\citenamefont
  {Marchetti}, \citenamefont {Joanny}, \citenamefont {Ramaswamy}, \citenamefont
  {Liverpool}, \citenamefont {Prost}, \citenamefont {Rao},\ and\ \citenamefont
  {Simha}}]{marchetti2013hydrodynamics}%
  \BibitemOpen
  \bibfield  {author} {\bibinfo {author} {\bibfnamefont {M.}~\bibnamefont
  {Marchetti}}, \bibinfo {author} {\bibfnamefont {J.}~\bibnamefont {Joanny}},
  \bibinfo {author} {\bibfnamefont {S.}~\bibnamefont {Ramaswamy}}, \bibinfo
  {author} {\bibfnamefont {T.}~\bibnamefont {Liverpool}}, \bibinfo {author}
  {\bibfnamefont {J.}~\bibnamefont {Prost}}, \bibinfo {author} {\bibfnamefont
  {M.}~\bibnamefont {Rao}}, \ and\ \bibinfo {author} {\bibfnamefont
  {R.}~\bibnamefont {Simha}},\ }\href {\doibase 10.1103/RevModPhys.85.1143}
  {\bibfield  {journal} {\bibinfo  {journal} {Rev. Mod. Phys.}\ }\textbf
  {\bibinfo {volume} {85}},\ \bibinfo {pages} {1143} (\bibinfo {year}
  {2013})}\BibitemShut {NoStop}%
\bibitem [{\citenamefont {Lauga}\ and\ \citenamefont
  {Powers}(2009)}]{lauga2009hydrodynamics}%
  \BibitemOpen
  \bibfield  {author} {\bibinfo {author} {\bibfnamefont {E.}~\bibnamefont
  {Lauga}}\ and\ \bibinfo {author} {\bibfnamefont {T.~R.}\ \bibnamefont
  {Powers}},\ }\href {\doibase 10.1088/0034-4885/72/9/096601} {\bibfield
  {journal} {\bibinfo  {journal} {Rep. Progr. Phys.}\ }\textbf {\bibinfo
  {volume} {72}},\ \bibinfo {pages} {096601} (\bibinfo {year}
  {2009})}\BibitemShut {NoStop}%
\bibitem [{\citenamefont {Koch}\ and\ \citenamefont
  {Subramanian}(2011)}]{koch2011collective}%
  \BibitemOpen
  \bibfield  {author} {\bibinfo {author} {\bibfnamefont {D.~L.}\ \bibnamefont
  {Koch}}\ and\ \bibinfo {author} {\bibfnamefont {G.}~\bibnamefont
  {Subramanian}},\ }\href {\doibase 10.1146/annurev-fluid-121108-145434}
  {\bibfield  {journal} {\bibinfo  {journal} {Ann. Rev. Fluid Mech.}\ }\textbf
  {\bibinfo {volume} {43}},\ \bibinfo {pages} {637} (\bibinfo {year}
  {2011})}\BibitemShut {NoStop}%
\bibitem [{\citenamefont {Elgeti}\ \emph {et~al.}(2015)\citenamefont {Elgeti},
  \citenamefont {Winkler},\ and\ \citenamefont {Gompper}}]{elgeti2015physics}%
  \BibitemOpen
  \bibfield  {author} {\bibinfo {author} {\bibfnamefont {J.}~\bibnamefont
  {Elgeti}}, \bibinfo {author} {\bibfnamefont {R.~G.}\ \bibnamefont {Winkler}},
  \ and\ \bibinfo {author} {\bibfnamefont {G.}~\bibnamefont {Gompper}},\ }\href
  {\doibase 10.1088/0034-4885/78/5/056601} {\bibfield  {journal} {\bibinfo
  {journal} {Rep. Progr. Phys.}\ }\textbf {\bibinfo {volume} {78}},\ \bibinfo
  {pages} {056601} (\bibinfo {year} {2015})}\BibitemShut {NoStop}%
\bibitem [{\citenamefont {Cates}\ and\ \citenamefont
  {Tailleur}(2015)}]{cates15a}%
  \BibitemOpen
  \bibfield  {author} {\bibinfo {author} {\bibfnamefont {M.}~\bibnamefont
  {Cates}}\ and\ \bibinfo {author} {\bibfnamefont {J.}~\bibnamefont
  {Tailleur}},\ }\href {\doibase 10.1209/0295-5075/108/56004} {\bibfield
  {journal} {\bibinfo  {journal} {Annu. Rev. Cond. Mat. Phys.}\ }\textbf
  {\bibinfo {volume} {6}},\ \bibinfo {pages} {219} (\bibinfo {year}
  {2015})}\BibitemShut {NoStop}%
\bibitem [{\citenamefont {Bechinger}\ \emph {et~al.}(2016)\citenamefont
  {Bechinger}, \citenamefont {Di~Leonardo}, \citenamefont {L{\"o}wen},
  \citenamefont {Reichhardt}, \citenamefont {Volpe},\ and\ \citenamefont
  {Volpe}}]{bechinger2016active}%
  \BibitemOpen
  \bibfield  {author} {\bibinfo {author} {\bibfnamefont {C.}~\bibnamefont
  {Bechinger}}, \bibinfo {author} {\bibfnamefont {R.}~\bibnamefont
  {Di~Leonardo}}, \bibinfo {author} {\bibfnamefont {H.}~\bibnamefont
  {L{\"o}wen}}, \bibinfo {author} {\bibfnamefont {C.}~\bibnamefont
  {Reichhardt}}, \bibinfo {author} {\bibfnamefont {G.}~\bibnamefont {Volpe}}, \
  and\ \bibinfo {author} {\bibfnamefont {G.}~\bibnamefont {Volpe}},\ }\href
  {\doibase 10.1103/RevModPhys.88.045006} {\bibfield  {journal} {\bibinfo
  {journal} {Rev. Mod. Phys.}\ }\textbf {\bibinfo {volume} {88}},\ \bibinfo
  {pages} {045006} (\bibinfo {year} {2016})}\BibitemShut {NoStop}%
\bibitem [{\citenamefont {Z{\"o}ttl}\ and\ \citenamefont
  {Stark}(2016)}]{zottl2016emergent}%
  \BibitemOpen
  \bibfield  {author} {\bibinfo {author} {\bibfnamefont {A.}~\bibnamefont
  {Z{\"o}ttl}}\ and\ \bibinfo {author} {\bibfnamefont {H.}~\bibnamefont
  {Stark}},\ }\href {\doibase 10.1088/0953-8984/28/25/253001} {\bibfield
  {journal} {\bibinfo  {journal} {J. Phys. Cond. Mat.}\ }\textbf {\bibinfo
  {volume} {28}},\ \bibinfo {pages} {253001} (\bibinfo {year}
  {2016})}\BibitemShut {NoStop}%
\bibitem [{\citenamefont {Needleman}\ and\ \citenamefont
  {Dogic}(2017)}]{needleman2017active}%
  \BibitemOpen
  \bibfield  {author} {\bibinfo {author} {\bibfnamefont {D.}~\bibnamefont
  {Needleman}}\ and\ \bibinfo {author} {\bibfnamefont {Z.}~\bibnamefont
  {Dogic}},\ }\href {\doibase 10.1038/natrevmats.2017.48} {\bibfield  {journal}
  {\bibinfo  {journal} {Nat. Rev. Materials}\ }\textbf {\bibinfo {volume}
  {2}},\ \bibinfo {pages} {17048} (\bibinfo {year} {2017})}\BibitemShut
  {NoStop}%
\bibitem [{\citenamefont {Magar}\ \emph {et~al.}(2003)\citenamefont {Magar},
  \citenamefont {Goto},\ and\ \citenamefont {Pedley}}]{magar2003nutrient}%
  \BibitemOpen
  \bibfield  {author} {\bibinfo {author} {\bibfnamefont {V.}~\bibnamefont
  {Magar}}, \bibinfo {author} {\bibfnamefont {T.}~\bibnamefont {Goto}}, \ and\
  \bibinfo {author} {\bibfnamefont {T.~J.}\ \bibnamefont {Pedley}},\ }\href
  {\doibase 10.1093/qjmam/56.1.65} {\bibfield  {journal} {\bibinfo  {journal}
  {Q. J. Mech. Appl. Math.}\ }\textbf {\bibinfo {volume} {56}},\ \bibinfo
  {pages} {65} (\bibinfo {year} {2003})}\BibitemShut {NoStop}%
\bibitem [{\citenamefont {Short}\ \emph {et~al.}(2006)\citenamefont {Short},
  \citenamefont {Solari}, \citenamefont {Ganguly}, \citenamefont {Powers},
  \citenamefont {Kessler},\ and\ \citenamefont {Goldstein}}]{short2006flows}%
  \BibitemOpen
  \bibfield  {author} {\bibinfo {author} {\bibfnamefont {M.~B.}\ \bibnamefont
  {Short}}, \bibinfo {author} {\bibfnamefont {C.~A.}\ \bibnamefont {Solari}},
  \bibinfo {author} {\bibfnamefont {S.}~\bibnamefont {Ganguly}}, \bibinfo
  {author} {\bibfnamefont {T.~R.}\ \bibnamefont {Powers}}, \bibinfo {author}
  {\bibfnamefont {J.~O.}\ \bibnamefont {Kessler}}, \ and\ \bibinfo {author}
  {\bibfnamefont {R.~E.}\ \bibnamefont {Goldstein}},\ }\href {\doibase
  10.1073/pnas.0600566103} {\bibfield  {journal} {\bibinfo  {journal} {Proc.
  Natl. Acad. Sci.}\ }\textbf {\bibinfo {volume} {103}},\ \bibinfo {pages}
  {8315} (\bibinfo {year} {2006})}\BibitemShut {NoStop}%
\bibitem [{\citenamefont {Michelin}\ and\ \citenamefont
  {Lauga}(2011)}]{michelin2011optimal}%
  \BibitemOpen
  \bibfield  {author} {\bibinfo {author} {\bibfnamefont {S.}~\bibnamefont
  {Michelin}}\ and\ \bibinfo {author} {\bibfnamefont {E.}~\bibnamefont
  {Lauga}},\ }\href {\doibase 10.1063/1.3642645} {\bibfield  {journal}
  {\bibinfo  {journal} {Phys. Fluids}\ }\textbf {\bibinfo {volume} {23}},\
  \bibinfo {pages} {101901} (\bibinfo {year} {2011})}\BibitemShut {NoStop}%
\bibitem [{\citenamefont {Tam}\ and\ \citenamefont
  {Hosoi}(2011)}]{tam2011optimal}%
  \BibitemOpen
  \bibfield  {author} {\bibinfo {author} {\bibfnamefont {D.}~\bibnamefont
  {Tam}}\ and\ \bibinfo {author} {\bibfnamefont {A.~E.}\ \bibnamefont
  {Hosoi}},\ }\href {\doibase 10.1073/pnas.1011185108} {\bibfield  {journal}
  {\bibinfo  {journal} {Proc. Natl. Acad. Sci.}\ }\textbf {\bibinfo {volume}
  {108}},\ \bibinfo {pages} {1001} (\bibinfo {year} {2011})}\BibitemShut
  {NoStop}%
\bibitem [{\citenamefont {Tuval}\ \emph {et~al.}(2005)\citenamefont {Tuval},
  \citenamefont {Cisneros}, \citenamefont {Dombrowski}, \citenamefont
  {Wolgemuth}, \citenamefont {Kessler},\ and\ \citenamefont
  {Goldstein}}]{tuval2005bacterial}%
  \BibitemOpen
  \bibfield  {author} {\bibinfo {author} {\bibfnamefont {I.}~\bibnamefont
  {Tuval}}, \bibinfo {author} {\bibfnamefont {L.}~\bibnamefont {Cisneros}},
  \bibinfo {author} {\bibfnamefont {C.}~\bibnamefont {Dombrowski}}, \bibinfo
  {author} {\bibfnamefont {C.~W.}\ \bibnamefont {Wolgemuth}}, \bibinfo {author}
  {\bibfnamefont {J.~O.}\ \bibnamefont {Kessler}}, \ and\ \bibinfo {author}
  {\bibfnamefont {R.~E.}\ \bibnamefont {Goldstein}},\ }\href {\doibase
  10.1073/pnas.0406724102} {\bibfield  {journal} {\bibinfo  {journal} {Proc.
  Nat. Acad. Sci.}\ }\textbf {\bibinfo {volume} {102}},\ \bibinfo {pages}
  {2277} (\bibinfo {year} {2005})}\BibitemShut {NoStop}%
\bibitem [{\citenamefont {Wu}\ and\ \citenamefont
  {Libchaber}(2000)}]{wu2000particle}%
  \BibitemOpen
  \bibfield  {author} {\bibinfo {author} {\bibfnamefont {X.-L.}\ \bibnamefont
  {Wu}}\ and\ \bibinfo {author} {\bibfnamefont {A.}~\bibnamefont {Libchaber}},\
  }\href {\doibase 10.1103/physrevlett.84.3017} {\bibfield  {journal} {\bibinfo
   {journal} {Phys. Rev. Lett.}\ }\textbf {\bibinfo {volume} {84}},\ \bibinfo
  {pages} {3017} (\bibinfo {year} {2000})}\BibitemShut {NoStop}%
\bibitem [{\citenamefont {Kim}\ and\ \citenamefont
  {Breuer}(2004)}]{kim2004enhanced}%
  \BibitemOpen
  \bibfield  {author} {\bibinfo {author} {\bibfnamefont {M.~J.}\ \bibnamefont
  {Kim}}\ and\ \bibinfo {author} {\bibfnamefont {K.~S.}\ \bibnamefont
  {Breuer}},\ }\href {\doibase 10.1063/1.1787527} {\bibfield  {journal}
  {\bibinfo  {journal} {Phys. Fluids}\ }\textbf {\bibinfo {volume} {16}},\
  \bibinfo {pages} {L78} (\bibinfo {year} {2004})}\BibitemShut {NoStop}%
\bibitem [{\citenamefont {Thiffeault}\ and\ \citenamefont
  {Childress}(2010)}]{thiffeault2010stirring}%
  \BibitemOpen
  \bibfield  {author} {\bibinfo {author} {\bibfnamefont {J.-L.}\ \bibnamefont
  {Thiffeault}}\ and\ \bibinfo {author} {\bibfnamefont {S.}~\bibnamefont
  {Childress}},\ }\href {\doibase 10.1016/j.physleta.2010.06.043} {\bibfield
  {journal} {\bibinfo  {journal} {Phys. Lett. A}\ }\textbf {\bibinfo {volume}
  {374}},\ \bibinfo {pages} {3487} (\bibinfo {year} {2010})}\BibitemShut
  {NoStop}%
\bibitem [{\citenamefont {Pedley}\ \emph {et~al.}(1988)\citenamefont {Pedley},
  \citenamefont {Hill},\ and\ \citenamefont {Kessler}}]{pedley1988growth}%
  \BibitemOpen
  \bibfield  {author} {\bibinfo {author} {\bibfnamefont {T.}~\bibnamefont
  {Pedley}}, \bibinfo {author} {\bibfnamefont {N.}~\bibnamefont {Hill}}, \ and\
  \bibinfo {author} {\bibfnamefont {J.}~\bibnamefont {Kessler}},\ }\href
  {\doibase 10.1017/S0022112088002393} {\bibfield  {journal} {\bibinfo
  {journal} {J. Fluid. Mech.}\ }\textbf {\bibinfo {volume} {195}},\ \bibinfo
  {pages} {223} (\bibinfo {year} {1988})}\BibitemShut {NoStop}%
\bibitem [{\citenamefont {Hill}\ and\ \citenamefont
  {Pedley}(2005)}]{hill2005bioconvection}%
  \BibitemOpen
  \bibfield  {author} {\bibinfo {author} {\bibfnamefont {N.}~\bibnamefont
  {Hill}}\ and\ \bibinfo {author} {\bibfnamefont {T.~J.}\ \bibnamefont
  {Pedley}},\ }\href {\doibase 10.1016/j.fluiddyn.2005.03.002} {\bibfield
  {journal} {\bibinfo  {journal} {Fluid Dyn. Res.}\ }\textbf {\bibinfo {volume}
  {37}},\ \bibinfo {pages} {1} (\bibinfo {year} {2005})}\BibitemShut {NoStop}%
\bibitem [{\citenamefont {Karimi}\ and\ \citenamefont
  {Ardekani}(2013)}]{karimi2013gyrotactic}%
  \BibitemOpen
  \bibfield  {author} {\bibinfo {author} {\bibfnamefont {A.}~\bibnamefont
  {Karimi}}\ and\ \bibinfo {author} {\bibfnamefont {A.}~\bibnamefont
  {Ardekani}},\ }\href {\doibase 10.1017/jfm.2013.415} {\bibfield  {journal}
  {\bibinfo  {journal} {J. Fluid Mech.}\ }\textbf {\bibinfo {volume} {733}},\
  \bibinfo {pages} {245} (\bibinfo {year} {2013})}\BibitemShut {NoStop}%
\bibitem [{\citenamefont {Pushkin}\ \emph {et~al.}(2013)\citenamefont
  {Pushkin}, \citenamefont {Shum},\ and\ \citenamefont
  {Yeomans}}]{pushkin2012fluid}%
  \BibitemOpen
  \bibfield  {author} {\bibinfo {author} {\bibfnamefont {D.~O.}\ \bibnamefont
  {Pushkin}}, \bibinfo {author} {\bibfnamefont {H.}~\bibnamefont {Shum}}, \
  and\ \bibinfo {author} {\bibfnamefont {J.~M.}\ \bibnamefont {Yeomans}},\
  }\href {\doibase 10.1017/jfm.2013.208} {\bibfield  {journal} {\bibinfo
  {journal} {J. Fluid Mech.}\ }\textbf {\bibinfo {volume} {726}},\ \bibinfo
  {pages} {5} (\bibinfo {year} {2013})}\BibitemShut {NoStop}%
\bibitem [{\citenamefont {Jeanneret}\ \emph {et~al.}(2016)\citenamefont
  {Jeanneret}, \citenamefont {Pushkin}, \citenamefont {Kantsler},\ and\
  \citenamefont {Polin}}]{jeanneret2016entrainment}%
  \BibitemOpen
  \bibfield  {author} {\bibinfo {author} {\bibfnamefont {R.}~\bibnamefont
  {Jeanneret}}, \bibinfo {author} {\bibfnamefont {D.~O.}\ \bibnamefont
  {Pushkin}}, \bibinfo {author} {\bibfnamefont {V.}~\bibnamefont {Kantsler}}, \
  and\ \bibinfo {author} {\bibfnamefont {M.}~\bibnamefont {Polin}},\ }\href
  {\doibase 10.1038/ncomms12518} {\bibfield  {journal} {\bibinfo  {journal}
  {Nat. Comm.}\ }\textbf {\bibinfo {volume} {7}},\ \bibinfo {pages} {12518}
  (\bibinfo {year} {2016})}\BibitemShut {NoStop}%
\bibitem [{\citenamefont {Mathijssen}\ \emph
  {et~al.}(2018{\natexlab{a}})\citenamefont {Mathijssen}, \citenamefont
  {Jeanneret},\ and\ \citenamefont {Polin}}]{mathijssen2018universal}%
  \BibitemOpen
  \bibfield  {author} {\bibinfo {author} {\bibfnamefont {A.~J. T.~M.}\
  \bibnamefont {Mathijssen}}, \bibinfo {author} {\bibfnamefont
  {R.}~\bibnamefont {Jeanneret}}, \ and\ \bibinfo {author} {\bibfnamefont
  {M.}~\bibnamefont {Polin}},\ }\href {\doibase 10.1103/PhysRevFluids.3.033103}
  {\bibfield  {journal} {\bibinfo  {journal} {Phys. Rev. Fluids}\ }\textbf
  {\bibinfo {volume} {3}},\ \bibinfo {pages} {033103} (\bibinfo {year}
  {2018}{\natexlab{a}})}\BibitemShut {NoStop}%
\bibitem [{\citenamefont {Vaccari}\ \emph {et~al.}(2018)\citenamefont
  {Vaccari}, \citenamefont {Molaei}, \citenamefont {Leheny},\ and\
  \citenamefont {Stebe}}]{vaccari2018cargo}%
  \BibitemOpen
  \bibfield  {author} {\bibinfo {author} {\bibfnamefont {L.}~\bibnamefont
  {Vaccari}}, \bibinfo {author} {\bibfnamefont {M.}~\bibnamefont {Molaei}},
  \bibinfo {author} {\bibfnamefont {R.~L.}\ \bibnamefont {Leheny}}, \ and\
  \bibinfo {author} {\bibfnamefont {K.~J.}\ \bibnamefont {Stebe}},\ }\href
  {\doibase 10.1039/C8SM00481A} {\bibfield  {journal} {\bibinfo  {journal}
  {Soft Matter}\ ,\ \bibinfo {pages} {C8SM00481A}} (\bibinfo {year}
  {2018})}\BibitemShut {NoStop}%
\bibitem [{\citenamefont {Berke}\ \emph {et~al.}(2008)\citenamefont {Berke},
  \citenamefont {Turner}, \citenamefont {Berg},\ and\ \citenamefont
  {Lauga}}]{berke2008hydrodynamic}%
  \BibitemOpen
  \bibfield  {author} {\bibinfo {author} {\bibfnamefont {A.~P.}\ \bibnamefont
  {Berke}}, \bibinfo {author} {\bibfnamefont {L.}~\bibnamefont {Turner}},
  \bibinfo {author} {\bibfnamefont {H.~C.}\ \bibnamefont {Berg}}, \ and\
  \bibinfo {author} {\bibfnamefont {E.}~\bibnamefont {Lauga}},\ }\href
  {\doibase 10.1103/PhysRevLett.101.038102} {\bibfield  {journal} {\bibinfo
  {journal} {Phys. Rev. Lett.}\ }\textbf {\bibinfo {volume} {101}},\ \bibinfo
  {pages} {038102} (\bibinfo {year} {2008})}\BibitemShut {NoStop}%
\bibitem [{\citenamefont {Li}\ and\ \citenamefont
  {Tang}(2009)}]{li2009accumulation}%
  \BibitemOpen
  \bibfield  {author} {\bibinfo {author} {\bibfnamefont {G.}~\bibnamefont
  {Li}}\ and\ \bibinfo {author} {\bibfnamefont {J.~X.}\ \bibnamefont {Tang}},\
  }\href {\doibase 10.1103/PhysRevLett.103.078101} {\bibfield  {journal}
  {\bibinfo  {journal} {Phys. Rev. Lett.}\ }\textbf {\bibinfo {volume} {103}},\
  \bibinfo {pages} {078101} (\bibinfo {year} {2009})}\BibitemShut {NoStop}%
\bibitem [{\citenamefont {Molaei}\ \emph {et~al.}(2014)\citenamefont {Molaei},
  \citenamefont {Barry}, \citenamefont {Stocker},\ and\ \citenamefont
  {Sheng}}]{molaei2014failed}%
  \BibitemOpen
  \bibfield  {author} {\bibinfo {author} {\bibfnamefont {M.}~\bibnamefont
  {Molaei}}, \bibinfo {author} {\bibfnamefont {M.}~\bibnamefont {Barry}},
  \bibinfo {author} {\bibfnamefont {R.}~\bibnamefont {Stocker}}, \ and\
  \bibinfo {author} {\bibfnamefont {J.}~\bibnamefont {Sheng}},\ }\href
  {\doibase 10.1103/PhysRevLett.113.068103} {\bibfield  {journal} {\bibinfo
  {journal} {Phys. Rev. Lett.}\ }\textbf {\bibinfo {volume} {113}},\ \bibinfo
  {pages} {068103} (\bibinfo {year} {2014})}\BibitemShut {NoStop}%
\bibitem [{\citenamefont {Sipos}\ \emph {et~al.}(2015)\citenamefont {Sipos},
  \citenamefont {Nagy}, \citenamefont {Di~Leonardo},\ and\ \citenamefont
  {Galajda}}]{sipos2015hydro}%
  \BibitemOpen
  \bibfield  {author} {\bibinfo {author} {\bibfnamefont {O.}~\bibnamefont
  {Sipos}}, \bibinfo {author} {\bibfnamefont {K.}~\bibnamefont {Nagy}},
  \bibinfo {author} {\bibfnamefont {R.}~\bibnamefont {Di~Leonardo}}, \ and\
  \bibinfo {author} {\bibfnamefont {P.}~\bibnamefont {Galajda}},\ }\href
  {\doibase 10.1103/PhysRevLett.114.258104} {\bibfield  {journal} {\bibinfo
  {journal} {Phys. Rev. Lett.}\ }\textbf {\bibinfo {volume} {114}},\ \bibinfo
  {pages} {258104} (\bibinfo {year} {2015})}\BibitemShut {NoStop}%
\bibitem [{\citenamefont {Elgeti}\ and\ \citenamefont
  {Gompper}(2015)}]{elgeti2015run}%
  \BibitemOpen
  \bibfield  {author} {\bibinfo {author} {\bibfnamefont {J.}~\bibnamefont
  {Elgeti}}\ and\ \bibinfo {author} {\bibfnamefont {G.}~\bibnamefont
  {Gompper}},\ }\href {\doibase 10.1209/0295-5075/109/58003} {\bibfield
  {journal} {\bibinfo  {journal} {Eur. Phys. Lett.}\ }\textbf {\bibinfo
  {volume} {109}},\ \bibinfo {pages} {58003} (\bibinfo {year}
  {2015})}\BibitemShut {NoStop}%
\bibitem [{\citenamefont {Mathijssen}\ \emph {et~al.}(2016)\citenamefont
  {Mathijssen}, \citenamefont {Doostmohammadi}, \citenamefont {Yeomans},\ and\
  \citenamefont {Shendruk}}]{mathijssen2015hotspots}%
  \BibitemOpen
  \bibfield  {author} {\bibinfo {author} {\bibfnamefont {A.~J. T.~M.}\
  \bibnamefont {Mathijssen}}, \bibinfo {author} {\bibfnamefont
  {A.}~\bibnamefont {Doostmohammadi}}, \bibinfo {author} {\bibfnamefont
  {J.~M.}\ \bibnamefont {Yeomans}}, \ and\ \bibinfo {author} {\bibfnamefont
  {T.~N.}\ \bibnamefont {Shendruk}},\ }\href {\doibase 10.1098/rsif.2015.0936}
  {\bibfield  {journal} {\bibinfo  {journal} {J. R. Soc. Interface}\ }\textbf
  {\bibinfo {volume} {13}},\ \bibinfo {pages} {20150936} (\bibinfo {year}
  {2016})}\BibitemShut {NoStop}%
\bibitem [{\citenamefont {Figueroa-Morales}\ \emph {et~al.}(2015)\citenamefont
  {Figueroa-Morales}, \citenamefont {Mi{\~n}o}, \citenamefont {Rivera},
  \citenamefont {Caballero}, \citenamefont {Cl{\'e}ment}, \citenamefont
  {Altshuler},\ and\ \citenamefont {Lindner}}]{figueroa2015living}%
  \BibitemOpen
  \bibfield  {author} {\bibinfo {author} {\bibfnamefont {N.}~\bibnamefont
  {Figueroa-Morales}}, \bibinfo {author} {\bibfnamefont {G.}~\bibnamefont
  {Mi{\~n}o}}, \bibinfo {author} {\bibfnamefont {A.}~\bibnamefont {Rivera}},
  \bibinfo {author} {\bibfnamefont {R.}~\bibnamefont {Caballero}}, \bibinfo
  {author} {\bibfnamefont {E.}~\bibnamefont {Cl{\'e}ment}}, \bibinfo {author}
  {\bibfnamefont {E.}~\bibnamefont {Altshuler}}, \ and\ \bibinfo {author}
  {\bibfnamefont {A.}~\bibnamefont {Lindner}},\ }\href {\doibase
  10.1039/C5SM00939A} {\bibfield  {journal} {\bibinfo  {journal} {Soft Matter}\
  }\textbf {\bibinfo {volume} {11}},\ \bibinfo {pages} {6284} (\bibinfo {year}
  {2015})}\BibitemShut {NoStop}%
\bibitem [{\citenamefont {Jin}\ \emph {et~al.}(2018)\citenamefont {Jin},
  \citenamefont {Hokmabad}, \citenamefont {Baldwin},\ and\ \citenamefont
  {Maass}}]{jin2018chemotactic}%
  \BibitemOpen
  \bibfield  {author} {\bibinfo {author} {\bibfnamefont {C.}~\bibnamefont
  {Jin}}, \bibinfo {author} {\bibfnamefont {B.~V.}\ \bibnamefont {Hokmabad}},
  \bibinfo {author} {\bibfnamefont {K.~A.}\ \bibnamefont {Baldwin}}, \ and\
  \bibinfo {author} {\bibfnamefont {C.~C.}\ \bibnamefont {Maass}},\ }\href
  {\doibase 10.1088/1361-648X/aaa208} {\bibfield  {journal} {\bibinfo
  {journal} {J. Phys. Cond. Matt.}\ }\textbf {\bibinfo {volume} {30}},\
  \bibinfo {pages} {054003} (\bibinfo {year} {2018})}\BibitemShut {NoStop}%
\bibitem [{\citenamefont {Daddi-Moussa-Ider}\ \emph {et~al.}(2018)\citenamefont
  {Daddi-Moussa-Ider}, \citenamefont {Lisicki}, \citenamefont {Mathijssen},
  \citenamefont {Hoell}, \citenamefont {Goh}, \citenamefont {B{\l}awzdziewicz},
  \citenamefont {Menzel},\ and\ \citenamefont {L{\"o}wen}}]{daddi2018state}%
  \BibitemOpen
  \bibfield  {author} {\bibinfo {author} {\bibfnamefont {A.}~\bibnamefont
  {Daddi-Moussa-Ider}}, \bibinfo {author} {\bibfnamefont {M.}~\bibnamefont
  {Lisicki}}, \bibinfo {author} {\bibfnamefont {A.~J. T.~M.}\ \bibnamefont
  {Mathijssen}}, \bibinfo {author} {\bibfnamefont {C.}~\bibnamefont {Hoell}},
  \bibinfo {author} {\bibfnamefont {S.}~\bibnamefont {Goh}}, \bibinfo {author}
  {\bibfnamefont {J.}~\bibnamefont {B{\l}awzdziewicz}}, \bibinfo {author}
  {\bibfnamefont {A.~M.}\ \bibnamefont {Menzel}}, \ and\ \bibinfo {author}
  {\bibfnamefont {H.}~\bibnamefont {L{\"o}wen}},\ }\href {\doibase
  10.1088/1361-648X/aac470} {\bibfield  {journal} {\bibinfo  {journal} {J.
  Phys. Cond. Mat.}\ }\textbf {\bibinfo {volume} {30}},\ \bibinfo {pages}
  {254004} (\bibinfo {year} {2018})}\BibitemShut {NoStop}%
\bibitem [{\citenamefont {Ohmura}\ \emph {et~al.}(2018)\citenamefont {Ohmura},
  \citenamefont {Nishigami}, \citenamefont {Taniguchi}, \citenamefont {Nonaka},
  \citenamefont {Manabe}, \citenamefont {Ishikawa},\ and\ \citenamefont
  {Ichikawa}}]{ohmura2018simple}%
  \BibitemOpen
  \bibfield  {author} {\bibinfo {author} {\bibfnamefont {T.}~\bibnamefont
  {Ohmura}}, \bibinfo {author} {\bibfnamefont {Y.}~\bibnamefont {Nishigami}},
  \bibinfo {author} {\bibfnamefont {A.}~\bibnamefont {Taniguchi}}, \bibinfo
  {author} {\bibfnamefont {S.}~\bibnamefont {Nonaka}}, \bibinfo {author}
  {\bibfnamefont {J.}~\bibnamefont {Manabe}}, \bibinfo {author} {\bibfnamefont
  {T.}~\bibnamefont {Ishikawa}}, \ and\ \bibinfo {author} {\bibfnamefont
  {M.}~\bibnamefont {Ichikawa}},\ }\href {\doibase 10.1073/pnas.1718294115}
  {\bibfield  {journal} {\bibinfo  {journal} {Proc. Nat. Acad. Sci.}\ }\textbf
  {\bibinfo {volume} {115}},\ \bibinfo {pages} {3231} (\bibinfo {year}
  {2018})}\BibitemShut {NoStop}%
\bibitem [{\citenamefont {Mathijssen}\ \emph
  {et~al.}(2018{\natexlab{b}})\citenamefont {Mathijssen}, \citenamefont
  {Figueroa-Morales}, \citenamefont {Junot}, \citenamefont {Cl{\'e}ment},
  \citenamefont {Lindner},\ and\ \citenamefont
  {Z{\"o}ttl}}]{mathijssen2018oscillatory}%
  \BibitemOpen
  \bibfield  {author} {\bibinfo {author} {\bibfnamefont {A.~J. T.~M.}\
  \bibnamefont {Mathijssen}}, \bibinfo {author} {\bibfnamefont
  {N.}~\bibnamefont {Figueroa-Morales}}, \bibinfo {author} {\bibfnamefont
  {G.}~\bibnamefont {Junot}}, \bibinfo {author} {\bibfnamefont
  {E.}~\bibnamefont {Cl{\'e}ment}}, \bibinfo {author} {\bibfnamefont
  {A.}~\bibnamefont {Lindner}}, \ and\ \bibinfo {author} {\bibfnamefont
  {A.}~\bibnamefont {Z{\"o}ttl}},\ }\href {https://arxiv.org/abs/1803.01743}
  {\bibfield  {journal} {\bibinfo  {journal} {arXiv:1803.01743}\ } (\bibinfo
  {year} {2018}{\natexlab{b}})}\BibitemShut {NoStop}%
\bibitem [{\citenamefont {Mathijssen}\ \emph
  {et~al.}(2015{\natexlab{a}})\citenamefont {Mathijssen}, \citenamefont
  {Pushkin},\ and\ \citenamefont {Yeomans}}]{mathijssen2015tracer}%
  \BibitemOpen
  \bibfield  {author} {\bibinfo {author} {\bibfnamefont {A.~J. T.~M.}\
  \bibnamefont {Mathijssen}}, \bibinfo {author} {\bibfnamefont {D.~O.}\
  \bibnamefont {Pushkin}}, \ and\ \bibinfo {author} {\bibfnamefont {J.~M.}\
  \bibnamefont {Yeomans}},\ }\href {\doibase 10.1017/jfm.2015.269} {\bibfield
  {journal} {\bibinfo  {journal} {J. Fluid Mech.}\ }\textbf {\bibinfo {volume}
  {773}},\ \bibinfo {pages} {498 } (\bibinfo {year}
  {2015}{\natexlab{a}})}\BibitemShut {NoStop}%
\bibitem [{\citenamefont {Toner}\ and\ \citenamefont
  {Tu}(1995)}]{tonertu1995long}%
  \BibitemOpen
  \bibfield  {author} {\bibinfo {author} {\bibfnamefont {J.}~\bibnamefont
  {Toner}}\ and\ \bibinfo {author} {\bibfnamefont {Y.}~\bibnamefont {Tu}},\
  }\href {\doibase 10.1103/PhysRevLett.75.4326} {\bibfield  {journal} {\bibinfo
   {journal} {Phys. Rev. Lett.}\ }\textbf {\bibinfo {volume} {75}},\ \bibinfo
  {pages} {4326} (\bibinfo {year} {1995})}\BibitemShut {NoStop}%
\bibitem [{\citenamefont {Riedel}\ \emph {et~al.}(2005)\citenamefont {Riedel},
  \citenamefont {Kruse},\ and\ \citenamefont {Howard}}]{riedel2005self}%
  \BibitemOpen
  \bibfield  {author} {\bibinfo {author} {\bibfnamefont {I.~H.}\ \bibnamefont
  {Riedel}}, \bibinfo {author} {\bibfnamefont {K.}~\bibnamefont {Kruse}}, \
  and\ \bibinfo {author} {\bibfnamefont {J.}~\bibnamefont {Howard}},\ }\href
  {\doibase 10.1126/science.1110329} {\bibfield  {journal} {\bibinfo  {journal}
  {Science}\ }\textbf {\bibinfo {volume} {309}},\ \bibinfo {pages} {300}
  (\bibinfo {year} {2005})}\BibitemShut {NoStop}%
\bibitem [{\citenamefont {Lushi}\ \emph {et~al.}(2014)\citenamefont {Lushi},
  \citenamefont {Wioland},\ and\ \citenamefont {Goldstein}}]{lushi2014fluid}%
  \BibitemOpen
  \bibfield  {author} {\bibinfo {author} {\bibfnamefont {E.}~\bibnamefont
  {Lushi}}, \bibinfo {author} {\bibfnamefont {H.}~\bibnamefont {Wioland}}, \
  and\ \bibinfo {author} {\bibfnamefont {R.~E.}\ \bibnamefont {Goldstein}},\
  }\href {\doibase 10.1073/pnas.1405698111} {\bibfield  {journal} {\bibinfo
  {journal} {Proc. Nat. Acad. Sci.}\ }\textbf {\bibinfo {volume} {111}},\
  \bibinfo {pages} {9733} (\bibinfo {year} {2014})}\BibitemShut {NoStop}%
\bibitem [{\citenamefont {Ingham}\ and\ \citenamefont
  {Jacob}(2008)}]{ingham2008swarming}%
  \BibitemOpen
  \bibfield  {author} {\bibinfo {author} {\bibfnamefont {C.~J.}\ \bibnamefont
  {Ingham}}\ and\ \bibinfo {author} {\bibfnamefont {E.~B.}\ \bibnamefont
  {Jacob}},\ }\href {\doibase 10.1186/1471-2180-8-36} {\bibfield  {journal}
  {\bibinfo  {journal} {BMC Microbiol.}\ }\textbf {\bibinfo {volume} {8}},\
  \bibinfo {pages} {36} (\bibinfo {year} {2008})}\BibitemShut {NoStop}%
\bibitem [{\citenamefont {Wioland}\ \emph {et~al.}(2013)\citenamefont
  {Wioland}, \citenamefont {Woodhouse}, \citenamefont {Dunkel}, \citenamefont
  {Kessler},\ and\ \citenamefont {Goldstein}}]{wioland2013confinement}%
  \BibitemOpen
  \bibfield  {author} {\bibinfo {author} {\bibfnamefont {H.}~\bibnamefont
  {Wioland}}, \bibinfo {author} {\bibfnamefont {F.~G.}\ \bibnamefont
  {Woodhouse}}, \bibinfo {author} {\bibfnamefont {J.}~\bibnamefont {Dunkel}},
  \bibinfo {author} {\bibfnamefont {J.~O.}\ \bibnamefont {Kessler}}, \ and\
  \bibinfo {author} {\bibfnamefont {R.~E.}\ \bibnamefont {Goldstein}},\ }\href
  {\doibase 10.1103/PhysRevLett.110.268102} {\bibfield  {journal} {\bibinfo
  {journal} {Phys. Rev. Lett.}\ }\textbf {\bibinfo {volume} {110}},\ \bibinfo
  {pages} {268102} (\bibinfo {year} {2013})}\BibitemShut {NoStop}%
\bibitem [{\citenamefont {Wioland}\ \emph {et~al.}(2016)\citenamefont
  {Wioland}, \citenamefont {Woodhouse}, \citenamefont {Dunkel},\ and\
  \citenamefont {Goldstein}}]{wioland2016ferromagnetic}%
  \BibitemOpen
  \bibfield  {author} {\bibinfo {author} {\bibfnamefont {H.}~\bibnamefont
  {Wioland}}, \bibinfo {author} {\bibfnamefont {F.~G.}\ \bibnamefont
  {Woodhouse}}, \bibinfo {author} {\bibfnamefont {J.}~\bibnamefont {Dunkel}}, \
  and\ \bibinfo {author} {\bibfnamefont {R.~E.}\ \bibnamefont {Goldstein}},\
  }\href {\doibase 10.1038/nphys3607} {\bibfield  {journal} {\bibinfo
  {journal} {Nat. Phys.}\ }\textbf {\bibinfo {volume} {12}},\ \bibinfo {pages}
  {341} (\bibinfo {year} {2016})}\BibitemShut {NoStop}%
\bibitem [{\citenamefont {Dombrowski}\ \emph {et~al.}(2004)\citenamefont
  {Dombrowski}, \citenamefont {Cisneros}, \citenamefont {Chatkaew},
  \citenamefont {Goldstein},\ and\ \citenamefont
  {Kessler}}]{dombrowski2004self}%
  \BibitemOpen
  \bibfield  {author} {\bibinfo {author} {\bibfnamefont {C.}~\bibnamefont
  {Dombrowski}}, \bibinfo {author} {\bibfnamefont {L.}~\bibnamefont
  {Cisneros}}, \bibinfo {author} {\bibfnamefont {S.}~\bibnamefont {Chatkaew}},
  \bibinfo {author} {\bibfnamefont {R.~E.}\ \bibnamefont {Goldstein}}, \ and\
  \bibinfo {author} {\bibfnamefont {J.~O.}\ \bibnamefont {Kessler}},\ }\href
  {\doibase 10.1103/PhysRevLett.93.098103} {\bibfield  {journal} {\bibinfo
  {journal} {Phys. Rev. Lett.}\ }\textbf {\bibinfo {volume} {93}},\ \bibinfo
  {pages} {098103} (\bibinfo {year} {2004})}\BibitemShut {NoStop}%
\bibitem [{\citenamefont {Sokolov}\ \emph {et~al.}(2007)\citenamefont
  {Sokolov}, \citenamefont {Aranson}, \citenamefont {Kessler},\ and\
  \citenamefont {Goldstein}}]{sokolov2007concentration}%
  \BibitemOpen
  \bibfield  {author} {\bibinfo {author} {\bibfnamefont {A.}~\bibnamefont
  {Sokolov}}, \bibinfo {author} {\bibfnamefont {I.}~\bibnamefont {Aranson}},
  \bibinfo {author} {\bibfnamefont {J.}~\bibnamefont {Kessler}}, \ and\
  \bibinfo {author} {\bibfnamefont {R.}~\bibnamefont {Goldstein}},\ }\href
  {\doibase 10.1103/physrevlett.98.158102} {\bibfield  {journal} {\bibinfo
  {journal} {Phys. Rev. Lett.}\ }\textbf {\bibinfo {volume} {98}},\ \bibinfo
  {pages} {158102} (\bibinfo {year} {2007})}\BibitemShut {NoStop}%
\bibitem [{\citenamefont {Zhang}\ \emph {et~al.}(2010)\citenamefont {Zhang},
  \citenamefont {Be’er}, \citenamefont {Florin},\ and\ \citenamefont
  {Swinney}}]{zhang2010collective}%
  \BibitemOpen
  \bibfield  {author} {\bibinfo {author} {\bibfnamefont {H.-P.}\ \bibnamefont
  {Zhang}}, \bibinfo {author} {\bibfnamefont {A.}~\bibnamefont {Be’er}},
  \bibinfo {author} {\bibfnamefont {E.-L.}\ \bibnamefont {Florin}}, \ and\
  \bibinfo {author} {\bibfnamefont {H.~L.}\ \bibnamefont {Swinney}},\ }\href
  {\doibase 10.1073/pnas.1001651107} {\bibfield  {journal} {\bibinfo  {journal}
  {Proc. Nat. Acad. Sci.}\ }\textbf {\bibinfo {volume} {107}},\ \bibinfo
  {pages} {13626} (\bibinfo {year} {2010})}\BibitemShut {NoStop}%
\bibitem [{\citenamefont {Cisneros}\ \emph {et~al.}(2010)\citenamefont
  {Cisneros}, \citenamefont {Cortez}, \citenamefont {Dombrowski}, \citenamefont
  {Goldstein},\ and\ \citenamefont {Kessler}}]{cisneros2010fluid}%
  \BibitemOpen
  \bibfield  {author} {\bibinfo {author} {\bibfnamefont {L.~H.}\ \bibnamefont
  {Cisneros}}, \bibinfo {author} {\bibfnamefont {R.}~\bibnamefont {Cortez}},
  \bibinfo {author} {\bibfnamefont {C.}~\bibnamefont {Dombrowski}}, \bibinfo
  {author} {\bibfnamefont {R.~E.}\ \bibnamefont {Goldstein}}, \ and\ \bibinfo
  {author} {\bibfnamefont {J.~O.}\ \bibnamefont {Kessler}},\ }\href {\doibase
  10.1007/978-3-642-11633-9_10} {\bibfield  {journal} {\bibinfo  {journal}
  {Anim. Locomot.}\ }\textbf {\bibinfo {volume} {43}},\ \bibinfo {pages} {737}
  (\bibinfo {year} {2010})}\BibitemShut {NoStop}%
\bibitem [{\citenamefont {Sokolov}\ and\ \citenamefont
  {Aranson}(2012)}]{sokolov2012physical}%
  \BibitemOpen
  \bibfield  {author} {\bibinfo {author} {\bibfnamefont {A.}~\bibnamefont
  {Sokolov}}\ and\ \bibinfo {author} {\bibfnamefont {I.~S.}\ \bibnamefont
  {Aranson}},\ }\href {\doibase 10.1103/PhysRevLett.109.248109} {\bibfield
  {journal} {\bibinfo  {journal} {Phys. Rev. Lett.}\ }\textbf {\bibinfo
  {volume} {109}},\ \bibinfo {pages} {248109} (\bibinfo {year}
  {2012})}\BibitemShut {NoStop}%
\bibitem [{\citenamefont {Dunkel}\ \emph {et~al.}(2013)\citenamefont {Dunkel},
  \citenamefont {Heidenreich}, \citenamefont {Drescher}, \citenamefont
  {Wensink}, \citenamefont {B\"ar},\ and\ \citenamefont
  {Goldstein}}]{dunkel2013fluid}%
  \BibitemOpen
  \bibfield  {author} {\bibinfo {author} {\bibfnamefont {J.}~\bibnamefont
  {Dunkel}}, \bibinfo {author} {\bibfnamefont {S.}~\bibnamefont {Heidenreich}},
  \bibinfo {author} {\bibfnamefont {K.}~\bibnamefont {Drescher}}, \bibinfo
  {author} {\bibfnamefont {H.~H.}\ \bibnamefont {Wensink}}, \bibinfo {author}
  {\bibfnamefont {M.}~\bibnamefont {B\"ar}}, \ and\ \bibinfo {author}
  {\bibfnamefont {R.~E.}\ \bibnamefont {Goldstein}},\ }\href {\doibase
  10.1103/PhysRevLett.110.228102} {\bibfield  {journal} {\bibinfo  {journal}
  {Phys. Rev. Lett.}\ }\textbf {\bibinfo {volume} {110}},\ \bibinfo {pages}
  {228102} (\bibinfo {year} {2013})}\BibitemShut {NoStop}%
\bibitem [{\citenamefont {Wensink}\ \emph {et~al.}(2012)\citenamefont
  {Wensink}, \citenamefont {Dunkel}, \citenamefont {Heidenreich}, \citenamefont
  {Drescher}, \citenamefont {Goldstein}, \citenamefont {L{\"o}wen},\ and\
  \citenamefont {Yeomans}}]{wensink2012meso}%
  \BibitemOpen
  \bibfield  {author} {\bibinfo {author} {\bibfnamefont {H.~H.}\ \bibnamefont
  {Wensink}}, \bibinfo {author} {\bibfnamefont {J.}~\bibnamefont {Dunkel}},
  \bibinfo {author} {\bibfnamefont {S.}~\bibnamefont {Heidenreich}}, \bibinfo
  {author} {\bibfnamefont {K.}~\bibnamefont {Drescher}}, \bibinfo {author}
  {\bibfnamefont {R.~E.}\ \bibnamefont {Goldstein}}, \bibinfo {author}
  {\bibfnamefont {H.}~\bibnamefont {L{\"o}wen}}, \ and\ \bibinfo {author}
  {\bibfnamefont {J.~M.}\ \bibnamefont {Yeomans}},\ }\href {\doibase
  10.1073/pnas.1202032109} {\bibfield  {journal} {\bibinfo  {journal} {Proc.
  Nat. Acad. Sci.}\ }\textbf {\bibinfo {volume} {109}},\ \bibinfo {pages}
  {14308} (\bibinfo {year} {2012})}\BibitemShut {NoStop}%
\bibitem [{\citenamefont {Narayan}\ \emph {et~al.}(2007)\citenamefont
  {Narayan}, \citenamefont {Ramaswamy},\ and\ \citenamefont
  {Menon}}]{narayan2007long}%
  \BibitemOpen
  \bibfield  {author} {\bibinfo {author} {\bibfnamefont {V.}~\bibnamefont
  {Narayan}}, \bibinfo {author} {\bibfnamefont {S.}~\bibnamefont {Ramaswamy}},
  \ and\ \bibinfo {author} {\bibfnamefont {N.}~\bibnamefont {Menon}},\ }\href
  {\doibase 10.1126/science.1140414} {\bibfield  {journal} {\bibinfo  {journal}
  {Science}\ }\textbf {\bibinfo {volume} {317}},\ \bibinfo {pages} {105}
  (\bibinfo {year} {2007})}\BibitemShut {NoStop}%
\bibitem [{\citenamefont {Fily}\ and\ \citenamefont
  {Marchetti}(2012)}]{yaouen2012athermal}%
  \BibitemOpen
  \bibfield  {author} {\bibinfo {author} {\bibfnamefont {Y.}~\bibnamefont
  {Fily}}\ and\ \bibinfo {author} {\bibfnamefont {M.~C.}\ \bibnamefont
  {Marchetti}},\ }\href {\doibase 10.1103/PhysRevLett.108.235702} {\bibfield
  {journal} {\bibinfo  {journal} {Phys. Rev. Lett.}\ }\textbf {\bibinfo
  {volume} {108}},\ \bibinfo {pages} {235702} (\bibinfo {year}
  {2012})}\BibitemShut {NoStop}%
\bibitem [{\citenamefont {Buttinoni}\ \emph {et~al.}(2013)\citenamefont
  {Buttinoni}, \citenamefont {Bialk\'e}, \citenamefont {K\"ummel},
  \citenamefont {L\"owen}, \citenamefont {Bechinger},\ and\ \citenamefont
  {Speck}}]{buttinoni2013dynamical}%
  \BibitemOpen
  \bibfield  {author} {\bibinfo {author} {\bibfnamefont {I.}~\bibnamefont
  {Buttinoni}}, \bibinfo {author} {\bibfnamefont {J.}~\bibnamefont {Bialk\'e}},
  \bibinfo {author} {\bibfnamefont {F.}~\bibnamefont {K\"ummel}}, \bibinfo
  {author} {\bibfnamefont {H.}~\bibnamefont {L\"owen}}, \bibinfo {author}
  {\bibfnamefont {C.}~\bibnamefont {Bechinger}}, \ and\ \bibinfo {author}
  {\bibfnamefont {T.}~\bibnamefont {Speck}},\ }\href {\doibase
  10.1103/PhysRevLett.110.238301} {\bibfield  {journal} {\bibinfo  {journal}
  {Phys. Rev. Lett.}\ }\textbf {\bibinfo {volume} {110}},\ \bibinfo {pages}
  {238301} (\bibinfo {year} {2013})}\BibitemShut {NoStop}%
\bibitem [{\citenamefont {Sep\'ulveda}\ and\ \citenamefont
  {Soto}(2017)}]{sepulveda2017wetting}%
  \BibitemOpen
  \bibfield  {author} {\bibinfo {author} {\bibfnamefont {N.}~\bibnamefont
  {Sep\'ulveda}}\ and\ \bibinfo {author} {\bibfnamefont {R.}~\bibnamefont
  {Soto}},\ }\href {\doibase 10.1103/PhysRevLett.119.078001} {\bibfield
  {journal} {\bibinfo  {journal} {Phys. Rev. Lett.}\ }\textbf {\bibinfo
  {volume} {119}},\ \bibinfo {pages} {078001} (\bibinfo {year}
  {2017})}\BibitemShut {NoStop}%
\bibitem [{\citenamefont {Reichhardt}\ and\ \citenamefont
  {Reichhardt}(2018)}]{reichhardt2018clogging}%
  \BibitemOpen
  \bibfield  {author} {\bibinfo {author} {\bibfnamefont {C.}~\bibnamefont
  {Reichhardt}}\ and\ \bibinfo {author} {\bibfnamefont {C.}~\bibnamefont
  {Reichhardt}},\ }\href {http://arxiv.org/abs/1803.08992} {\bibfield
  {journal} {\bibinfo  {journal} {arXiv preprint 1803.08992}\ } (\bibinfo
  {year} {2018})}\BibitemShut {NoStop}%
\bibitem [{\citenamefont {T\'oth}\ \emph {et~al.}(2002)\citenamefont {T\'oth},
  \citenamefont {Denniston},\ and\ \citenamefont
  {Yeomans}}]{toth2002hydrodynamics}%
  \BibitemOpen
  \bibfield  {author} {\bibinfo {author} {\bibfnamefont {G.}~\bibnamefont
  {T\'oth}}, \bibinfo {author} {\bibfnamefont {C.}~\bibnamefont {Denniston}}, \
  and\ \bibinfo {author} {\bibfnamefont {J.~M.}\ \bibnamefont {Yeomans}},\
  }\href {\doibase 10.1103/PhysRevLett.88.105504} {\bibfield  {journal}
  {\bibinfo  {journal} {Phys. Rev. Lett.}\ }\textbf {\bibinfo {volume} {88}},\
  \bibinfo {pages} {105504} (\bibinfo {year} {2002})}\BibitemShut {NoStop}%
\bibitem [{\citenamefont {Elgeti}\ \emph {et~al.}(2011)\citenamefont {Elgeti},
  \citenamefont {Cates},\ and\ \citenamefont {Marenduzzo}}]{elgeti2011defect}%
  \BibitemOpen
  \bibfield  {author} {\bibinfo {author} {\bibfnamefont {J.}~\bibnamefont
  {Elgeti}}, \bibinfo {author} {\bibfnamefont {M.}~\bibnamefont {Cates}}, \
  and\ \bibinfo {author} {\bibfnamefont {D.}~\bibnamefont {Marenduzzo}},\
  }\href {\doibase 10.1039/C0SM01097A} {\bibfield  {journal} {\bibinfo
  {journal} {Soft Matter}\ }\textbf {\bibinfo {volume} {7}},\ \bibinfo {pages}
  {3177} (\bibinfo {year} {2011})}\BibitemShut {NoStop}%
\bibitem [{\citenamefont {Sanchez}\ \emph {et~al.}(2012)\citenamefont
  {Sanchez}, \citenamefont {Chen}, \citenamefont {DeCamp}, \citenamefont
  {Heymann},\ and\ \citenamefont {Dogic}}]{sanchez2012spontaneous}%
  \BibitemOpen
  \bibfield  {author} {\bibinfo {author} {\bibfnamefont {T.}~\bibnamefont
  {Sanchez}}, \bibinfo {author} {\bibfnamefont {D.~T.}\ \bibnamefont {Chen}},
  \bibinfo {author} {\bibfnamefont {S.~J.}\ \bibnamefont {DeCamp}}, \bibinfo
  {author} {\bibfnamefont {M.}~\bibnamefont {Heymann}}, \ and\ \bibinfo
  {author} {\bibfnamefont {Z.}~\bibnamefont {Dogic}},\ }\href {\doibase
  10.1038/nature11591} {\bibfield  {journal} {\bibinfo  {journal} {Nature}\
  }\textbf {\bibinfo {volume} {491}},\ \bibinfo {pages} {431} (\bibinfo {year}
  {2012})}\BibitemShut {NoStop}%
\bibitem [{\citenamefont {Keber}\ \emph {et~al.}(2014)\citenamefont {Keber},
  \citenamefont {Loiseau}, \citenamefont {Sanchez}, \citenamefont {DeCamp},
  \citenamefont {Giomi}, \citenamefont {Bowick}, \citenamefont {Marchetti},
  \citenamefont {Dogic},\ and\ \citenamefont {Bausch}}]{keber2014topology}%
  \BibitemOpen
  \bibfield  {author} {\bibinfo {author} {\bibfnamefont {F.~C.}\ \bibnamefont
  {Keber}}, \bibinfo {author} {\bibfnamefont {E.}~\bibnamefont {Loiseau}},
  \bibinfo {author} {\bibfnamefont {T.}~\bibnamefont {Sanchez}}, \bibinfo
  {author} {\bibfnamefont {S.~J.}\ \bibnamefont {DeCamp}}, \bibinfo {author}
  {\bibfnamefont {L.}~\bibnamefont {Giomi}}, \bibinfo {author} {\bibfnamefont
  {M.~J.}\ \bibnamefont {Bowick}}, \bibinfo {author} {\bibfnamefont {M.~C.}\
  \bibnamefont {Marchetti}}, \bibinfo {author} {\bibfnamefont {Z.}~\bibnamefont
  {Dogic}}, \ and\ \bibinfo {author} {\bibfnamefont {A.~R.}\ \bibnamefont
  {Bausch}},\ }\href {\doibase 10.1126/science.1254784} {\bibfield  {journal}
  {\bibinfo  {journal} {Science}\ }\textbf {\bibinfo {volume} {345}},\ \bibinfo
  {pages} {1135} (\bibinfo {year} {2014})}\BibitemShut {NoStop}%
\bibitem [{\citenamefont {Giomi}(2015)}]{giomi2015geometry}%
  \BibitemOpen
  \bibfield  {author} {\bibinfo {author} {\bibfnamefont {L.}~\bibnamefont
  {Giomi}},\ }\href {\doibase 10.1103/PhysRevX.5.031003} {\bibfield  {journal}
  {\bibinfo  {journal} {Phys. Rev. X}\ }\textbf {\bibinfo {volume} {5}},\
  \bibinfo {pages} {031003} (\bibinfo {year} {2015})}\BibitemShut {NoStop}%
\bibitem [{\citenamefont {Saw}\ \emph {et~al.}(2017)\citenamefont {Saw},
  \citenamefont {Doostmohammadi}, \citenamefont {Nier}, \citenamefont
  {Kocgozlu}, \citenamefont {Thampi}, \citenamefont {Toyama}, \citenamefont
  {Marcq}, \citenamefont {Lim}, \citenamefont {Yeomans},\ and\ \citenamefont
  {Ladoux}}]{saw2017topological}%
  \BibitemOpen
  \bibfield  {author} {\bibinfo {author} {\bibfnamefont {T.~B.}\ \bibnamefont
  {Saw}}, \bibinfo {author} {\bibfnamefont {A.}~\bibnamefont {Doostmohammadi}},
  \bibinfo {author} {\bibfnamefont {V.}~\bibnamefont {Nier}}, \bibinfo {author}
  {\bibfnamefont {L.}~\bibnamefont {Kocgozlu}}, \bibinfo {author}
  {\bibfnamefont {S.}~\bibnamefont {Thampi}}, \bibinfo {author} {\bibfnamefont
  {Y.}~\bibnamefont {Toyama}}, \bibinfo {author} {\bibfnamefont
  {P.}~\bibnamefont {Marcq}}, \bibinfo {author} {\bibfnamefont {C.~T.}\
  \bibnamefont {Lim}}, \bibinfo {author} {\bibfnamefont {J.~M.}\ \bibnamefont
  {Yeomans}}, \ and\ \bibinfo {author} {\bibfnamefont {B.}~\bibnamefont
  {Ladoux}},\ }\href {\doibase 10.1038/nature21718} {\bibfield  {journal}
  {\bibinfo  {journal} {Nature}\ }\textbf {\bibinfo {volume} {544}},\ \bibinfo
  {pages} {212} (\bibinfo {year} {2017})}\BibitemShut {NoStop}%
\bibitem [{\citenamefont {Drescher}\ \emph {et~al.}(2011)\citenamefont
  {Drescher}, \citenamefont {Dunkel}, \citenamefont {Cisneros}, \citenamefont
  {Ganguly},\ and\ \citenamefont {Goldstein}}]{drescher2011fluid}%
  \BibitemOpen
  \bibfield  {author} {\bibinfo {author} {\bibfnamefont {K.}~\bibnamefont
  {Drescher}}, \bibinfo {author} {\bibfnamefont {J.}~\bibnamefont {Dunkel}},
  \bibinfo {author} {\bibfnamefont {L.~H.}\ \bibnamefont {Cisneros}}, \bibinfo
  {author} {\bibfnamefont {S.}~\bibnamefont {Ganguly}}, \ and\ \bibinfo
  {author} {\bibfnamefont {R.~E.}\ \bibnamefont {Goldstein}},\ }\href {\doibase
  10.1073/pnas.1019079108} {\bibfield  {journal} {\bibinfo  {journal} {Proc.
  Nat. Acad. Sci.}\ }\textbf {\bibinfo {volume} {108}},\ \bibinfo {pages}
  {10940} (\bibinfo {year} {2011})}\BibitemShut {NoStop}%
\bibitem [{\citenamefont {Blake}(1971)}]{blake1971note}%
  \BibitemOpen
  \bibfield  {author} {\bibinfo {author} {\bibfnamefont {J.~R.}\ \bibnamefont
  {Blake}},\ }\href {\doibase 10.1017/S0305004100049902} {\bibfield  {journal}
  {\bibinfo  {journal} {Math. Proc. Camb. Phil. Soc}\ }\textbf {\bibinfo
  {volume} {70}},\ \bibinfo {pages} {303 } (\bibinfo {year}
  {1971})}\BibitemShut {NoStop}%
\bibitem [{\citenamefont {Mathijssen}\ \emph
  {et~al.}(2015{\natexlab{b}})\citenamefont {Mathijssen}, \citenamefont
  {Doostmohammadi}, \citenamefont {Yeomans},\ and\ \citenamefont
  {Shendruk}}]{mathijssen2015hydrodynamics}%
  \BibitemOpen
  \bibfield  {author} {\bibinfo {author} {\bibfnamefont {A.~J. T.~M.}\
  \bibnamefont {Mathijssen}}, \bibinfo {author} {\bibfnamefont
  {A.}~\bibnamefont {Doostmohammadi}}, \bibinfo {author} {\bibfnamefont
  {J.~M.}\ \bibnamefont {Yeomans}}, \ and\ \bibinfo {author} {\bibfnamefont
  {T.~N.}\ \bibnamefont {Shendruk}},\ }\href {\doibase 10.1017/jfm.2016.479}
  {\bibfield  {journal} {\bibinfo  {journal} {J. Fluid Mech.}\ }\textbf
  {\bibinfo {volume} {806}},\ \bibinfo {pages} {35} (\bibinfo {year}
  {2015}{\natexlab{b}})}\BibitemShut {NoStop}%
\bibitem [{\citenamefont {Steager}\ \emph {et~al.}(2007)\citenamefont
  {Steager}, \citenamefont {Kim}, \citenamefont {Patel}, \citenamefont {Bith},
  \citenamefont {Naik}, \citenamefont {Reber},\ and\ \citenamefont
  {Kim}}]{steager2007control}%
  \BibitemOpen
  \bibfield  {author} {\bibinfo {author} {\bibfnamefont {E.}~\bibnamefont
  {Steager}}, \bibinfo {author} {\bibfnamefont {C.-B.}\ \bibnamefont {Kim}},
  \bibinfo {author} {\bibfnamefont {J.}~\bibnamefont {Patel}}, \bibinfo
  {author} {\bibfnamefont {S.}~\bibnamefont {Bith}}, \bibinfo {author}
  {\bibfnamefont {C.}~\bibnamefont {Naik}}, \bibinfo {author} {\bibfnamefont
  {L.}~\bibnamefont {Reber}}, \ and\ \bibinfo {author} {\bibfnamefont {M.~J.}\
  \bibnamefont {Kim}},\ }\href {\doibase 10.1063/1.2752721} {\bibfield
  {journal} {\bibinfo  {journal} {Appl. Phys. Lett.}\ }\textbf {\bibinfo
  {volume} {90}},\ \bibinfo {pages} {263901} (\bibinfo {year}
  {2007})}\BibitemShut {NoStop}%
\bibitem [{\citenamefont {Palacci}\ \emph {et~al.}(2013)\citenamefont
  {Palacci}, \citenamefont {Sacanna}, \citenamefont {Steinberg}, \citenamefont
  {Pine},\ and\ \citenamefont {Chaikin}}]{palacci2013living}%
  \BibitemOpen
  \bibfield  {author} {\bibinfo {author} {\bibfnamefont {J.}~\bibnamefont
  {Palacci}}, \bibinfo {author} {\bibfnamefont {S.}~\bibnamefont {Sacanna}},
  \bibinfo {author} {\bibfnamefont {A.~P.}\ \bibnamefont {Steinberg}}, \bibinfo
  {author} {\bibfnamefont {D.~J.}\ \bibnamefont {Pine}}, \ and\ \bibinfo
  {author} {\bibfnamefont {P.~M.}\ \bibnamefont {Chaikin}},\ }\href {\doibase
  10.1126/science.1230020} {\bibfield  {journal} {\bibinfo  {journal}
  {Science}\ }\textbf {\bibinfo {volume} {339}},\ \bibinfo {pages} {936}
  (\bibinfo {year} {2013})}\BibitemShut {NoStop}%
\bibitem [{\citenamefont {Arlt}\ \emph {et~al.}(2018)\citenamefont {Arlt},
  \citenamefont {Martinez}, \citenamefont {Dawson}, \citenamefont {Pilizota},\
  and\ \citenamefont {Poon}}]{arlt2018painting}%
  \BibitemOpen
  \bibfield  {author} {\bibinfo {author} {\bibfnamefont {J.}~\bibnamefont
  {Arlt}}, \bibinfo {author} {\bibfnamefont {V.~A.}\ \bibnamefont {Martinez}},
  \bibinfo {author} {\bibfnamefont {A.}~\bibnamefont {Dawson}}, \bibinfo
  {author} {\bibfnamefont {T.}~\bibnamefont {Pilizota}}, \ and\ \bibinfo
  {author} {\bibfnamefont {W.~C.}\ \bibnamefont {Poon}},\ }\href {\doibase
  10.1038/s41467-018-03161-8} {\bibfield  {journal} {\bibinfo  {journal} {Nat.
  Comm.}\ }\textbf {\bibinfo {volume} {9}},\ \bibinfo {pages} {768} (\bibinfo
  {year} {2018})}\BibitemShut {NoStop}%
\bibitem [{\citenamefont {Frangipane}\ \emph {et~al.}(2018)\citenamefont
  {Frangipane}, \citenamefont {Dell'Arciprete}, \citenamefont {Petracchini},
  \citenamefont {Maggi}, \citenamefont {Saglimbeni}, \citenamefont {Bianchi},
  \citenamefont {Vizsnyiczai}, \citenamefont {Bernardini},\ and\ \citenamefont
  {Di~Leonardo}}]{frangipane2018dynamic}%
  \BibitemOpen
  \bibfield  {author} {\bibinfo {author} {\bibfnamefont {G.}~\bibnamefont
  {Frangipane}}, \bibinfo {author} {\bibfnamefont {D.}~\bibnamefont
  {Dell'Arciprete}}, \bibinfo {author} {\bibfnamefont {S.}~\bibnamefont
  {Petracchini}}, \bibinfo {author} {\bibfnamefont {C.}~\bibnamefont {Maggi}},
  \bibinfo {author} {\bibfnamefont {F.}~\bibnamefont {Saglimbeni}}, \bibinfo
  {author} {\bibfnamefont {S.}~\bibnamefont {Bianchi}}, \bibinfo {author}
  {\bibfnamefont {G.}~\bibnamefont {Vizsnyiczai}}, \bibinfo {author}
  {\bibfnamefont {M.~L.}\ \bibnamefont {Bernardini}}, \ and\ \bibinfo {author}
  {\bibfnamefont {R.}~\bibnamefont {Di~Leonardo}},\ }\href {\doibase
  10.7554/eLife.36608.001} {\bibfield  {journal} {\bibinfo  {journal}
  {e{L}ife}\ }\textbf {\bibinfo {volume} {7}},\ \bibinfo {pages} {e36608}
  (\bibinfo {year} {2018})}\BibitemShut {NoStop}%
\bibitem [{\citenamefont {de~Gennes}\ and\ \citenamefont
  {Prost}(1993)}]{degennes1993physics}%
  \BibitemOpen
  \bibfield  {author} {\bibinfo {author} {\bibfnamefont {P.~G.}\ \bibnamefont
  {de~Gennes}}\ and\ \bibinfo {author} {\bibfnamefont {J.}~\bibnamefont
  {Prost}},\ }\href@noop {} {\emph {\bibinfo {title} {The physics of liquid
  crystals}}}\ (\bibinfo  {publisher} {Oxford University Press},\ \bibinfo
  {year} {1993})\BibitemShut {NoStop}%
\bibitem [{\citenamefont {Genkin}\ \emph {et~al.}(2017)\citenamefont {Genkin},
  \citenamefont {Sokolov}, \citenamefont {Lavrentovich},\ and\ \citenamefont
  {Aranson}}]{genkin2017topological}%
  \BibitemOpen
  \bibfield  {author} {\bibinfo {author} {\bibfnamefont {M.~M.}\ \bibnamefont
  {Genkin}}, \bibinfo {author} {\bibfnamefont {A.}~\bibnamefont {Sokolov}},
  \bibinfo {author} {\bibfnamefont {O.~D.}\ \bibnamefont {Lavrentovich}}, \
  and\ \bibinfo {author} {\bibfnamefont {I.~S.}\ \bibnamefont {Aranson}},\
  }\href {\doibase 10.1103/PhysRevX.7.011029} {\bibfield  {journal} {\bibinfo
  {journal} {Phys. Rev. X}\ }\textbf {\bibinfo {volume} {7}},\ \bibinfo {pages}
  {011029} (\bibinfo {year} {2017})}\BibitemShut {NoStop}%
\bibitem [{\citenamefont {Doostmohammadi}\ \emph {et~al.}(2016)\citenamefont
  {Doostmohammadi}, \citenamefont {Adamer}, \citenamefont {Thampi},\ and\
  \citenamefont {Yeomans}}]{doostmohammadi2016stabilization}%
  \BibitemOpen
  \bibfield  {author} {\bibinfo {author} {\bibfnamefont {A.}~\bibnamefont
  {Doostmohammadi}}, \bibinfo {author} {\bibfnamefont {M.~F.}\ \bibnamefont
  {Adamer}}, \bibinfo {author} {\bibfnamefont {S.~P.}\ \bibnamefont {Thampi}},
  \ and\ \bibinfo {author} {\bibfnamefont {J.~M.}\ \bibnamefont {Yeomans}},\
  }\href {\doibase 10.1038/ncomms10557} {\bibfield  {journal} {\bibinfo
  {journal} {Nature Comm.}\ }\textbf {\bibinfo {volume} {7}},\ \bibinfo {pages}
  {10557} (\bibinfo {year} {2016})}\BibitemShut {NoStop}%
\bibitem [{\citenamefont {Ishikawa}\ \emph {et~al.}(2011)\citenamefont
  {Ishikawa}, \citenamefont {Yoshida}, \citenamefont {Ueno}, \citenamefont
  {Wiedeman}, \citenamefont {Imai},\ and\ \citenamefont
  {Yamaguchi}}]{ishikawa2011energy}%
  \BibitemOpen
  \bibfield  {author} {\bibinfo {author} {\bibfnamefont {T.}~\bibnamefont
  {Ishikawa}}, \bibinfo {author} {\bibfnamefont {N.}~\bibnamefont {Yoshida}},
  \bibinfo {author} {\bibfnamefont {H.}~\bibnamefont {Ueno}}, \bibinfo {author}
  {\bibfnamefont {M.}~\bibnamefont {Wiedeman}}, \bibinfo {author}
  {\bibfnamefont {Y.}~\bibnamefont {Imai}}, \ and\ \bibinfo {author}
  {\bibfnamefont {T.}~\bibnamefont {Yamaguchi}},\ }\href {\doibase
  10.1103/PhysRevLett.107.028102} {\bibfield  {journal} {\bibinfo  {journal}
  {Phys. Rev. Lett.}\ }\textbf {\bibinfo {volume} {107}},\ \bibinfo {pages}
  {028102} (\bibinfo {year} {2011})}\BibitemShut {NoStop}%
\bibitem [{\citenamefont {Shendruk}\ \emph {et~al.}(2018)\citenamefont
  {Shendruk}, \citenamefont {Thijssen}, \citenamefont {Yeomans},\ and\
  \citenamefont {Doostmohammadi}}]{shendruk2018twist}%
  \BibitemOpen
  \bibfield  {author} {\bibinfo {author} {\bibfnamefont {T.~N.}\ \bibnamefont
  {Shendruk}}, \bibinfo {author} {\bibfnamefont {K.}~\bibnamefont {Thijssen}},
  \bibinfo {author} {\bibfnamefont {J.~M.}\ \bibnamefont {Yeomans}}, \ and\
  \bibinfo {author} {\bibfnamefont {A.}~\bibnamefont {Doostmohammadi}},\ }\href
  {\doibase 10.1103/PhysRevE.98.010601} {\bibfield  {journal} {\bibinfo
  {journal} {Phys. Rev. E}\ }\textbf {\bibinfo {volume} {98}},\ \bibinfo
  {pages} {010601} (\bibinfo {year} {2018})}\BibitemShut {NoStop}%
\bibitem [{\citenamefont {Elgeti}\ and\ \citenamefont
  {Gompper}(2013)}]{elgeti2013emergence}%
  \BibitemOpen
  \bibfield  {author} {\bibinfo {author} {\bibfnamefont {J.}~\bibnamefont
  {Elgeti}}\ and\ \bibinfo {author} {\bibfnamefont {G.}~\bibnamefont
  {Gompper}},\ }\href {\doibase 10.1073/pnas.1218869110} {\bibfield  {journal}
  {\bibinfo  {journal} {Proc. Nat. Acad. Sci.}\ }\textbf {\bibinfo {volume}
  {110}},\ \bibinfo {pages} {4470} (\bibinfo {year} {2013})}\BibitemShut
  {NoStop}%
\bibitem [{\citenamefont {Ding}\ \emph {et~al.}(2014)\citenamefont {Ding},
  \citenamefont {Nawroth}, \citenamefont {McFall-Ngai},\ and\ \citenamefont
  {Kanso}}]{ding2014mixing}%
  \BibitemOpen
  \bibfield  {author} {\bibinfo {author} {\bibfnamefont {Y.}~\bibnamefont
  {Ding}}, \bibinfo {author} {\bibfnamefont {J.~C.}\ \bibnamefont {Nawroth}},
  \bibinfo {author} {\bibfnamefont {M.~J.}\ \bibnamefont {McFall-Ngai}}, \ and\
  \bibinfo {author} {\bibfnamefont {E.}~\bibnamefont {Kanso}},\ }\href
  {\doibase 10.1017/jfm.2014.36} {\bibfield  {journal} {\bibinfo  {journal} {J.
  Fluid Mech.}\ }\textbf {\bibinfo {volume} {743}},\ \bibinfo {pages} {124}
  (\bibinfo {year} {2014})}\BibitemShut {NoStop}%
\bibitem [{\citenamefont {Uchida}\ and\ \citenamefont
  {Golestanian}(2010{\natexlab{a}})}]{uchida2010synchronization}%
  \BibitemOpen
  \bibfield  {author} {\bibinfo {author} {\bibfnamefont {N.}~\bibnamefont
  {Uchida}}\ and\ \bibinfo {author} {\bibfnamefont {R.}~\bibnamefont
  {Golestanian}},\ }\href {\doibase 10.1103/PhysRevLett.104.178103} {\bibfield
  {journal} {\bibinfo  {journal} {Phys. Rev. Lett.}\ }\textbf {\bibinfo
  {volume} {104}},\ \bibinfo {pages} {178103} (\bibinfo {year}
  {2010}{\natexlab{a}})}\BibitemShut {NoStop}%
\bibitem [{\citenamefont {Uchida}\ and\ \citenamefont
  {Golestanian}(2010{\natexlab{b}})}]{uchida2010bsynchronization}%
  \BibitemOpen
  \bibfield  {author} {\bibinfo {author} {\bibfnamefont {N.}~\bibnamefont
  {Uchida}}\ and\ \bibinfo {author} {\bibfnamefont {R.}~\bibnamefont
  {Golestanian}},\ }\href {\doibase 10.1209/0295-5075/89/50011} {\bibfield
  {journal} {\bibinfo  {journal} {Europhys. Lett.}\ }\textbf {\bibinfo {volume}
  {89}},\ \bibinfo {pages} {50011} (\bibinfo {year}
  {2010}{\natexlab{b}})}\BibitemShut {NoStop}%
\bibitem [{\citenamefont {Darnton}\ \emph {et~al.}(2004)\citenamefont
  {Darnton}, \citenamefont {Turner}, \citenamefont {Breuer},\ and\
  \citenamefont {Berg}}]{darnton2004moving}%
  \BibitemOpen
  \bibfield  {author} {\bibinfo {author} {\bibfnamefont {N.}~\bibnamefont
  {Darnton}}, \bibinfo {author} {\bibfnamefont {L.}~\bibnamefont {Turner}},
  \bibinfo {author} {\bibfnamefont {K.}~\bibnamefont {Breuer}}, \ and\ \bibinfo
  {author} {\bibfnamefont {H.~C.}\ \bibnamefont {Berg}},\ }\href {\doibase
  10.1016/S0006-3495(04)74253-8} {\bibfield  {journal} {\bibinfo  {journal}
  {Biophys. J.}\ }\textbf {\bibinfo {volume} {86}},\ \bibinfo {pages} {1863}
  (\bibinfo {year} {2004})}\BibitemShut {NoStop}%
\bibitem [{\citenamefont {Kim}\ and\ \citenamefont
  {Breuer}(2008)}]{kim2008microfluidic}%
  \BibitemOpen
  \bibfield  {author} {\bibinfo {author} {\bibfnamefont {M.~J.}\ \bibnamefont
  {Kim}}\ and\ \bibinfo {author} {\bibfnamefont {K.~S.}\ \bibnamefont
  {Breuer}},\ }\href {\doibase 10.1002/smll.200700641} {\bibfield  {journal}
  {\bibinfo  {journal} {Small}\ }\textbf {\bibinfo {volume} {4}},\ \bibinfo
  {pages} {111} (\bibinfo {year} {2008})}\BibitemShut {NoStop}%
\bibitem [{\citenamefont {Hsiao}\ \emph {et~al.}(2014)\citenamefont {Hsiao},
  \citenamefont {Wang}, \citenamefont {Wu}, \citenamefont {Tsai}, \citenamefont
  {Chang},\ and\ \citenamefont {Woon}}]{hsiao2014collective}%
  \BibitemOpen
  \bibfield  {author} {\bibinfo {author} {\bibfnamefont {Y.-T.}\ \bibnamefont
  {Hsiao}}, \bibinfo {author} {\bibfnamefont {J.-H.}\ \bibnamefont {Wang}},
  \bibinfo {author} {\bibfnamefont {K.-T.}\ \bibnamefont {Wu}}, \bibinfo
  {author} {\bibfnamefont {J.}~\bibnamefont {Tsai}}, \bibinfo {author}
  {\bibfnamefont {C.-H.}\ \bibnamefont {Chang}}, \ and\ \bibinfo {author}
  {\bibfnamefont {W.-Y.}\ \bibnamefont {Woon}},\ }\href {\doibase
  10.1063/1.4902111} {\bibfield  {journal} {\bibinfo  {journal} {Appl. Phys.
  Lett.}\ }\textbf {\bibinfo {volume} {105}},\ \bibinfo {pages} {203702}
  (\bibinfo {year} {2014})}\BibitemShut {NoStop}%
\bibitem [{\citenamefont {Hsiao}\ \emph {et~al.}(2016)\citenamefont {Hsiao},
  \citenamefont {Wu}, \citenamefont {Uchida},\ and\ \citenamefont
  {Woon}}]{hsiao2016impurity}%
  \BibitemOpen
  \bibfield  {author} {\bibinfo {author} {\bibfnamefont {Y.-T.}\ \bibnamefont
  {Hsiao}}, \bibinfo {author} {\bibfnamefont {K.-T.}\ \bibnamefont {Wu}},
  \bibinfo {author} {\bibfnamefont {N.}~\bibnamefont {Uchida}}, \ and\ \bibinfo
  {author} {\bibfnamefont {W.-Y.}\ \bibnamefont {Woon}},\ }\href {\doibase
  10.1063/1.4948766} {\bibfield  {journal} {\bibinfo  {journal} {Appl. Phys.
  Lett.}\ }\textbf {\bibinfo {volume} {108}},\ \bibinfo {pages} {183701}
  (\bibinfo {year} {2016})}\BibitemShut {NoStop}%
\bibitem [{\citenamefont {Marcos}\ \emph {et~al.}(2012)\citenamefont {Marcos},
  \citenamefont {Powers},\ and\ \citenamefont {Stocker}}]{marcos2012bacterial}%
  \BibitemOpen
  \bibfield  {author} {\bibinfo {author} {\bibfnamefont {H.~C.~F.}\
  \bibnamefont {Marcos}}, \bibinfo {author} {\bibfnamefont {T.~R.}\
  \bibnamefont {Powers}}, \ and\ \bibinfo {author} {\bibfnamefont
  {R.}~\bibnamefont {Stocker}},\ }\href {\doibase 10.1073/pnas.1120955109}
  {\bibfield  {journal} {\bibinfo  {journal} {Proc. Nat. Acad. Sci.}\ }\textbf
  {\bibinfo {volume} {109}},\ \bibinfo {pages} {4780} (\bibinfo {year}
  {2012})}\BibitemShut {NoStop}%
\bibitem [{\citenamefont {Aditi~Simha}\ and\ \citenamefont
  {Ramaswamy}(2002)}]{simha2002instabilities}%
  \BibitemOpen
  \bibfield  {author} {\bibinfo {author} {\bibfnamefont {R.}~\bibnamefont
  {Aditi~Simha}}\ and\ \bibinfo {author} {\bibfnamefont {S.}~\bibnamefont
  {Ramaswamy}},\ }\href {\doibase 10.1103/PhysRevLett.89.058101} {\bibfield
  {journal} {\bibinfo  {journal} {Phys. Rev. Lett.}\ }\textbf {\bibinfo
  {volume} {89}},\ \bibinfo {pages} {058101} (\bibinfo {year}
  {2002})}\BibitemShut {NoStop}%
\bibitem [{\citenamefont {Vladescu}\ \emph {et~al.}(2014)\citenamefont
  {Vladescu}, \citenamefont {Marsden}, \citenamefont {Schwarz-Linek},
  \citenamefont {Martinez}, \citenamefont {Arlt}, \citenamefont {Morozov},
  \citenamefont {Marenduzzo}, \citenamefont {Cates},\ and\ \citenamefont
  {Poon}}]{vladescu2014filling}%
  \BibitemOpen
  \bibfield  {author} {\bibinfo {author} {\bibfnamefont {I.}~\bibnamefont
  {Vladescu}}, \bibinfo {author} {\bibfnamefont {E.}~\bibnamefont {Marsden}},
  \bibinfo {author} {\bibfnamefont {J.}~\bibnamefont {Schwarz-Linek}}, \bibinfo
  {author} {\bibfnamefont {V.}~\bibnamefont {Martinez}}, \bibinfo {author}
  {\bibfnamefont {J.}~\bibnamefont {Arlt}}, \bibinfo {author} {\bibfnamefont
  {A.}~\bibnamefont {Morozov}}, \bibinfo {author} {\bibfnamefont
  {D.}~\bibnamefont {Marenduzzo}}, \bibinfo {author} {\bibfnamefont
  {M.}~\bibnamefont {Cates}}, \ and\ \bibinfo {author} {\bibfnamefont
  {W.}~\bibnamefont {Poon}},\ }\href {\doibase 10.1103/physrevlett.113.268101}
  {\bibfield  {journal} {\bibinfo  {journal} {Phys. Rev. Lett.}\ }\textbf
  {\bibinfo {volume} {113}},\ \bibinfo {pages} {268101} (\bibinfo {year}
  {2014})}\BibitemShut {NoStop}%
\bibitem [{\citenamefont {Janssen}\ \emph {et~al.}(2017)\citenamefont
  {Janssen}, \citenamefont {Kaiser},\ and\ \citenamefont
  {L{\"o}wen}}]{janssen2017aging}%
  \BibitemOpen
  \bibfield  {author} {\bibinfo {author} {\bibfnamefont {L.~M.}\ \bibnamefont
  {Janssen}}, \bibinfo {author} {\bibfnamefont {A.}~\bibnamefont {Kaiser}}, \
  and\ \bibinfo {author} {\bibfnamefont {H.}~\bibnamefont {L{\"o}wen}},\ }\href
  {\doibase 10.1038/s41598-017-05569-6} {\bibfield  {journal} {\bibinfo
  {journal} {Sci. Rep.}\ }\textbf {\bibinfo {volume} {7}},\ \bibinfo {pages}
  {5667} (\bibinfo {year} {2017})}\BibitemShut {NoStop}%
\bibitem [{\citenamefont {Galajda}\ \emph {et~al.}(2007)\citenamefont
  {Galajda}, \citenamefont {Keymer}, \citenamefont {Chaikin},\ and\
  \citenamefont {Austin}}]{galajda2007wall}%
  \BibitemOpen
  \bibfield  {author} {\bibinfo {author} {\bibfnamefont {P.}~\bibnamefont
  {Galajda}}, \bibinfo {author} {\bibfnamefont {J.}~\bibnamefont {Keymer}},
  \bibinfo {author} {\bibfnamefont {P.}~\bibnamefont {Chaikin}}, \ and\
  \bibinfo {author} {\bibfnamefont {R.}~\bibnamefont {Austin}},\ }\href
  {\doibase 10.1128/JB.01033-07} {\bibfield  {journal} {\bibinfo  {journal} {J.
  Bacteriol.}\ }\textbf {\bibinfo {volume} {189}},\ \bibinfo {pages} {8704}
  (\bibinfo {year} {2007})}\BibitemShut {NoStop}%
\bibitem [{\citenamefont {Wan}\ \emph {et~al.}(2008)\citenamefont {Wan},
  \citenamefont {Olson~Reichhardt}, \citenamefont {Nussinov},\ and\
  \citenamefont {Reichhardt}}]{wan2008rectification}%
  \BibitemOpen
  \bibfield  {author} {\bibinfo {author} {\bibfnamefont {M.~B.}\ \bibnamefont
  {Wan}}, \bibinfo {author} {\bibfnamefont {C.~J.}\ \bibnamefont
  {Olson~Reichhardt}}, \bibinfo {author} {\bibfnamefont {Z.}~\bibnamefont
  {Nussinov}}, \ and\ \bibinfo {author} {\bibfnamefont {C.}~\bibnamefont
  {Reichhardt}},\ }\href {\doibase 10.1103/PhysRevLett.101.018102} {\bibfield
  {journal} {\bibinfo  {journal} {Phys. Rev. Lett.}\ }\textbf {\bibinfo
  {volume} {101}},\ \bibinfo {pages} {018102} (\bibinfo {year}
  {2008})}\BibitemShut {NoStop}%
\bibitem [{\citenamefont {Koumakis}\ \emph {et~al.}(2013)\citenamefont
  {Koumakis}, \citenamefont {Lepore}, \citenamefont {Maggi},\ and\
  \citenamefont {Di~Leonardo}}]{koumakis2013targeted}%
  \BibitemOpen
  \bibfield  {author} {\bibinfo {author} {\bibfnamefont {N.}~\bibnamefont
  {Koumakis}}, \bibinfo {author} {\bibfnamefont {A.}~\bibnamefont {Lepore}},
  \bibinfo {author} {\bibfnamefont {C.}~\bibnamefont {Maggi}}, \ and\ \bibinfo
  {author} {\bibfnamefont {R.}~\bibnamefont {Di~Leonardo}},\ }\href {\doibase
  10.1038/ncomms3588} {\bibfield  {journal} {\bibinfo  {journal} {Nat. Comm.}\
  }\textbf {\bibinfo {volume} {4}},\ \bibinfo {pages} {2588} (\bibinfo {year}
  {2013})}\BibitemShut {NoStop}%
\bibitem [{\citenamefont {Simmchen}\ \emph {et~al.}(2016)\citenamefont
  {Simmchen}, \citenamefont {Katuri}, \citenamefont {Uspal}, \citenamefont
  {Popescu}, \citenamefont {Tasinkevych},\ and\ \citenamefont
  {S{\'a}nchez}}]{simmchen2016topographical}%
  \BibitemOpen
  \bibfield  {author} {\bibinfo {author} {\bibfnamefont {J.}~\bibnamefont
  {Simmchen}}, \bibinfo {author} {\bibfnamefont {J.}~\bibnamefont {Katuri}},
  \bibinfo {author} {\bibfnamefont {W.~E.}\ \bibnamefont {Uspal}}, \bibinfo
  {author} {\bibfnamefont {M.~N.}\ \bibnamefont {Popescu}}, \bibinfo {author}
  {\bibfnamefont {M.}~\bibnamefont {Tasinkevych}}, \ and\ \bibinfo {author}
  {\bibfnamefont {S.}~\bibnamefont {S{\'a}nchez}},\ }\href {\doibase
  10.1038/ncomms10598} {\bibfield  {journal} {\bibinfo  {journal} {Nat. Comm.}\
  }\textbf {\bibinfo {volume} {7}},\ \bibinfo {pages} {10598} (\bibinfo {year}
  {2016})}\BibitemShut {NoStop}%
\bibitem [{\citenamefont {Woodhouse}\ \emph {et~al.}(2016)\citenamefont
  {Woodhouse}, \citenamefont {Forrow}, \citenamefont {Fawcett},\ and\
  \citenamefont {Dunkel}}]{woodhouse2016stochastic}%
  \BibitemOpen
  \bibfield  {author} {\bibinfo {author} {\bibfnamefont {F.~G.}\ \bibnamefont
  {Woodhouse}}, \bibinfo {author} {\bibfnamefont {A.}~\bibnamefont {Forrow}},
  \bibinfo {author} {\bibfnamefont {J.~B.}\ \bibnamefont {Fawcett}}, \ and\
  \bibinfo {author} {\bibfnamefont {J.}~\bibnamefont {Dunkel}},\ }\href
  {\doibase 10.1073/pnas.1603351113} {\bibfield  {journal} {\bibinfo  {journal}
  {Proc. Nat. Acad. Sci.}\ }\textbf {\bibinfo {volume} {113}},\ \bibinfo
  {pages} {8200} (\bibinfo {year} {2016})}\BibitemShut {NoStop}%
\bibitem [{\citenamefont {Vidakovic}\ \emph {et~al.}(2018)\citenamefont
  {Vidakovic}, \citenamefont {Singh}, \citenamefont {Hartmann}, \citenamefont
  {Nadell},\ and\ \citenamefont {Drescher}}]{vidakovic2018dynamic}%
  \BibitemOpen
  \bibfield  {author} {\bibinfo {author} {\bibfnamefont {L.}~\bibnamefont
  {Vidakovic}}, \bibinfo {author} {\bibfnamefont {P.~K.}\ \bibnamefont
  {Singh}}, \bibinfo {author} {\bibfnamefont {R.}~\bibnamefont {Hartmann}},
  \bibinfo {author} {\bibfnamefont {C.~D.}\ \bibnamefont {Nadell}}, \ and\
  \bibinfo {author} {\bibfnamefont {K.}~\bibnamefont {Drescher}},\ }\href
  {\doibase 10.1038/s41564-017-0050-1} {\bibfield  {journal} {\bibinfo
  {journal} {Nat. Microbiol.}\ }\textbf {\bibinfo {volume} {3}},\ \bibinfo
  {pages} {26} (\bibinfo {year} {2018})}\BibitemShut {NoStop}%
\bibitem [{\citenamefont {Kim}\ and\ \citenamefont
  {Breuer}(2007)}]{kim2007controlled}%
  \BibitemOpen
  \bibfield  {author} {\bibinfo {author} {\bibfnamefont {M.~J.}\ \bibnamefont
  {Kim}}\ and\ \bibinfo {author} {\bibfnamefont {K.~S.}\ \bibnamefont
  {Breuer}},\ }\href {\doibase 10.1021/ac0614691} {\bibfield  {journal}
  {\bibinfo  {journal} {Anal. Chem.}\ }\textbf {\bibinfo {volume} {79}},\
  \bibinfo {pages} {955} (\bibinfo {year} {2007})}\BibitemShut {NoStop}%
\bibitem [{\citenamefont {Lin}\ \emph {et~al.}(2011)\citenamefont {Lin},
  \citenamefont {Thiffeault},\ and\ \citenamefont
  {Childress}}]{lin2011stirring}%
  \BibitemOpen
  \bibfield  {author} {\bibinfo {author} {\bibfnamefont {Z.}~\bibnamefont
  {Lin}}, \bibinfo {author} {\bibfnamefont {J.-L.}\ \bibnamefont {Thiffeault}},
  \ and\ \bibinfo {author} {\bibfnamefont {S.}~\bibnamefont {Childress}},\
  }\href {\doibase 10.1017/s002211201000563x} {\bibfield  {journal} {\bibinfo
  {journal} {J. Fluid Mech.}\ }\textbf {\bibinfo {volume} {669}},\ \bibinfo
  {pages} {167} (\bibinfo {year} {2011})}\BibitemShut {NoStop}%
\bibitem [{\citenamefont {Pushkin}\ and\ \citenamefont
  {Yeomans}(2013)}]{pushkin2013fluid}%
  \BibitemOpen
  \bibfield  {author} {\bibinfo {author} {\bibfnamefont {D.~O.}\ \bibnamefont
  {Pushkin}}\ and\ \bibinfo {author} {\bibfnamefont {J.~M.}\ \bibnamefont
  {Yeomans}},\ }\href {\doibase 10.1103/PhysRevLett.111.188101} {\bibfield
  {journal} {\bibinfo  {journal} {Phys. Rev. Lett.}\ }\textbf {\bibinfo
  {volume} {111}},\ \bibinfo {pages} {188101} (\bibinfo {year}
  {2013})}\BibitemShut {NoStop}%
\bibitem [{\citenamefont {Mueller}\ and\ \citenamefont
  {Thiffeault}(2017)}]{mueller2017fluid}%
  \BibitemOpen
  \bibfield  {author} {\bibinfo {author} {\bibfnamefont {P.}~\bibnamefont
  {Mueller}}\ and\ \bibinfo {author} {\bibfnamefont {J.-L.}\ \bibnamefont
  {Thiffeault}},\ }\href {\doibase 10.1103/PhysRevFluids.2.013103} {\bibfield
  {journal} {\bibinfo  {journal} {Phys. Rev. Fluids}\ }\textbf {\bibinfo
  {volume} {2}},\ \bibinfo {pages} {013103} (\bibinfo {year}
  {2017})}\BibitemShut {NoStop}%
\bibitem [{\citenamefont {Chandrasekhar}(1943)}]{chandrasekhar1943stochastic}%
  \BibitemOpen
  \bibfield  {author} {\bibinfo {author} {\bibfnamefont {S.}~\bibnamefont
  {Chandrasekhar}},\ }\href {\doibase 10.1103/RevModPhys.15.1} {\bibfield
  {journal} {\bibinfo  {journal} {Rev. Mod. Phys.}\ }\textbf {\bibinfo {volume}
  {15}},\ \bibinfo {pages} {1} (\bibinfo {year} {1943})}\BibitemShut {NoStop}%
\bibitem [{\citenamefont {King}\ \emph {et~al.}(2008)\citenamefont {King},
  \citenamefont {Westbrook}, \citenamefont {Young}, \citenamefont {Kuo},
  \citenamefont {Abedin}, \citenamefont {Chapman}, \citenamefont {Fairclough},
  \citenamefont {Hellsten}, \citenamefont {Isogai}, \citenamefont {Letunic}
  \emph {et~al.}}]{king2008genome}%
  \BibitemOpen
  \bibfield  {author} {\bibinfo {author} {\bibfnamefont {N.}~\bibnamefont
  {King}}, \bibinfo {author} {\bibfnamefont {M.~J.}\ \bibnamefont {Westbrook}},
  \bibinfo {author} {\bibfnamefont {S.~L.}\ \bibnamefont {Young}}, \bibinfo
  {author} {\bibfnamefont {A.}~\bibnamefont {Kuo}}, \bibinfo {author}
  {\bibfnamefont {M.}~\bibnamefont {Abedin}}, \bibinfo {author} {\bibfnamefont
  {J.}~\bibnamefont {Chapman}}, \bibinfo {author} {\bibfnamefont
  {S.}~\bibnamefont {Fairclough}}, \bibinfo {author} {\bibfnamefont
  {U.}~\bibnamefont {Hellsten}}, \bibinfo {author} {\bibfnamefont
  {Y.}~\bibnamefont {Isogai}}, \bibinfo {author} {\bibfnamefont
  {I.}~\bibnamefont {Letunic}},  \emph {et~al.},\ }\href {\doibase
  10.1038/nature06617} {\bibfield  {journal} {\bibinfo  {journal} {Nature}\
  }\textbf {\bibinfo {volume} {451}},\ \bibinfo {pages} {783} (\bibinfo {year}
  {2008})}\BibitemShut {NoStop}%
\bibitem [{\citenamefont {Roper}\ \emph {et~al.}(2013)\citenamefont {Roper},
  \citenamefont {Dayel}, \citenamefont {Pepper},\ and\ \citenamefont
  {Koehl}}]{roper2013coop}%
  \BibitemOpen
  \bibfield  {author} {\bibinfo {author} {\bibfnamefont {M.}~\bibnamefont
  {Roper}}, \bibinfo {author} {\bibfnamefont {M.~J.}\ \bibnamefont {Dayel}},
  \bibinfo {author} {\bibfnamefont {R.~E.}\ \bibnamefont {Pepper}}, \ and\
  \bibinfo {author} {\bibfnamefont {M.~A.~R.}\ \bibnamefont {Koehl}},\ }\href
  {\doibase 10.1103/PhysRevLett.110.228104} {\bibfield  {journal} {\bibinfo
  {journal} {Phys. Rev. Lett.}\ }\textbf {\bibinfo {volume} {110}},\ \bibinfo
  {pages} {228104} (\bibinfo {year} {2013})}\BibitemShut {NoStop}%
\bibitem [{\citenamefont {Nielsen}\ \emph {et~al.}(2017)\citenamefont
  {Nielsen}, \citenamefont {Asadzadeh}, \citenamefont {D{\"o}lger},
  \citenamefont {Walther}, \citenamefont {Ki{\o}rboe},\ and\ \citenamefont
  {Andersen}}]{nielsen2017hydrodynamics}%
  \BibitemOpen
  \bibfield  {author} {\bibinfo {author} {\bibfnamefont {L.~T.}\ \bibnamefont
  {Nielsen}}, \bibinfo {author} {\bibfnamefont {S.~S.}\ \bibnamefont
  {Asadzadeh}}, \bibinfo {author} {\bibfnamefont {J.}~\bibnamefont
  {D{\"o}lger}}, \bibinfo {author} {\bibfnamefont {J.~H.}\ \bibnamefont
  {Walther}}, \bibinfo {author} {\bibfnamefont {T.}~\bibnamefont {Ki{\o}rboe}},
  \ and\ \bibinfo {author} {\bibfnamefont {A.}~\bibnamefont {Andersen}},\
  }\href {\doibase 10.1073/pnas.1708873114} {\bibfield  {journal} {\bibinfo
  {journal} {Proc. Nat. Acad. Sci.}\ }\textbf {\bibinfo {volume} {114}},\
  \bibinfo {pages} {9373} (\bibinfo {year} {2017})}\BibitemShut {NoStop}%
\bibitem [{\citenamefont {Kirkegaard}\ and\ \citenamefont
  {Goldstein}(2016)}]{kirkegaard2016filter}%
  \BibitemOpen
  \bibfield  {author} {\bibinfo {author} {\bibfnamefont {J.~B.}\ \bibnamefont
  {Kirkegaard}}\ and\ \bibinfo {author} {\bibfnamefont {R.~E.}\ \bibnamefont
  {Goldstein}},\ }\href {\doibase 10.1103/PhysRevE.94.052401} {\bibfield
  {journal} {\bibinfo  {journal} {Phys. Rev. E}\ }\textbf {\bibinfo {volume}
  {94}},\ \bibinfo {pages} {052401} (\bibinfo {year} {2016})}\BibitemShut
  {NoStop}%
\bibitem [{\citenamefont {Kirkegaard}\ \emph {et~al.}(2016)\citenamefont
  {Kirkegaard}, \citenamefont {Marron},\ and\ \citenamefont
  {Goldstein}}]{kirkegaard2016motility}%
  \BibitemOpen
  \bibfield  {author} {\bibinfo {author} {\bibfnamefont {J.~B.}\ \bibnamefont
  {Kirkegaard}}, \bibinfo {author} {\bibfnamefont {A.~O.}\ \bibnamefont
  {Marron}}, \ and\ \bibinfo {author} {\bibfnamefont {R.~E.}\ \bibnamefont
  {Goldstein}},\ }\href {\doibase 10.1103/PhysRevLett.116.038102} {\bibfield
  {journal} {\bibinfo  {journal} {Phys. Rev. Lett.}\ }\textbf {\bibinfo
  {volume} {116}},\ \bibinfo {pages} {038102} (\bibinfo {year}
  {2016})}\BibitemShut {NoStop}%
\bibitem [{\citenamefont {D{\"o}lger}\ \emph {et~al.}(2017)\citenamefont
  {D{\"o}lger}, \citenamefont {Tor~Nielsen}, \citenamefont {Ki{\o}rboe},\ and\
  \citenamefont {Andersen}}]{dolger2017swimming}%
  \BibitemOpen
  \bibfield  {author} {\bibinfo {author} {\bibfnamefont {J.}~\bibnamefont
  {D{\"o}lger}}, \bibinfo {author} {\bibfnamefont {L.}~\bibnamefont
  {Tor~Nielsen}}, \bibinfo {author} {\bibfnamefont {T.}~\bibnamefont
  {Ki{\o}rboe}}, \ and\ \bibinfo {author} {\bibfnamefont {A.}~\bibnamefont
  {Andersen}},\ }\href {\doibase 10.1038/srep39892} {\bibfield  {journal}
  {\bibinfo  {journal} {Sci. Rep.}\ }\textbf {\bibinfo {volume} {7}},\ \bibinfo
  {pages} {39892} (\bibinfo {year} {2017})}\BibitemShut {NoStop}%
\bibitem [{\citenamefont {Gilpin}\ \emph {et~al.}(2017)\citenamefont {Gilpin},
  \citenamefont {Prakash},\ and\ \citenamefont {Prakash}}]{gilpin2017vortex}%
  \BibitemOpen
  \bibfield  {author} {\bibinfo {author} {\bibfnamefont {W.}~\bibnamefont
  {Gilpin}}, \bibinfo {author} {\bibfnamefont {V.~N.}\ \bibnamefont {Prakash}},
  \ and\ \bibinfo {author} {\bibfnamefont {M.}~\bibnamefont {Prakash}},\ }\href
  {\doibase 10.1038/nphys3981} {\bibfield  {journal} {\bibinfo  {journal} {Nat.
  Phys.}\ }\textbf {\bibinfo {volume} {13}},\ \bibinfo {pages} {380} (\bibinfo
  {year} {2017})}\BibitemShut {NoStop}%
\bibitem [{\citenamefont {Stocker}\ \emph {et~al.}(2008)\citenamefont
  {Stocker}, \citenamefont {Seymour}, \citenamefont {Samadani}, \citenamefont
  {Hunt},\ and\ \citenamefont {Polz}}]{stocker2008rapid}%
  \BibitemOpen
  \bibfield  {author} {\bibinfo {author} {\bibfnamefont {R.}~\bibnamefont
  {Stocker}}, \bibinfo {author} {\bibfnamefont {J.~R.}\ \bibnamefont
  {Seymour}}, \bibinfo {author} {\bibfnamefont {A.}~\bibnamefont {Samadani}},
  \bibinfo {author} {\bibfnamefont {D.~E.}\ \bibnamefont {Hunt}}, \ and\
  \bibinfo {author} {\bibfnamefont {M.~F.}\ \bibnamefont {Polz}},\ }\href
  {\doibase 10.1073/pnas.0709765105} {\bibfield  {journal} {\bibinfo  {journal}
  {Proc. Nat. Acad. Sci.}\ }\textbf {\bibinfo {volume} {105}},\ \bibinfo
  {pages} {4209} (\bibinfo {year} {2008})}\BibitemShut {NoStop}%
\bibitem [{\citenamefont {Eisenbach}(2004)}]{eisenbach2004chemotaxis}%
  \BibitemOpen
  \bibfield  {author} {\bibinfo {author} {\bibfnamefont {M.}~\bibnamefont
  {Eisenbach}},\ }\href {\doibase 10.1142/p303} {\emph {\bibinfo {title}
  {Chemotaxis}}}\ (\bibinfo  {publisher} {World Scientific},\ \bibinfo {year}
  {2004})\BibitemShut {NoStop}%
\bibitem [{\citenamefont {Wadhams}\ and\ \citenamefont
  {Armitage}(2004)}]{wadhams2004making}%
  \BibitemOpen
  \bibfield  {author} {\bibinfo {author} {\bibfnamefont {G.~H.}\ \bibnamefont
  {Wadhams}}\ and\ \bibinfo {author} {\bibfnamefont {J.~P.}\ \bibnamefont
  {Armitage}},\ }\href {\doibase 10.1038/nrm1524} {\bibfield  {journal}
  {\bibinfo  {journal} {Nat. Rev. Mol. Cell Biol.}\ }\textbf {\bibinfo {volume}
  {5}},\ \bibinfo {pages} {1024} (\bibinfo {year} {2004})}\BibitemShut
  {NoStop}%
\bibitem [{\citenamefont {Hall-Stoodley}\ \emph {et~al.}(2004)\citenamefont
  {Hall-Stoodley}, \citenamefont {Costerton},\ and\ \citenamefont
  {Stoodley}}]{hall-stoodley04}%
  \BibitemOpen
  \bibfield  {author} {\bibinfo {author} {\bibfnamefont {L.}~\bibnamefont
  {Hall-Stoodley}}, \bibinfo {author} {\bibfnamefont {J.~W.}\ \bibnamefont
  {Costerton}}, \ and\ \bibinfo {author} {\bibfnamefont {P.}~\bibnamefont
  {Stoodley}},\ }\href {\doibase 10.1038/nrmicro821} {\bibfield  {journal}
  {\bibinfo  {journal} {Nat. Rev. Microbiol.}\ }\textbf {\bibinfo {volume}
  {2}},\ \bibinfo {pages} {95} (\bibinfo {year} {2004})}\BibitemShut {NoStop}%
\bibitem [{\citenamefont {Stocker}\ and\ \citenamefont
  {Seymour}(2012)}]{stocker2012ecology}%
  \BibitemOpen
  \bibfield  {author} {\bibinfo {author} {\bibfnamefont {R.}~\bibnamefont
  {Stocker}}\ and\ \bibinfo {author} {\bibfnamefont {J.~R.}\ \bibnamefont
  {Seymour}},\ }\href {\doibase 10.1128/MMBR.00029-12} {\bibfield  {journal}
  {\bibinfo  {journal} {Microbiol. Molec. Biol. Rev.}\ }\textbf {\bibinfo
  {volume} {76}},\ \bibinfo {pages} {792} (\bibinfo {year} {2012})}\BibitemShut
  {NoStop}%
\bibitem [{\citenamefont {Romanczuk}\ \emph {et~al.}(2012)\citenamefont
  {Romanczuk}, \citenamefont {B{\"a}r}, \citenamefont {Ebeling}, \citenamefont
  {Lindner},\ and\ \citenamefont {Schimansky-Geier}}]{romanczuk2012active}%
  \BibitemOpen
  \bibfield  {author} {\bibinfo {author} {\bibfnamefont {P.}~\bibnamefont
  {Romanczuk}}, \bibinfo {author} {\bibfnamefont {M.}~\bibnamefont {B{\"a}r}},
  \bibinfo {author} {\bibfnamefont {W.}~\bibnamefont {Ebeling}}, \bibinfo
  {author} {\bibfnamefont {B.}~\bibnamefont {Lindner}}, \ and\ \bibinfo
  {author} {\bibfnamefont {L.}~\bibnamefont {Schimansky-Geier}},\ }\href
  {\doibase 10.1140/epjst/e2012-01529-y} {\bibfield  {journal} {\bibinfo
  {journal} {Eur. Phys. J. Spec. Top.}\ }\textbf {\bibinfo {volume} {202}},\
  \bibinfo {pages} {1} (\bibinfo {year} {2012})}\BibitemShut {NoStop}%
\bibitem [{\citenamefont {Ben-Jacob}\ \emph {et~al.}(2000)\citenamefont
  {Ben-Jacob}, \citenamefont {Cohen},\ and\ \citenamefont
  {Levine}}]{ben2000cooperative}%
  \BibitemOpen
  \bibfield  {author} {\bibinfo {author} {\bibfnamefont {E.}~\bibnamefont
  {Ben-Jacob}}, \bibinfo {author} {\bibfnamefont {I.}~\bibnamefont {Cohen}}, \
  and\ \bibinfo {author} {\bibfnamefont {H.}~\bibnamefont {Levine}},\ }\href
  {\doibase 10.1080/000187300405228} {\bibfield  {journal} {\bibinfo  {journal}
  {Adv. Phys.}\ }\textbf {\bibinfo {volume} {49}},\ \bibinfo {pages} {395}
  (\bibinfo {year} {2000})}\BibitemShut {NoStop}%
\bibitem [{\citenamefont {Berg}\ and\ \citenamefont
  {Brown}(1972)}]{berg1972chemotaxis}%
  \BibitemOpen
  \bibfield  {author} {\bibinfo {author} {\bibfnamefont {H.~C.}\ \bibnamefont
  {Berg}}\ and\ \bibinfo {author} {\bibfnamefont {D.~A.}\ \bibnamefont
  {Brown}},\ }\href {\doibase 10.1038/239500a0} {\bibfield  {journal} {\bibinfo
   {journal} {Nature}\ }\textbf {\bibinfo {volume} {239}},\ \bibinfo {pages}
  {500} (\bibinfo {year} {1972})}\BibitemShut {NoStop}%
\bibitem [{\citenamefont {Mittal}\ \emph {et~al.}(2003)\citenamefont {Mittal},
  \citenamefont {Budrene}, \citenamefont {Brenner},\ and\ \citenamefont
  {Van~Oudenaarden}}]{mittal2003motility}%
  \BibitemOpen
  \bibfield  {author} {\bibinfo {author} {\bibfnamefont {N.}~\bibnamefont
  {Mittal}}, \bibinfo {author} {\bibfnamefont {E.~O.}\ \bibnamefont {Budrene}},
  \bibinfo {author} {\bibfnamefont {M.~P.}\ \bibnamefont {Brenner}}, \ and\
  \bibinfo {author} {\bibfnamefont {A.}~\bibnamefont {Van~Oudenaarden}},\
  }\href {\doibase 10.1073/pnas.2233626100} {\bibfield  {journal} {\bibinfo
  {journal} {Proc. Nat. Acad. Sci.}\ }\textbf {\bibinfo {volume} {100}},\
  \bibinfo {pages} {13259} (\bibinfo {year} {2003})}\BibitemShut {NoStop}%
\bibitem [{\citenamefont {Saragosti}\ \emph {et~al.}(2012)\citenamefont
  {Saragosti}, \citenamefont {Silberzan},\ and\ \citenamefont
  {Buguin}}]{saragosti2012modeling}%
  \BibitemOpen
  \bibfield  {author} {\bibinfo {author} {\bibfnamefont {J.}~\bibnamefont
  {Saragosti}}, \bibinfo {author} {\bibfnamefont {P.}~\bibnamefont
  {Silberzan}}, \ and\ \bibinfo {author} {\bibfnamefont {A.}~\bibnamefont
  {Buguin}},\ }\href {\doibase 10.1371/journal.pone.0035412} {\bibfield
  {journal} {\bibinfo  {journal} {Plo{S} one}\ }\textbf {\bibinfo {volume}
  {7}},\ \bibinfo {pages} {e35412} (\bibinfo {year} {2012})}\BibitemShut
  {NoStop}%
\bibitem [{\citenamefont {Cates}\ and\ \citenamefont
  {Tailleur}(2013)}]{cates2013when}%
  \BibitemOpen
  \bibfield  {author} {\bibinfo {author} {\bibfnamefont {M.~E.}\ \bibnamefont
  {Cates}}\ and\ \bibinfo {author} {\bibfnamefont {J.}~\bibnamefont
  {Tailleur}},\ }\href {\doibase 10.1209/0295-5075/101/20010} {\bibfield
  {journal} {\bibinfo  {journal} {Eur. Phys. Lett.}\ }\textbf {\bibinfo
  {volume} {101}},\ \bibinfo {pages} {20010} (\bibinfo {year}
  {2013})}\BibitemShut {NoStop}%
\bibitem [{\citenamefont {Romanczuk}\ \emph {et~al.}(2008)\citenamefont
  {Romanczuk}, \citenamefont {Erdmann}, \citenamefont {Engel},\ and\
  \citenamefont {Schimansky-Geier}}]{romanczuk2008beyond}%
  \BibitemOpen
  \bibfield  {author} {\bibinfo {author} {\bibfnamefont {P.}~\bibnamefont
  {Romanczuk}}, \bibinfo {author} {\bibfnamefont {U.}~\bibnamefont {Erdmann}},
  \bibinfo {author} {\bibfnamefont {H.}~\bibnamefont {Engel}}, \ and\ \bibinfo
  {author} {\bibfnamefont {L.}~\bibnamefont {Schimansky-Geier}},\ }\href
  {\doibase 10.1140/epjst/e2008-00631-1} {\bibfield  {journal} {\bibinfo
  {journal} {Eur. Phys. J. Spec. Top.}\ }\textbf {\bibinfo {volume} {157}},\
  \bibinfo {pages} {61} (\bibinfo {year} {2008})}\BibitemShut {NoStop}%
\bibitem [{\citenamefont {Sevilla}\ and\ \citenamefont
  {Sandoval}(2015)}]{sevilla2015smoluchowski}%
  \BibitemOpen
  \bibfield  {author} {\bibinfo {author} {\bibfnamefont {F.~J.}\ \bibnamefont
  {Sevilla}}\ and\ \bibinfo {author} {\bibfnamefont {M.}~\bibnamefont
  {Sandoval}},\ }\href {\doibase 10.1103/PhysRevE.91.052150} {\bibfield
  {journal} {\bibinfo  {journal} {Phys. Rev. E}\ }\textbf {\bibinfo {volume}
  {91}},\ \bibinfo {pages} {052150} (\bibinfo {year} {2015})}\BibitemShut
  {NoStop}%
\bibitem [{\citenamefont {Wensink}\ and\ \citenamefont
  {L{\"o}wen}(2012)}]{wensink2012emergent}%
  \BibitemOpen
  \bibfield  {author} {\bibinfo {author} {\bibfnamefont {H.~H.}\ \bibnamefont
  {Wensink}}\ and\ \bibinfo {author} {\bibfnamefont {H.}~\bibnamefont
  {L{\"o}wen}},\ }\href {\doibase 10.1088/0953-8984/24/46/464130} {\bibfield
  {journal} {\bibinfo  {journal} {J. Phys. Cond. Matt.}\ }\textbf {\bibinfo
  {volume} {24}},\ \bibinfo {pages} {464130} (\bibinfo {year}
  {2012})}\BibitemShut {NoStop}%
\bibitem [{\citenamefont {Kirchhoff}\ \emph {et~al.}(1996)\citenamefont
  {Kirchhoff}, \citenamefont {L\"owen},\ and\ \citenamefont
  {Klein}}]{kirchhoff1996dynamical}%
  \BibitemOpen
  \bibfield  {author} {\bibinfo {author} {\bibfnamefont {T.}~\bibnamefont
  {Kirchhoff}}, \bibinfo {author} {\bibfnamefont {H.}~\bibnamefont {L\"owen}},
  \ and\ \bibinfo {author} {\bibfnamefont {R.}~\bibnamefont {Klein}},\ }\href
  {\doibase 10.1103/PhysRevE.53.5011} {\bibfield  {journal} {\bibinfo
  {journal} {Phys. Rev. E}\ }\textbf {\bibinfo {volume} {53}},\ \bibinfo
  {pages} {5011} (\bibinfo {year} {1996})}\BibitemShut {NoStop}%
\bibitem [{\citenamefont {Tirado}\ \emph {et~al.}(1984)\citenamefont {Tirado},
  \citenamefont {Mart{\'\i}nez},\ and\ \citenamefont {de~la
  Torre}}]{tirado1984comparison}%
  \BibitemOpen
  \bibfield  {author} {\bibinfo {author} {\bibfnamefont {M.~M.}\ \bibnamefont
  {Tirado}}, \bibinfo {author} {\bibfnamefont {C.~L.}\ \bibnamefont
  {Mart{\'\i}nez}}, \ and\ \bibinfo {author} {\bibfnamefont {J.~G.}\
  \bibnamefont {de~la Torre}},\ }\href {\doibase 10.1063/1.447827} {\bibfield
  {journal} {\bibinfo  {journal} {J. Chem. Phys.}\ }\textbf {\bibinfo {volume}
  {81}},\ \bibinfo {pages} {2047} (\bibinfo {year} {1984})}\BibitemShut
  {NoStop}%
\end{thebibliography}%

\clearpage
\clearpage

\onecolumngrid
\renewcommand{\thesection}{SI~\S \arabic{section}}   
\renewcommand{\thefigure}{S\arabic{figure}} 
\renewcommand{\theequation}{S\arabic{equation}} 
\setcounter{figure}{0}  
\setcounter{equation}{0}  
\setcounter{section}{0}

\section{Model}

The average flow $\langle \vec{v} (\vec{r}) \rangle$ due to a distribution of swimmers $f(\vec{r}_s, \vec{p}_s)$ at positions $\vec{r}_s = (x_s, y_s, z_s)$ and orientations  $\vec{p}_s = (p_x, p_y, p_z)$, all in standard Cartesian coordinates and evaluated at position $\vec{r}=(x,y,z)$, is
\begin{align}
\label{SIEq:AverageTotalFlow}
\langle \vec{v} (\vec{r}) \rangle
&= \int \vec{u} (\vec{r}, \vec{r}_s, \vec{p}_s) f(\vec{r}_s, \vec{p}_s) d\vec{r}_s d\vec{p}_s,
\end{align}
where $\vec{u} (\vec{r}, \vec{r}_s, \vec{p}_s)$ is the flow due to an individual swimmer,
\begin{align}
\label{SIEq:IndividualDipoleFlow}
\vec{u} (\vec{r}, \vec{r}_s, \vec{p}_s) = \kappa ( \vec{p}_s \cdot \vec{\nabla}_s) \mathcal{B}(\vec{r}, \vec{r}_s) \cdot \vec{p}_s,
\end{align}
where $\kappa$ is the dipole coefficient and the Blake tensor \cite{blake1971note} expressed in terms of the Oseen tensor (Stokeslet) \cite{mathijssen2015hydrodynamics} is
\begin{align}
\label{SIEq:BlakeTensor}
\mathcal{B}_{ij}(\vec{r}, \vec{r}_s) &= (- \delta_{jk} + 2 \swh \delta_{k3} (\partial_s)_j + \swh^2 \mbox{M}_{jk} \nabla_s^2) \mathcal{J}_{ik}(\vec{r}, \vec{r}_s),
\\
\mathcal{J}_{ij}(\vec{r}, \vec{r}_s) &=  \frac{1}{8\pi \eta} \left( \frac{\delta_{ij}}{|\vec{d}|} + \frac{d_i d_j}{|\vec{d}|^3} \right), 
\end{align}
with indices $i,j,k \in \{1,2,3\}$, mirror matrix $\mbox{M}_{jk} = \text{diag}(1,1,-1)$, distance $\vec{d} = \vec{r} - \vec{r}_s$, and all derivatives of the Oseen tensor $\mathcal{J}_{ij}$ are with respect to the swimmer position $\vec{r}_s$.
Combining these equations, the currents due to the active carpets can be derived by integrating Eq.~\ref{SIEq:AverageTotalFlow} analytically.
These exact results are given in \S \ref{SISec:ExactFlowsDensity} and \S \ref{SISec:ExactFlowsOrientations} and \S \ref{SISec:ExactFlowsTopo} below.

Equivalently, these solutions can be verified by computing the flows numerically.
Simulations are performed by summing the (exact) flows due to $N$ individual swimmers, 
\begin{align}
\label{SIEq:SimulatedTotalFlow}
\langle \vec{v} (\vec{r}) \rangle
= \sum_{i=1}^N \vec{u}_i (\vec{r}, \vec{r}_i, \vec{p}_i),
\end{align}
where the positions $\vec{r}_i$ and orientations $\vec{p}_i$ are distributed such that they satisfy the probability density $f(\vec{r}_s, \vec{p}_s)$.
All simulations were carried out using Wolfram Mathematica (version 10.0.1.0) on a desktop PC (operating on Windows 10).  
To expedite the evaluation of currents due large active carpets, $N\sim250,000$, the flows generated by an individual swimmer (Eq.~\ref{SIEq:IndividualDipoleFlow}) were derived, maximally simplified and compiled with the built-in \texttt{Compile[]} function, and summed over in parallel with the \texttt{ParallelSum[]} function.
Detailed descriptions of all simulation procedures are provided in \S \ref{SISec:SimFlows} and \S \ref{SISec:SimFlowsTurbulence} below.

\section{Derivation of flows due to density gradients}
\label{SISec:ExactFlowsDensity}

\subsection{Uniform cluster}

Consider a cluster of $N$ bacteria swimming at height $z_s=\swh$ over a surface, located at $z=0$.
They are uniformly distributed within a disk of radius $R$, so that the surface concentration is constant, $n=N/(\pi R^2)$.
The bacteria are oriented randomly, parallel to the surface, $p_z=0$, according to a uniform distribution, $\phi_s \in [-\pi,\pi]$, where $\phi_s = \text{arctan}(p_y / p_x)$.
Without loss of generality, we take the cluster to be centred at the origin.
The carpet distribution is then given by
\begin{align}
\label{SIEq:ClusterProfile}
f(\vec{r}_s, \vec{p}_s) &= N \frac{\delta(z_s-\swh) \Theta(R-\rho_s) }{\pi R^2} \frac{\delta(|\vec{p}_s|-1) \delta(\vec{p}_s\cdot  \hat{\vec{z}})}{2 \pi},
\end{align}
where $\Theta(x)$ and $\delta(x)$ are the Heaviside and Dirac delta functions, and cylindrical symmetry about the $z$ axis gives  $\rho_s^2 = x_s^2 + y_s^2$ and $\theta_s = \text{arctan}(y_s / x_s)$.
For $\rho_s \in [0,R]$, $\theta_s \in [-\pi, \pi]$ and $\phi_s  \in [-\pi, \pi]$ this profile simplifies to $f(\rho_s, \theta_s, \phi_s) = n/(2\pi)$, and it is normalised so that the integrated distribution gives the number of swimmers,
\begin{align}
\label{SIEq:ClusterProfileNormalisation}
 \int  f(\vec{r}_s, \vec{p}_s) d\vec{r}_s d\vec{p}_s 
& = \int_0^R \int_{-\pi}^\pi \int_{-\pi}^\pi   \frac{N}{\pi R^2} \frac{1}{2 \pi} d\phi_s  d\theta_s  \rho_s d \rho_s 
= N.
\end{align}

\subsubsection{Directly above cluster, $\rho=0$, in limit $h \ll z$}

To find the overall flow due to the cluster, we substitute this distribution (\ref{SIEq:ClusterProfile}) into (\ref{SIEq:AverageTotalFlow}) and evaluate the integral.
In general this is not trivial, but progress can be made by considering flows far from the surface, $z_s = \epsilon z$ with $\epsilon \ll1$. 
We Taylor-expand the individual swimmer flows (\ref{SIEq:IndividualDipoleFlow}) to first order in $\epsilon$,
\begin{align}
\label{SIEq:TaylorSeriesSmallH}
\vec{u}(\vec{r}, \vec{r}_s, \vec{p}_s) 
&= \left. \frac{\partial \vec{u}}{\partial \epsilon} \right |_{\epsilon=0} \epsilon + \mathcal{O}\left( \epsilon^2 \right)
\\
&= \tilde{\vec{u}}(\vec{r},  \vec{r}_s, \vec{p}_s) + \mathcal{O}\left( \frac{z_s}{z} \right)^2.
\end{align}
To be explicit, on the $z$ axis this gives the individual flows
\begin{align}
\left . \tilde{u}_x  \right|_{\rho=0}
&=
\frac{3 \kappa  \rho_s z^2 \epsilon  \left(-5 \rho_s^2 \cos (3 \theta_s-2 \phi_s )+\left(2 z^2-3 \rho_s^2\right) \cos (\theta_s-2 \phi_s )+\cos (\theta_s) \left(6 z^2-4 \rho_s^2\right)\right)}{\left(\rho_s^2+z^2\right)^{7/2}},
\\
\left . \tilde{u}_y  \right|_{\rho=0}
&=
-\frac{3 \kappa  \rho_s z^2 \epsilon  \left(5 \rho_s^2 \sin (3 \theta_s-2 \phi_s )+\left(2 z^2-3 \rho_s^2\right) \sin (\theta_s-2 \phi_s )+\sin (\theta_s) \left(4 \rho_s^2-6 z^2\right)\right)}{\left(\rho_s^2+z^2\right)^{7/2}},
\\
\left . \tilde{u}_z  \right|_{\rho=0}
&=
-\frac{6 \kappa  z^3  \left(-5 \rho_s^2 \cos (2 (\theta_s-\phi_s ))-3 \rho_s^2+2 z^2\right)}{\left(\rho_s^2+z^2\right)^{7/2}} \epsilon + \mathcal{O}\left( \epsilon^2 \right).
\end{align}
Inserting these into  (\ref{SIEq:AverageTotalFlow}) gives the average flows directly above a bacterial cluster,
\begin{align}
\left . \langle \vec{v}(\vec{r}) \rangle \right|_{\rho=0}
&= 
\int \tilde{\vec{u}} (\rho=0) f(\vec{r}_s, \vec{p}_s) d\vec{r}_s d\vec{p}_s 
\\
&= 
\int_0^R \int_{-\pi}^\pi \int_{-\pi}^\pi   \left . \tilde{\vec{u}} (z,\rho_s,\theta_s,\phi_s)  \right|_{\rho=0}  
\frac{N}{\pi R^2} \frac{1}{2 \pi} d\phi_s  d\theta_s  \rho_s d \rho_s 
\\
\label{SIEq:ClusterFlowZ}
&=
-12\pi n \swh \kappa\frac{z^2 R^2}{(R^2+z^2)^{5/2}} \hat{\vec{z}},
\end{align}
which corresponds to Eq.~(3) of the Main Text.

\subsubsection{Everywhere above cluster, $\rho \neq 0$, in limit $h \ll z$}

For all other positions, $\rho \neq 0$, the integral can be performed to give the complete cluster flow
\begin{align}
\label{SIEq:ClusterFlowFullZ}
\langle v_z \rangle
&=
\frac{4 \kappa  z^2 h}{\left((R-\rho)^2+z^2\right)^2 \left((R+\rho)^2+z^2\right)^{3/2}} 
\\
\nonumber 
&\Bigg(
\left(2 \rho z^2 (R-\rho)+(R-\rho)^3 (R+\rho)-z^4\right) \mathcal{K}\left[\frac{4 R \rho}{(R+\rho)^2+z^2}\right]+ 
\\ 
\nonumber 
&
\left(-7 R^4+6 R^2 (\rho-z) (\rho+z)+\left(\rho^2+z^2\right)^2\right) \mathcal{E}\left[\frac{4 R \rho}{(R+\rho)^2+z^2}\right]
\Bigg),
\\
\label{SIEq:ClusterFlowFullRho}
\langle v_\rho \rangle
&=
-\frac{4 \kappa  z h }{\rho \left((R-\rho)^2+z^2\right)^{3/2} \left((R+\rho)^2+z^2\right)^2}  
\\
\nonumber 
&\Bigg(
\left((R+\rho)^2+z^2\right) \left(3 z^2 \left(R^2+\rho^2\right)+2 \left(R^2-\rho^2\right)^2+z^4\right) \mathcal{K}\left[-\frac{4 R \rho}{(R-\rho)^2+z^2}\right] -
\\
\nonumber 
&
\left(4 z^4 \left(R^2+\rho^2\right)+5 z^2 \left(R^2-\rho^2\right)^2+2 \left(R^2-\rho^2\right)^2 \left(R^2+\rho^2\right)+z^6\right) \mathcal{E}\left[-\frac{4 R \rho}{(R-\rho)^2+z^2}\right]
\Bigg).
\end{align}
Here the complete elliptic integrals of the first and second kind are defined as
\begin{align}
\label{SIEq:ellipticIntegrals}
\mathcal{K}[z] = \int _0^{\pi/2}\frac{1}{\sqrt{1-z\sin^2\theta}} d\theta, \qquad
\mathcal{E}[z] = \int_0^{\pi/2} \sqrt{1-z\sin^2\theta} d\theta
\end{align}
It is convenient to notice that these functions obey the following identities,
\begin{align}
\label{SIEq:ellipticIntegralsIdentities1}
\frac{\mathcal{K}\left[ \frac{-4\zeta}{(1-\zeta)^2 + \xi^2} \right]}{\sqrt{(1-\zeta)^2 + \xi^2}} =&
\frac{\mathcal{K}\left[ \frac{+4\zeta}{(1+\zeta)^2 + \xi^2} \right]}{\sqrt{(1+\zeta)^2 + \xi^2}},
\\
\frac{\mathcal{E}\left[ \frac{-4\zeta}{(1-\zeta)^2 + \xi^2} \right]}{\sqrt{(1-\zeta)^2 + \xi^2}((1+\zeta)^2 + \xi^2)} =&
\frac{\mathcal{E}\left[ \frac{+4\zeta}{(1+\zeta)^2 + \xi^2} \right]}{\sqrt{(1+\zeta)^2 + \xi^2}((1-\zeta)^2 + \xi^2)}.
\end{align}
where $\xi$ and $\zeta$ are real variables.
As a verification, note that we recover (\ref{SIEq:ClusterFlowZ}) when evaluating (\ref{SIEq:ClusterFlowFullZ},\ref{SIEq:ClusterFlowFullRho}) in the limit $\rho \to 0$. 
This result is plotted in Fig.~1(b-d) of the main text.

\subsubsection{Directly above cluster, $\rho=0$, without taking limit}

Moreover, we also consider the full solution along $\rho=0$ without taking the limit $h\ll z$ in the Taylor series (\ref{SIEq:TaylorSeriesSmallH}). 
Starting from the dipole solution at $z_s = h$ and its image system at $z_i = -h$, and averaging over the swimmer positions and orientations yields
\begin{align}
\left . \langle \vec{v}^*(\vec{r}) \rangle \right|_{\rho=0}
&= 
\int {\vec{u}} (\rho=0) f(\vec{r}_s, \vec{p}_s) d\vec{r}_s d\vec{p}_s 
\\
&= 
\int_0^R \int_{-\pi}^\pi \int_{-\pi}^\pi   \left . {\vec{u}} (z,\rho_s,\theta_s,\phi_s)  \right|_{\rho=0}  \frac{N}{\pi R^2} \frac{1}{2 \pi} d\phi_s  d\theta_s  \rho_s d \rho_s 
\\
\label{SIEq:ClusterFlowZfull}
&=
-\pi  \kappa n R^2 \left(\frac{h^3+7 h^2 z+h \left(R^2+5 z^2\right)-z \left(R^2+z^2\right)}{\left((h+z)^2+R^2\right)^{5/2}}+\frac{z-h}{\left((h-z)^2+R^2\right)^{3/2}}\right).
\end{align}
In the limit $h \ll z$ we recover equation (\ref{SIEq:ClusterFlowZfull}). Note that this solution is not singular at $z=h$ because the averaging regularises the flow at this point.
Another important property is that it vanishes if $z=0$ or if $h=0$, reflecting the no-slip condition.

\subsubsection{Shear rate induced above cluster}

To estimate whether the active carpet can induce shear flows that are strong enough to detach bacteria, we compute the shear rate directly above the surface.
Because the flow is radial by symmetry, $v_\rho(\rho, z)$, we define the shear rate at the surface as 
\begin{align}
\dot{\gamma} = \left . \frac{\partial v_\rho}{\partial z} \right |_{z=0},
\end{align}
where the full dipole flow (\ref{SIEq:IndividualDipoleFlow}) is integrated over all carpet configurations. Hence,
 \begin{align}
\dot{\gamma}
= 
\int_0^R \int_{-\pi}^\pi \int_{-\pi}^\pi
\left. \frac{\partial \vec{u}}{\partial z} \right |_{z=0} 
 \frac{N}{\pi R^2} \frac{1}{2 \pi} d\phi_s  d\theta_s  \rho_s d \rho_s 
\end{align}
This shear rate can be evaluated analytically, and results in another very long expression in terms of elliptic integrals like (\ref{SIEq:ClusterFlowFullRho}).
Evaluated in the middle of the carpet, at $\rho = R/2$, it simplifies to
\begin{align}
\dot{\gamma}(R, h, n, \kappa, \rho = R/2) &=
\frac{32 h \kappa n}{R \left(4 h^2+R^2\right)^{3/2} \left(4 h^2+9 R^2\right)^2} \Bigg (
\\
&\left(32 h^6+160 h^4 R^2+90 h^2 R^4+45 R^6\right) \mathcal{E}\left(-\frac{8 R^2}{4 h^2+R^2}\right)
\\
&-
\left(32 h^6+192 h^4 R^2+306 h^2 R^4+81 R^6\right) \mathcal{K}\left(-\frac{8 R^2}{4 h^2+R^2}\right)
\Bigg ).
\end{align}
Therefore, even at very high organism densities,  $n=1 \mummy^{-2}$, and using the parameters as before, $h=1\mu$m,  $R=10\mu$m,  $\kappa=30\mu\text{m}^3$/s, we find the shear rate $\dot{\gamma} \approx 8.8 \text{s}^{-1}$. This is still much smaller than the erosion shear rate measured \cite{figueroa2015living} for \textit{E. coli} bacteria, $\dot{\gamma}_e \approx 100 \text{s}^{-1}$.
So we conclude that the collective flows are not strong enough to detach cells from the surface.
One should rather expect a subtle but steady fluid recirculation that results in appreciable particle transport when integrated over time.

\subsection{Gaussian density profile}

Instead of a cluster with a `sharp' density gradient at the cluster edge, the Heaviside function in (\ref{SIEq:ClusterProfile}), we next consider a Gaussian density profile,
\begin{align}
\label{SIEq:GaussianClusterProfile}
f(\vec{r}_s, \vec{p}_s) &= N \frac{\delta(z_s-\swh)}{2 \pi R^2} \exp \left(- \frac{\rho_s^2}{2 R^2} \right) 
\frac{\delta(|\vec{p}_s|-1) \delta(\vec{p}_s\cdot  \hat{\vec{z}})}{2 \pi},
\end{align}
which is normalised to the number of swimmers as in (\ref{SIEq:ClusterProfileNormalisation}).
Inserting this profile into  (\ref{SIEq:AverageTotalFlow}) gives the average flows directly above a bacterial cluster,
\begin{align}
\left . \langle \vec{v}(\vec{r}) \rangle \right|_{\rho=0}
&= 
\int \tilde{\vec{u}} (\rho=0) f(\vec{r}_s, \vec{p}_s) d\vec{r}_s d\vec{p}_s 
\\
&= 
\int_0^\infty  \left . \tilde{\vec{u}} (z,\rho_s,\theta_s,\phi_s)  \right|_{\rho=0}  
\frac{N}{R^2} \exp \left(- \frac{\rho_s^2}{2 R^2} \right)
\rho_s d \rho_s 
\\
\label{SIEq:GaussianClusterFlowZ}
&=
N \left(
\frac{\sqrt{2 \pi } \swh  \kappa  z^2  e^{\frac{z^2}{2 R^2}} \left(3 R^2+z^2\right) \text{erfc}\left(\frac{z}{\sqrt{2} R}\right)}{R^7}-\frac{2 \swh \kappa  z  \left(2 R^2+z^2\right)}{R^6}
\right)
\hat{\vec{z}}.
\end{align}
This expression is a little more complicated than (\ref{SIEq:ClusterFlowZ}), but has exactly the same features: 
\begin{itemize}
\item The function is always negative for pushers, $\kappa >0$, representing attraction of nutrients towards the cluster,
\item It has a minimum around $z\sim R$, and 
\item It has same decay with distance from the active carpet, $v_z(z) \sim -12 \pi n \swh \kappa R^2 /z^3$.
\end{itemize}
However, this flow is a factor of $\sim2$ stronger because the swimmer density gradient is already present for small $\rho_s$ values. 

\subsection{Carpet of Stokeslets}

Above we found that a carpet of randomly oriented dipoles \textit{does generate a net drift}, for any finite carpet size.
However, in ciliary arrays \cite{elgeti2013emergence, ding2014mixing, uchida2010synchronization, uchida2010bsynchronization} and grafted cells \cite{darnton2004moving, kim2008microfluidic, hsiao2014collective, hsiao2016impurity} the force alignment is essential for microbiological transport.
To highlight this, note that a carpet of Stokeslets (\ref{SIEq:BlakeTensor}), $\vec{u}_S(\vec{p}_s) = \mathcal{B}(\vec{r}, \vec{r}_s) \cdot \vec{p}_s$, oriented randomly in the $x$-$y$ directions, $\vec{p}_s= \cos \phi_s \hat {\vec x} + \sin \phi_s \hat {\vec y}$, \textit{does not generate a net drift}, for any carpet size;
\begin{align}
\langle \vec{v}_S(\vec{r}) \rangle 
&= 
\int_0^R \int_{-\pi}^\pi \int_{-\pi}^\pi   \vec{u}_S (z,\rho,\rho_s,\theta_s,\phi_s)  \frac{N}{\pi R^2} \frac{1}{2 \pi} d\phi_s  d\theta_s  \rho_s d \rho_s 
\nonumber \\
&=
0.
\end{align}
Indeed, by symmetry we have $ \vec{u}_S(\hat{\vec{x}}) = - \vec{u}_S(-\hat{\vec{x}})$, so when averaging over $\vec{p}_s$ the flows must vanish. 
For dipoles the reflection is additive, $ \vec{u}_D(\hat{\vec{x}}) = + \vec{u}_D(-\hat{\vec{x}})$, so random orientations in the $x$-$y$ directions may lead to net flow. 

\subsection{Dipoles in bulk}

Finally, note that dipoles in the bulk feature orientations in three dimensions, $\vec{p}_s= \sin \psi_s \cos \phi_s \hat {\vec x} + \sin \psi_s \sin \phi_s \hat {\vec y} + \cos \psi_s  {\vec z}$. Then, averaging the dipole over the 3D orientations gives
\begin{align}
\langle \vec{v}_D^{3d}(\vec{r}) \rangle 
&= 
\int_0^R \int_{-\pi}^\pi \int_{-\pi}^\pi   \vec{u}_D (z,\rho,\rho_s,\theta_s,\phi_s, \psi_s)  \frac{d\phi_s \sin \psi_s d\psi_s}{4 \pi}  \frac{N}{\pi R^2} d\theta_s  \rho_s d \rho_s 
\nonumber \\
&=
0,
\end{align}
This yields zero to satisfy the incompressibility condition, $\vec{\nabla} \cdot  \vec{u} =0$.
Therefore the bulk swimmers do not lead to net currents on average, but may still contribute to enhanced diffusive flows  \cite{wu2000particle, kim2004enhanced, thiffeault2010stirring, kim2007controlled, lin2011stirring, pushkin2013fluid, jeanneret2016entrainment, pushkin2012fluid, mueller2017fluid, mathijssen2018universal}.

\section{Derivation of flows due to orientation gradients}
\label{SISec:ExactFlowsOrientations}

\subsection{Laning swimmers}

In the absence of density gradients, the currents cancel on average if all swimmers are oriented in the same direction, i.e. `laning'.
Using the profile
\begin{align}
f(\vec{r}_s, \vec{p}_s) &= N \frac{\delta(z_s-\swh) \Theta(R-\rho_s) }{\pi R^2} \delta^3(\vec{p}_s - \hat{\vec{x}}),
\end{align}
we obtain the average flow
\begin{align}
\left . \langle \vec{v}(\vec{r}) \rangle \right|_{\rho=0}
&= 
\int_0^R \int_{-\pi}^\pi \int_{-\pi}^\pi   \tilde{\vec{u}} (z,\rho_s,\theta_s,\phi_s; \rho=0) \frac{N}{\pi R^2} \delta(\phi_s)  d\phi_s  d\theta_s  \rho_s d \rho_s 
\\
&=
-12 \swh \kappa N \frac{z^2}{(R^2+z^2)^{5/2}} \hat{\vec{z}},
\end{align}
which vanishes in the thermodynamic limit, where $N,R\to \infty$ with constant $n = N / (\pi R^2)$.

\subsection{Bend gradients}

To understand the effect of bend gradients, we consider swimmer orientations $\vec{p}_s(\vec{r}_s)$ that are arranged along circle tangents, $\phi_s = \theta + \pi/2$, which gives the carpet profile
\begin{align}
f(\rho_s, \theta_s, \phi_s) &= \frac{N}{\pi R^2} \delta(\phi_s - \theta_s - \pi/2).
\end{align}
This corresponds to a bend gradient of
\begin{align}
B(\rho) &= (\vec{p}_s \times (\vec{\nabla}_s \times \vec{p}_s))^2 = \frac{1}{\rho^2}.
\end{align}
In the thermodynamic limit this yields the average flow
\begin{align}
\langle \vec{v}(\vec{r}) \rangle
&= 
\int_0^\infty \int_{-\pi}^\pi \int_{-\pi}^\pi   n ~ \delta(\phi_s - \theta_s - \pi/2) \tilde{\vec{u}} (\rho, z,\rho_s,\theta_s,\phi_s)  d\phi_s  d\theta_s  \rho_s d \rho_s 
\\
\label{SIEq:BendFlowFull}
&=
- \frac{8 \pi \swh n \kappa}{(\rho^2+z^2)^{3/2}}
\Bigg( z^2 \hat{\vec{z}}
+
\frac{\left(z^2 \left(z-\sqrt{\rho^2+z^2}\right)-\rho^2 \left(\sqrt{\rho^2+z^2}-2 z\right)\right)}{\rho} \hat{\vec{\rho}}
\Bigg).
\end{align}
The vertical component is always negative for bend gradients, but incompressibility demands that the horizontal component switches sign at $\rho /z = \sqrt{(1+\sqrt{5})/2}$, with outward flows for small $z$.

In the limit $\rho \gg z$ the bend gradients are approximately constant, and therefore it is possible to write the flow as
\begin{align}
\label{SIEq:BendFlows}
\langle v_\rho \rangle
&= \frac{8\pi n \swh \kappa}{\rho} + \mathcal{O}\left(\frac{1}{\rho^{2}} \right)
\approx 8\pi n \swh \kappa \sqrt{B(\rho)},
\\
\label{SIEq:BendFlowVertical}
\langle v_z \rangle &=-\frac{8\pi n \swh \kappa z^2}{\rho^3} + \mathcal{O}\left(\frac{1}{\rho^{4}} \right) 
\approx - 8\pi n \swh \kappa z^2 [B(\rho)]^{3/2}.
\end{align}

\subsection{Splay gradients}

To understand the effect of splay gradients, we consider swimmer orientations $\vec{p}_s(\vec{r}_s)$ that are arranged along circle radii, $\phi_s = \theta$, which gives the carpet profile
\begin{align}
f(\rho_s, \theta_s, \phi_s) &= \frac{N}{\pi R^2} \delta(\phi_s - \theta_s).
\end{align}
This corresponds to a splay gradient of
\begin{align}
S(\rho) &= (\vec{\nabla}_s \cdot \vec{p}_s)^2 = \frac{1}{\rho^2}.
\end{align}
In the thermodynamic limit this yields the average flow
\begin{align}
\langle \vec{v}(\vec{r}) \rangle
&= 
\int_0^\infty \int_{-\pi}^\pi \int_{-\pi}^\pi   n ~ \delta(\phi_s - \theta_s) \tilde{\vec{u}} (\rho, z,\rho_s,\theta_s,\phi_s)  d\phi_s  d\theta_s  \rho_s d \rho_s 
\\
\label{SIEq:SplayFlowFull}
&=
+ \frac{8 \pi \swh n \kappa}{(\rho^2+z^2)^{3/2}}
\Bigg( z^2 \hat{\vec{z}}
+
\frac{\left(z^2 \left(z-\sqrt{\rho^2+z^2}\right)-\rho^2 \left(\sqrt{\rho^2+z^2}-2 z\right)\right)}{\rho} \hat{\vec{\rho}}
\Bigg).
\end{align}
As opposed to bend gradients, the vertical component is always positive for splay gradients, and the horizontal component the still switches at $\rho /z = \sqrt{(1+\sqrt{5})/2}$, with inward flows for small $z$.

As before, in the limit $\rho \gg z$ the splay gradients are approximately constant, and therefore it is possible to write the flow as
\begin{align}
\label{SIEq:SplayFlows}
\langle v_\rho \rangle
&= -\frac{8\pi n \swh \kappa}{\rho} + \mathcal{O}\left(\frac{1}{\rho^{2}} \right)
\approx - 8\pi n \swh \kappa \sqrt{S(\rho)},
\\
\label{SIEq:SplayFlowVertical}
\langle v_z \rangle &=\frac{8\pi n \swh \kappa z^2}{\rho^3} + \mathcal{O}\left(\frac{1}{\rho^{4}} \right) 
\approx  8\pi n \swh \kappa z^2 [S(\rho)]^{3/2}.
\end{align}
Combining the equations (\ref{SIEq:BendFlows}-\ref{SIEq:SplayFlowVertical}) yields an expression for $\langle \vec{v} (\vec{r}) \rangle = g(B,S)$, which corresponds to Eqs.~(6,7) in the Main Text.

\section{Derivation of flows due to topological defects}
\label{SISec:ExactFlowsTopo}

Here we consider the flows due to swimmers arranged with a disinclination or defect at the origin, $x=y=0$, defined as 
\begin{align}
\label{SIEq:defectDefinition}
\phi_s = m \theta_s + \phi_0, 
\end{align}
where $m$ is the topological charge, $\phi_s = \text{arctan}(y_s / x_s)$ and $\theta_s = \text{arctan}(p_y / p_x)$.

\subsection{Vortex defect}
This is the same calculation as the one for bend gradients above, equation (\ref{SIEq:BendFlowFull}).

\subsection{Aster defect}
This is the same calculation as the one for splay gradients above, equation (\ref{SIEq:SplayFlowFull}).

\subsection{$+1/2$ defect}

The cylindrical symmetry that we employed earlier can no longer be used in the case of defects with $m \neq 1$.
Therefore we revert to standard Cartesian coordinates, with swimmer positions $(x_s, y_s, z_s=h)$.
We consider a +1/2 topological defect in the swimmer orientations along the $x$ axis, so that $\phi_s = \frac 1 2 (\theta_s + \pi)$. 
Note that the offset $\phi_0$ in this case only contributes to a rotation of the defect about the origin (\ref{SIEq:defectDefinition}), and we choose $\phi_0 = \pi/2$ so that the convex end points towards the positive $x$ direction.

Hence, using the carpet profile,
\begin{align}
f(\rho_s, \theta_s, \phi_s) &= \frac{N}{\pi R^2} \delta \left(\phi_s -  \frac{\theta_s + \pi}{2} \right),
\end{align}
we find the flow in the plane along the +1/2 defect direction
\begin{eqnarray}
\left. \langle \vec{v}(\vec{r}) \rangle \right|_{y=0}
&&= 
\int_0^\infty \int_{-\pi}^\pi \int_{-\pi}^\pi   n ~ \delta \left(\phi_s -  \frac{\theta_s + \pi}{2} \right) \tilde{\vec{u}} (x, y=0, z,\rho_s,\theta_s,\phi_s)  d\phi_s  d\theta_s  \rho_s d \rho_s 
\\
&&=
\frac{4 \pi  n \swh \kappa}{x^2 \left(x^2+z^2\right)^2 \sqrt{2 x \left(x-\sqrt{x^2+z^2}\right)+z^2}}  \Bigg( \left.
\right. \nonumber \\ && \left.
x^6-2 x^2 z^4+x^2 z \left(x^2+z^2\right) \sqrt{2 x \left(x-\sqrt{x^2+z^2}\right)+z^2}
\right. \nonumber \\ && \left.
+ x z^4 \sqrt{x^2+z^2}+z^3 \left(x^2+z^2\right) \sqrt{2 x \left(x-\sqrt{x^2+z^2}\right)+z^2}
\right. \nonumber \\ && \left.
-x^5 \sqrt{x^2+z^2}+x^3 z^2 \sqrt{x^2+z^2}-z^6 \Bigg) \hat{\vec{x}} \right.
 \nonumber \\ &&
+  \Bigg( \frac{4 \pi  n \swh \kappa  x  \left(\sqrt{x^2+z^2}+x\right) \sqrt{2 x \left(x-\sqrt{x^2+z^2}\right)+z^2}}{z \left(x^2+z^2\right)^{3/2}}  \Bigg) \hat{\vec{z}}.
\end{eqnarray}
Directly above the defect this simplifies to a purely longitudinal flow,
\begin{eqnarray}
\left. \langle \vec{v}(\vec{r}) \rangle \right|_{x=y=0}
&&= \frac{2\pi n h \kappa}{z}  \hat{\vec{x}}.
\end{eqnarray}

\subsection{$-1/2$ defect}

Similarly as for its positive counterpart, we use the carpet profile
\begin{align}
f(\rho_s, \theta_s, \phi_s) &= \frac{N}{\pi R^2} \delta \left(\phi_s -  \frac{\pi-\theta_s}{2} \right),
\end{align}
to compute the flows in the plane along the -1/2 defect direction,
\begin{eqnarray}
\left. \langle \vec{v}(\vec{r}) \rangle \right|_{y=0}
&&= 
\int_0^\infty \int_{-\pi}^\pi \int_{-\pi}^\pi   n ~ \delta \left(\phi_s -  \frac{\pi-\theta_s}{2} \right) \tilde{\vec{u}} (x, y=0, z,\rho_s,\theta_s,\phi_s)  d\phi_s  d\theta_s  \rho_s d \rho_s 
\\
&&=
-\frac{4 \pi n \swh \kappa}{x^4 \left(x^2+z^2\right)^{3/2} \sqrt{2 x \left(x-\sqrt{x^2+z^2}\right)+z^2}}
\Bigg(
-x^7+3 x^5 z^2+17 x^3 z^4
 \nonumber \\ &&
+12 z^5 \left(\sqrt{\left(x^2+z^2\right) \left(2 x \left(x-\sqrt{x^2+z^2}\right)+z^2\right)}-z \sqrt{x^2+z^2}\right)
 \nonumber \\ &&
+x^2 z^3 \left(11 \sqrt{\left(x^2+z^2\right) \left(2 x \left(x-\sqrt{x^2+z^2}\right)+z^2\right)}-17 z \sqrt{x^2+z^2}\right)
 \nonumber \\ &&
+x^6 \sqrt{x^2+z^2}-x^4 z \left(3 z \sqrt{x^2+z^2}+\sqrt{\left(x^2+z^2\right) \left(2 x \left(x-\sqrt{x^2+z^2}\right)+z^2\right)}\right)
 \nonumber \\ &&
+12 x z^6 
\Bigg)  \hat{\vec{x}}
\nonumber \\
&& +
\frac{4 \pi  n \swh \kappa  z }{x^3 \left(x^2+z^2\right)^{3/2} \sqrt{2 x \left(x-\sqrt{x^2+z^2}\right)+z^2}}
\Bigg(
3 x^5+12 x^3 z^2
 \nonumber \\ &&
+4 x^2 z \left(2 \sqrt{\left(x^2+z^2\right) \left(2 x \left(x-\sqrt{x^2+z^2}\right)+z^2\right)}-3 z \sqrt{x^2+z^2}\right)
 \nonumber \\ &&
+8 z^3 \left(\sqrt{\left(x^2+z^2\right) \left(2 x \left(x-\sqrt{x^2+z^2}\right)+z^2\right)}-z \sqrt{x^2+z^2}\right)
 \nonumber \\ &&
-3 x^4 \sqrt{x^2+z^2}+8 x z^4
\Bigg) \hat{\vec{z}}.
\end{eqnarray}
Directly above the defect this simplifies to
\begin{eqnarray}
\left. \langle \vec{v}(\vec{r}) \rangle \right|_{x=y=0}
&&= 0.
\end{eqnarray}

\subsection{Saddle defect}

Lastly, using the carpet profile
\begin{align}
f(\rho_s, \theta_s, \phi_s) &= \frac{N}{\pi R^2} \delta \left(\phi_s + \theta_s \right),
\end{align}
we find the flows for a saddle defect,
\begin{eqnarray}
\left. \langle \vec{v}(\vec{r}) \rangle \right|_{y=0}
&&= 
\int_0^\infty \int_{-\pi}^\pi \int_{-\pi}^\pi   n ~ \delta \left(\phi_s + \theta_s \right) \tilde{\vec{u}} (x, y=0, z,\rho_s,\theta_s,\phi_s)  d\phi_s  d\theta_s  \rho_s d \rho_s 
\\
&&=
-\frac{8 \pi  n \swh \kappa  }{x^5 \left(x^2+z^2\right)^{3/2} \sqrt{2 x \left(x-\sqrt{x^2+z^2}\right)+z^2}}
 \Bigg(
 \nonumber \\ &&
-x^8 +32 z^7 \left(z-\sqrt{2 x \left(x-\sqrt{x^2+z^2}\right)+z^2}\right)
 \nonumber \\ &&
-32 x z^6 \sqrt{x^2+z^2}+4 x^2 z^5 \left(17 z-13 \sqrt{2 x \left(x-\sqrt{x^2+z^2}\right)+z^2}\right)
 \nonumber \\ &&
+x^7 \sqrt{x^2+z^2}+2 x^6 z \left(\sqrt{2 x \left(x-\sqrt{x^2+z^2}\right)+z^2}+z\right)
 \nonumber \\ &&
-3 x^5 z^2 \sqrt{x^2+z^2}+x^4 z^3 \left(39 z-17 \sqrt{2 x \left(x-\sqrt{x^2+z^2}\right)+z^2}\right)
 \nonumber \\ &&
-36 x^3 z^4 \sqrt{x^2+z^2}
\Bigg)  \hat{\vec{x}}
\nonumber \\
&& -\frac{8 \pi  n \swh \kappa  z}{x^4 \left(x^2+z^2\right)^{3/2} \sqrt{2 x \left(x-\sqrt{x^2+z^2}\right)+z^2}}
\Bigg(
 \nonumber \\ &&
-4 x^6+24 z^5 \left(\sqrt{2 x \left(x-\sqrt{x^2+z^2}\right)+z^2}-z\right)
 \nonumber \\ &&
+24 x z^4 \sqrt{x^2+z^2}+4 x^2 z^3 \left(10 \sqrt{2 x \left(x-\sqrt{x^2+z^2}\right)+z^2}-13 z\right)
 \nonumber \\ &&
+4 x^5 \sqrt{x^2+z^2}+x^4 z \left(15 \sqrt{2 x \left(x-\sqrt{x^2+z^2}\right)+z^2}-32 z\right)
 \nonumber \\ &&
+28 x^3 z^2 \sqrt{x^2+z^2}
\Bigg) \hat{\vec{z}}.
\end{eqnarray}
Again, directly above the defect this simplifies to
\begin{eqnarray}
\left. \langle \vec{v}(\vec{r}) \rangle \right|_{x=y=0}
&&= 0.
\end{eqnarray}

\section{Simulations of active carpet flows}
\label{SISec:SimFlows}

Next to analytical integration, the flows due to an active carpet may also be approximated in simulations.
To determine the average flow we place $N$ swimmers on a surface and compute the sum
\begin{align}
\label{SIEq:SimulatedTotalFlow}
\langle \vec{v} (\vec{r}) \rangle
= \sum_{i=1}^N \vec{u}_i (\vec{r}, \vec{r}_i, \vec{p}_i),
\end{align}
where the positions $\vec{r}_i$ and orientations $\vec{p}_i$ are found numerically via inverse transform sampling (Smirnov transform) in order to satisfy the probability distribution $f(\vec{r}_s, \vec{p}_s)$.

\subsection{Uniform cluster}

To see this explicitly, we first consider the case of a uniform cluster profile (\ref{SIEq:ClusterProfile}), for which we aim to sample the random variates $\rho_s \in [0,R], \theta_s \in [-\pi, \pi], \phi_s  \in [-\pi, \pi]$ in terms of three random variates, $w_i \in [0,1]$ with $i \in [1,2,3]$, drawn from the standard uniform distribution.
This profile (\ref{SIEq:ClusterProfile}) is separable in the three variables, $f(\rho_s, \theta_s, \phi_s) = N f_{\rho_s} f_{\theta_s} f_{\rho_s}$, with angular distributions $f_{\theta_s} = f_{\phi_s} = 1/(2\pi)$ and the radial distribution $f_{\rho_s} = 2/R^2$.
Therefore we immediately find that the angles $\theta_s, \phi_s$ can be sampled by taking
\begin{align}
\label{SIEq:AngularVariate1}
\theta_s &= -\pi + 2\pi w_1,
\\
\label{SIEq:AngularVariate2}
\phi_s &= -\pi + 2\pi w_2.
\end{align}
To sample the distance $\rho_s$, we compute the cumulative distribution function (CDF),
\begin{align}
F_{\rho_s}({\rho_s}) = \int_0^{\rho_s} f_{\rho_s}(\tau) \tau d\tau
&=
\frac{\rho_s^2}{R^2}.
\end{align}
Solving the inverse transform, $w_3 = F_{\rho_s}({\rho_s})$, then gives the sampling
\begin{align}
\rho_s &= R \sqrt{w_3}.
\end{align}
Note that the numerical sampling must be better when simulating flows of bacterial clusters at low $z$ values. At high $z$ many swimmer flows contribute approximately equally, but at low $z$ only a few swimmers are nearby. Therefore many simulation samples are needed to achieve an equivalent averaging over the swimmer positions and orientations. This can be quantified by comparing $z$ with the average nearest-neighbour distance \cite{chandrasekhar1943stochastic}, $\langle r \rangle = 1/ \sqrt{4n}$, similar to the Wigner-Seitz radius.

\begin{figure}
\includegraphics[width=1\linewidth]{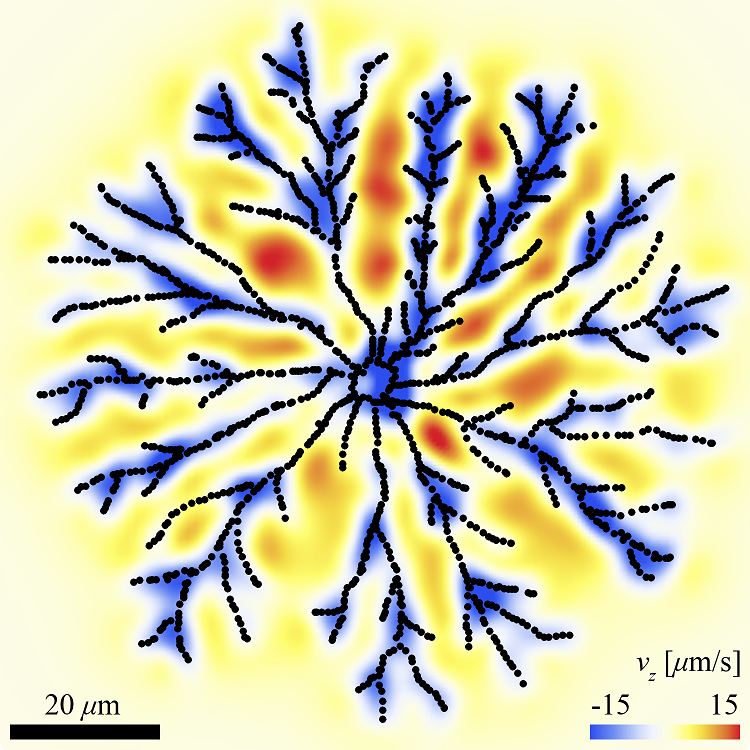}
\caption{\label{SIFig:Branch} 
Enlargement of Fig.~1f of the Main Text. 
Flows driven by a colony of bacteria arranged in a branching pattern, simulated with $N=1800$ cells (black points) using Eq.~\ref{SIEq:SimulatedTotalFlow}, shown for the plane $z=5 \mummy$ (top view).
}
\end{figure}

\subsection{Gaussian cluster}

Similarly, a cluster with a Gaussian cluster profile (\ref{SIEq:GaussianClusterProfile}) can be integrate to get the CDF
\begin{align}
F_{\rho_s}({\rho_s}) = \int_0^{\rho_s}  \frac{1}{R^2} \exp \left ( - \frac{\tau^2}{2 R^2} \right) \tau d\tau
&=
1 - \exp \left( - \frac{\rho_s^2}{2 R^2} \right) = w_3.
\end{align}
This is inverted directly to obtain
\begin{align}
\rho_s &= R \sqrt{2 \log \frac{1}{1-w}}.
\end{align}

\subsection{Density gradients}

To model a linear density gradient we consider the profile
\begin{align}
\label{SIEq:DensityGradientProfile}
f(\vec{r}_s, \vec{p}_s) &=  \frac{3N(1-\rho_s / R) \Theta(R-\rho_s) \delta(z_s-\swh)}{\pi R^2} \frac{\delta(|\vec{p}_s|-1) \delta(\vec{p}_s\cdot  \hat{\vec{z}})}{2 \pi},
\end{align}
which for the random variates $\rho_s \in [0,R], \theta_s \in [-\pi, \pi], \phi_s  \in [-\pi, \pi]$ simplifies to
\begin{align}
\label{SIEq:DensityGradientProfile1}
f(\rho_s, \theta_s, \phi_s) &= 
N f_{\rho_s} f_{\theta_s} f_{\rho_s}
=N \frac{6(1-\rho_s / R)}{R^2} \frac{1}{2\pi}  \frac{1}{2\pi},
\end{align}
and which is again normalised with respect to the number of swimmers,
\begin{align}
\int  f(\vec{r}_s, \vec{p}_s) d\vec{r}_s d\vec{p}_s 
& = \int_0^R \int_{-\pi}^\pi \int_{-\pi}^\pi  N \frac{6(1-\rho_s / R)}{R^2} \frac{1}{2\pi}  \frac{1}{2\pi} d\phi_s  d\theta_s  \rho_s d \rho_s 
= N.
\end{align}
Hence, it follows that the angular variates are still given by (\ref{SIEq:AngularVariate1},\ref{SIEq:AngularVariate2}), and for radial variate we have the CDF
\begin{align}
F_{\rho_s}({\rho_s}) = \int_0^{\rho_s}  f_{\rho_s}(\tau) \tau d\tau
&=
\frac{\rho_s^2 (3 R-2 \rho_s )}{R^3}
\end{align}
The radial variate is then given by the cubic expression
\begin{align}
\rho_s &= F^{-1}_{\rho_s}(w_3),
\end{align}
which can be solved numerically or by using the Cardano formula.
In Fig.~1e of the main text we show the flows generated by $N=12,500$ swimmers sampled in this manner, with $R=200 \mummy$ so that $N/(\pi R^2) = 0.1 / \mummy^2$, as computed using Eq.~\ref{SIEq:SimulatedTotalFlow}.
We focus on the area in the middle of the gradient, around $x=100$.

\subsection{Branching pattern}

The flows that attract nutrients down towards a colony do not depend strongly on the cluster morphology. 
To demonstrate this we manually arrange $N=1800$ bacteria in the shape of a branching pattern.
These swimmer positions are visualised in the enlarged SI Fig.~\ref{SIFig:Branch}.
In order to minimise orientation gradients and add focus on the density gradients at the edges of the cluster, the swimmers are given uniformly distributed orientations, $\phi_s \in [-\pi, \pi]$.
The flows are then computed using Eq.~\ref{SIEq:SimulatedTotalFlow}.

\begin{figure}
\includegraphics[width=1\linewidth]{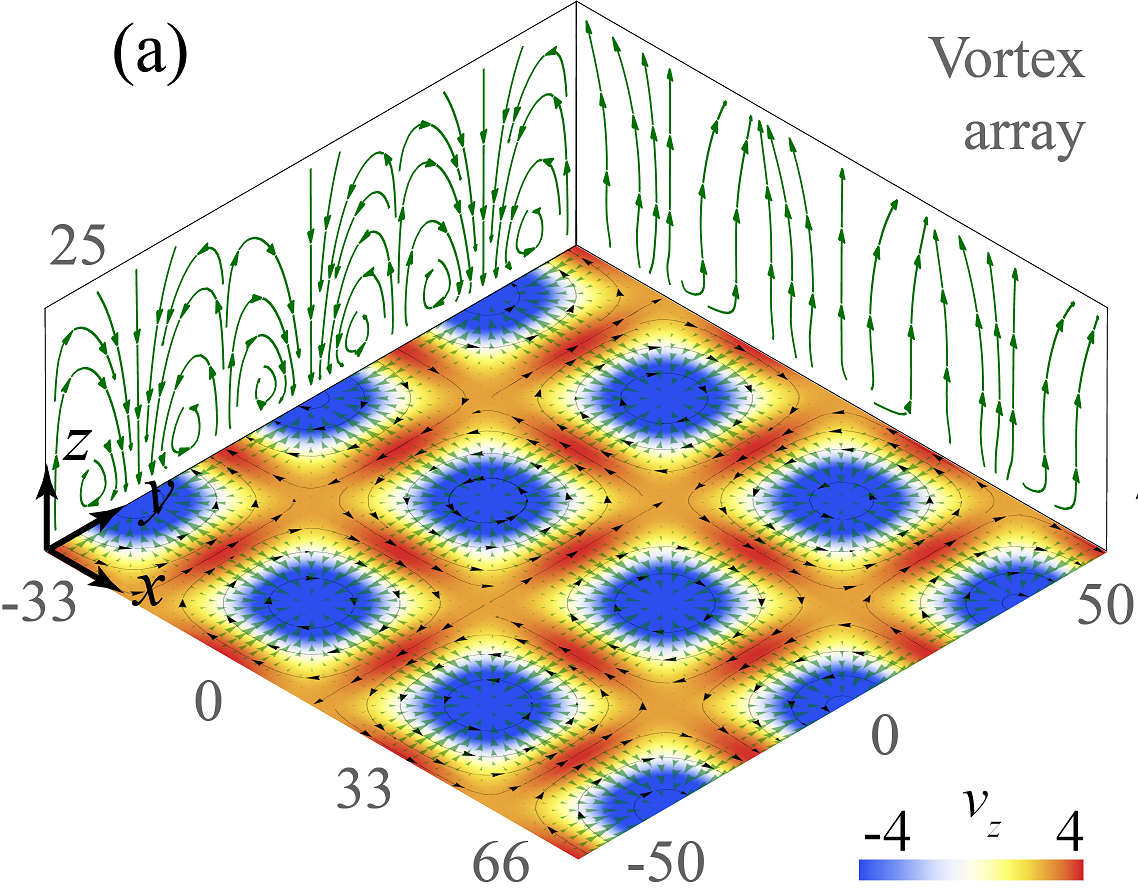}
\caption{\label{SIFig:Vortex} 
Enlargement of Fig.~4a of the Main Text. 
Flows above a bacterial vortex array, simulated as a Taylor-Green pattern with unit cell size $\lambda = 33 \mummy$, and uniform swimmer density $n=0.25 /\mummy^2$.
Colours indicate vertical flows in $\mummys$, simulated for $z=10 \mummy$ and side views at $x=- 33 \mummy$ and  $y=50 \mummy$. Green arrows are stream lines and black arrows the swimmer orientations.
}
\end{figure}

\subsection{Bend \& splay gradients, topological defects}

To simulate the flows due to bend and splay gradients [Main text Fig.~2], and topological defects [Main text Fig.~3], we employ the carpet profile
\begin{align}
f(\rho_s, \theta_s, \phi_s) &= \frac{N}{\pi R^2} \Theta(R - \rho_s) \delta(\phi_s - \theta_s - \phi_0).
\end{align}
In order to avoid density gradients, we place the swimmers on a dense regular lattice, 
\begin{align}
x_s = (1+2 i) \mummy, \quad \quad \quad
y_s = (1+2 j) \mummy, \quad \quad \quad i,j \in [-250, \dots, 0, \dots 249],
\end{align}
so that $n=0.25/\mummy^2$ and the offset is chosen to keep symmetry about the $x$ and $y$ axes and to avoid conflicts at the origin.
Next, the swimmers outside the radius $R$ are removed to retain axial symmetry, and $N=195,502$ swimmers remain.
Once the positions are set, the orientations are set by $\phi_s = \theta_s + \phi_0$.
The flows are then computed using Eq.~\ref{SIEq:SimulatedTotalFlow}.

\subsection{Vortex array}

The currents due to a bacterial vortex array is modelled using the Taylor-Green Vortex (TGV) model.
As for the bend \& splay gradients, the swimmers positions are determined by a dense regular mesh,
\begin{align}
x_s = (1+2 i) \mummy, \quad \quad \quad
y_s = (1+2 j) \mummy, \quad \quad \quad i,j \in [-250, \dots, 0, \dots 249],
\end{align}
so that the density is uniform with $n=0.25/\mummy^2$. No swimmers are removed for symmetry reasons, so $N=250,000$ swimmers.

Next, the swimmer orientations are set by the TGV director field,
\begin{align}
p_x = - \cos \left( \frac{\pi x_s}{\lambda} \right) \sin \left( \frac{\pi y_s}{\lambda} \right),
 \quad \quad \quad
p_y = \sin \left( \frac{\pi x_s}{\lambda} \right) \cos \left( \frac{\pi y_s}{\lambda} \right),
\end{align}
where the unit cell size (half-wavelength) is $\lambda = 33 \mummy$.
The flows are then computed using Eq.~\ref{SIEq:SimulatedTotalFlow}, as shown enlarged in SI Fig.~\ref{SIFig:Vortex}.

\section{Cluster stability}
\label{SISec:Cluster stability}

In the absence of an aggregation mechanism, clusters of swimming cells will disperse. 
Therefore, we discuss in the section the stability of active carpet clusters and consequences for the generation of long-ranged flows.

\subsection{Gradients in swimmer density or activity itself}

It should first be noted that collective flows need not be driven necessarily by gradients in swimmer number density, but rather by \textit{gradients in activity}. 
That is, the same amount of swimmers can locally generate more flows. 
For example, the metabolism could be increased locally by higher nutrient concentrations or temperature, so that the activity increases there, indeed fuelling a positive feedback loop.
These activity gradients are completely disconnected from the director field, or cell dispersion, so do not affect stability issues as long as metabolism is locally sustained.

Indeed, plenty of filter-feeding organisms generate flows without swimming much, as their multiple flagella or cilia drive flows in different directions. 
One of the most beautiful examples, which also highlights the importance of multicellular cooperation, is the rosette structure in many species of choanoflagellates \cite{king2008genome}. These free-living cell groups generate dipolar flows \cite{roper2013coop, nielsen2017hydrodynamics} when not sessile and also accumulate as carpets on surfaces \cite{kirkegaard2016filter}, acting like aggregate random walkers \cite{kirkegaard2016motility}. To our knowledge, the idea that accrued groups of such organisms could optimise the attraction of nutrients collectively has not been explored much in the current literature.
In conjunction with that point, there is often a trade-off between swimming faster or capturing more prey \cite{tam2011optimal, michelin2011optimal, dolger2017swimming, gilpin2017vortex, mathijssen2018universal}.
\\ \\
Having said that, many natural mechanisms exist that do cause cell accumulation;
\begin{enumerate}
\item Clustering due to local variations in the environment, by chemotaxis, thermotaxis, phototaxis, rheotaxis, et cetera;
\item Rather than an external source, organisms can attract one another by pheromones or other chemical signals;
\item Mechanical interactions, leading to self-assembly or motility-induced phase separation \cite{cates15a}; Swimmers reducing their swimming speed at locally higher organism densities.
\end{enumerate}
In the next sections we look further into cluster stability due to chemotaxis.

\begin{figure}
\includegraphics[width=\linewidth]{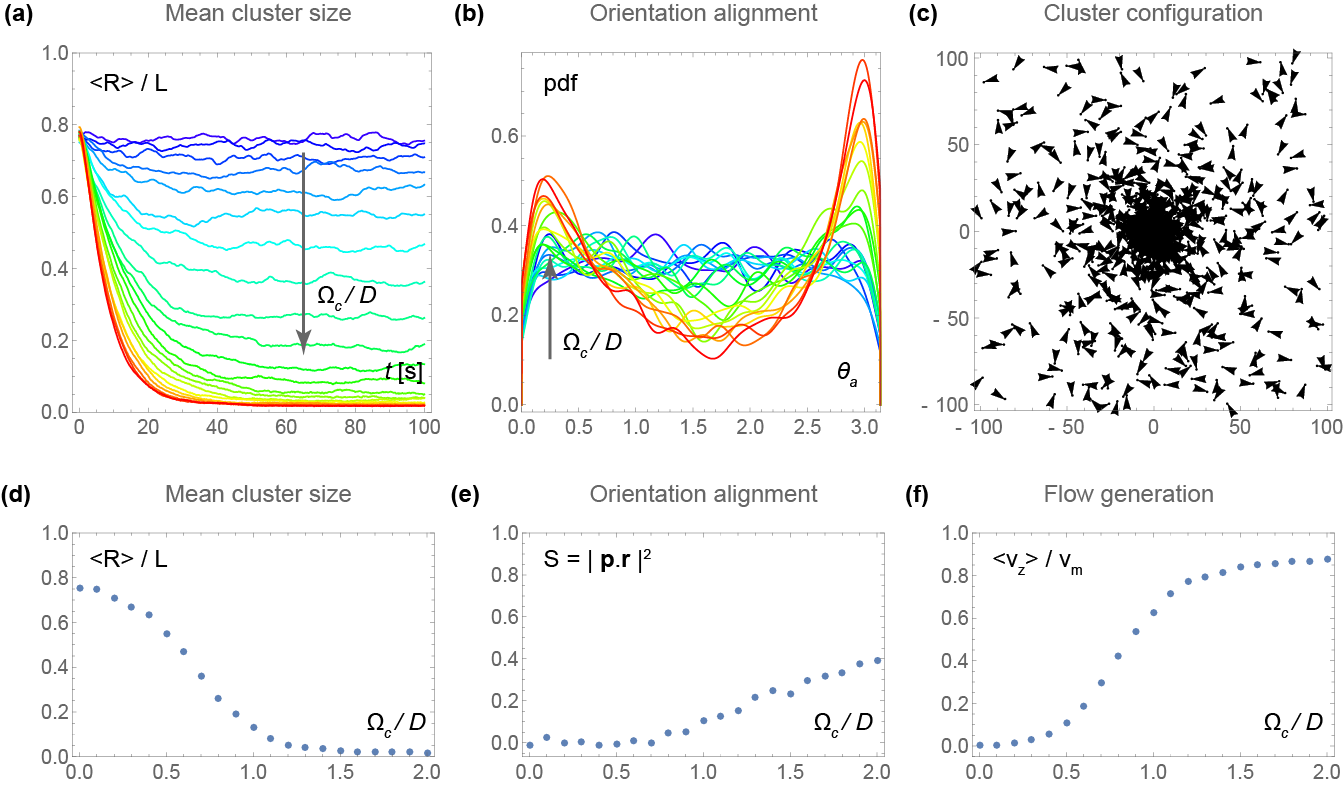}
\caption{\label{SIFig:Chemotaxis} 
Stability of bacterial clusters as a function of chemotactic strength $\Omega_a$ compared to rotational diffusion $D_r$.
(a) Mean cluster size over time [s], normalised by the system size $L$, for increasing $\Omega_c/D_r$.
(b) Alignment of the swimmer orientation towards the chemoattractant source. Shown are distributions of the alignment angle $\theta_a = \text{arccos}(\vec{p}_s \cdot \vec{r}_s) \in [0,\pi]$.
(c) Typical cluster configuration at $\Omega_c/D_r = 0.8$. Black arrows indicate the bacterial positions and orientations.
(d) Mean cluster size after steady state is reached, as a function of $\Omega_c/D_r$.
(e) Orientation alignment after steady state is reached. Shown is the order parameter $S = \langle |\vec{p}_s \cdot \vec{r}_s|^2 \rangle \in [0,1]$.
(f) Flow generated by the cluster after steady state is reached. Shown is $\langle v_z \rangle$ evaluated at the point $(0,0,z=50 \mummy)$, normalised by the flow at optimal clustering, $v_m = -12 N h \kappa / z^3$.
}
\end{figure}

\subsection{Fixed source of chemoattractant}

Bacteria rapidly respond to chemical gradients to exploit micro-scale nutrient patches \cite{stocker2008rapid}, and thus aggregate around these nutrient rich areas  \cite{eisenbach2004chemotaxis, wadhams2004making, hall-stoodley04, stocker2012ecology}.
Here we discuss the stability of these still freely swimming (non-sessile) aggregates.
We consider the case of $N$ independent bacteria being attracted to an external source of chemoattractant.
This source could also represent a thermal or other attraction point.
We describe the swimmer dynamics with the over-damped limit of the following Langevin equations \cite{romanczuk2012active},
\begin{align}
\label{SIEq:ChemDyn1}
\dot{\vec{r}_s} &= \vec{v}_s,
\quad \quad
\dot{\vec{v}_s} = - \gamma_0 \vec{v}_s + k(c) \vec{\nabla}_r c(\vec{r}_s) + \vec{\xi}(t),
\end{align}
where $\gamma_0$ is the friction coefficient, $c(\vec{r}) = M/r$ is the chemoattractant concentration field centred at the origin and the sensitivity to gradients $k(c)$ is modelled with the receptor law \cite{ben2000cooperative},
\begin{align}
k(c) = \frac{k_0}{(1+Sc)^2},
\end{align}
because at very high concentrations the chemotactic membrane receptors are saturated and gradients cannot be climbed.
At low Reynolds numbers, in the over-damped limit, these dynamics (\ref{SIEq:ChemDyn1}) reduce to
\begin{align}
\label{SIEq:ChemDyn2}
\dot{\phi_s} = \frac{A}{(r_s + r_0)^2} \sin (\phi_s - \theta_s) + \xi_\phi(t),
\quad \quad
\dot{\vec{r}_s} = v_s \vec{p}_s,
\end{align}
for the swimmer orientation $\phi_s$ and the radial and angular positions $(r_s, \theta_s)$. 
Here $A$ [$\text{rad~m}^2/\text{s}$] is the chemotactic bias, $r_0 = r_0(M,S)$ is the receptor saturation radius, $v_s = 20 \mu \text{m}/\text{s}$ is the swimming speed, and the fluctuations are modelled with a white noise obeying $\langle \xi_\phi(t) \rangle = 0$ and $\langle \xi_\phi(t) \xi_\phi(t') \rangle = 2 D_r \delta(t-t')$.
The time decorrelation is set to $D_r = 1 \text{rad}^2/\text{s}$, originating from rotational fluctuations or run-tumble dynamics \cite{berg1972chemotaxis, mittal2003motility, saragosti2012modeling, cates2013when}.

We model a system of size $2L \times 2L$ with periodic boundary conditions, where $L = 100 \mummy$ and $r_0 = 10 \mummy$. 
We non-dimensionalise the dynamics by considering the typical chemotactic strength, $\Omega_s = A/(0.5L)^2, [1/\text{s}]$, compared to the rotational diffusion coefficient $D_r [1/\text{s}]$.
A constant number of swimmers, $N=1,000$, are initiated from uniformly distributed positions and orientations, and are integrated numerically with time step $\delta t = 10$ms until a steady state is reached.
During these dynamics we evaluate the mean cluster size $\langle R \rangle = \langle | \vec{r}_s | \rangle$ averaged over all swimmers, and their orientation alignment with respect to the origin, $\theta_a = \text{arccos}(\vec{p}_s \cdot \vec{r}_s) \in [0,\pi]$. 
To quantify the amount of radial alignment we also define an order parameter $S = \langle |\vec{p}_s \cdot \vec{r}_s|^2 \rangle \in [0,1]$ averaged over all swimmers, where $S=0$ corresponds to no correlation between the swimmer orientation and position, and $S=1$ corresponds to all swimmers pointing towards or away from the chemoattractant source.
Then, we also compute the flow generated by all the organisms using Eq.~\ref{SIEq:SimulatedTotalFlow}, evaluated directly above the cluster at position $(0,0, z = 50\mummy)$, and non-dimensionalise with respect to the flow for optimal clustering, $v_m = -12 N h \kappa / z^3$, using Eq.~\ref{SIEq:ClusterFlowZ} in the limit $R\to0$ with $N=n \pi R^2 = 1,000$ constant.
To avoid finite system size effects we sum the swimmer flow over 10 periodic systems in each direction, so considering swimmers over an area of $20L\times20L$. This ensures that the average flow disappears in the absence of swimmer density gradients, when $\Omega_c=0$.

Figure \ref{SIFig:Chemotaxis}a shows the evolution of the mean cluster radius over time, for different values of $\Omega_c/D_r$ ranging from no (blue) to strong (red) chemotaxis. 
A steady state is reached for all strengths, when the run-tumble swimming is balanced by the chemotaxis, after a typical time scale larger than $L/v_S$ and $D_r^{-1}$. 
The positions and orientations of this steady state, at $t=100s$, are shown in Fig.~\ref{SIFig:Chemotaxis}c for strength $\Omega_c/D_r = 0.8$.
Here the cells have accumulated into a cluster with mean radius $\langle R \rangle / L \approx 0.3$.
Indeed, the mean cluster size reduces with increasing chemotaxis [Fig.~\ref{SIFig:Chemotaxis}a,d].
Moreover, the cells in Fig.~\ref{SIFig:Chemotaxis}c have mostly random orientations, with only a slight bias to the source.
To quantify this we consider the distribution of orientation alignment angles, $\text{pdf}(\theta_a)$, and the order parameter $S$ averaged over all swimmers [Fig.~\ref{SIFig:Chemotaxis}b,e]. 
At low strength (blue) the distributions are flat and $S\approx 0$, whereas at higher strengths (red) preferred orientations towards and away from the cluster emerge, $S\approx 0.4$, for values $\Omega_c/D_r >1$.
Still, the order parameter $S$ remains relatively small compared to unity.

Finally, we discuss the liquid transport generated by the cells.
No flows are produced in the absence of swimmer density gradients ($\Omega_c=0$), but currents appear due to cluster formation with increasing chemotaxis [Fig.~\ref{SIFig:Chemotaxis}f].
At intermediate values, $\Omega_c/D_r = 1$, we observe significant drifts, $\langle v_z \rangle \approx 0.65 v_m$, even if the swimmer orientations are not completely random but slightly correlated with position, $S\approx 0.2$.
At stronger chemotaxis the mean cluster size reduces further and the flows increase.
Only at higher chemotactic strengths we expect more radial orientations that could reduce the flows again.

\subsection{Movie of dynamic cluster}

We consider a dynamic cluster of swimmers that move around a chemoattractant source located at the origin.
The swimmers are modelled as above (Eq.~\ref{SIEq:ChemDyn2}), initially distributed randomly and converging to a cluster with chemotactic strength $\Omega_c/D_r = 1$.
Tracer particles ($N_T = 50$) are initiated with random positions, $x \in [-30, 30] \mummy$, $y \in [-2, 2] \mummy$ and $z \in [15, 25] \mummy$.
Then, the swimmer and tracer dynamics are integrated numerically with time step $\delta t = 10$ms, again using Eq.~\ref{SIEq:SimulatedTotalFlow}, and thermal Brownian noise is added with tracer diffusivity $D_t = 1\mummy^2 / \text{s}$.
The tracers are not allowed to pass the plane $z=2 \mummy$ to avoid contact with the near-field swimmer flows. 
The resulting Movie S1 shows that the tracer particles directly above the cluster are attracted downwards. Then they move sideways, down the swimmer concentration gradient, and finally back up again to complete the recirculation.

\subsection{Autochemotaxis}

Instead of being attracted to a fixed external source of chemoattractants, another common situation is bacteria attracting each other through the excretion of pheromones while they swim.
Here the chemoattractant concentration profile is no longer steady in time ($c(\vec{r}) = M/r$) but instead evolves according to
\begin{align}
\dot{c}(\vec{r},t) = \sum_{i=1}^N Q_i \delta(\vec{r} - \vec{r}_i) - d_c + D_c \nabla^2 c,
\end{align}
in terms of individual attractant excretion rates $Q_i$, the attractant decay rate $d_c$ and the attractant diffusion coefficient $D_c$.
We will not simulate these equations here but refer to the excellent reviews by Romanczuk et al. \cite{romanczuk2008beyond, romanczuk2012active} for a comprehensive discussion and summary of analytical solutions.
Moreover, another recent work has found solutions of the Smoluchowski diffusion equation for active Brownian swimmers \cite{sevilla2015smoluchowski}.
Qualitatively these show similar dynamics as before, that the mean size of a cluster is a decreasing function of the chemotactic strength, leading to stronger flows.

\section{Bacterial turbulence}
\label{SISec:SimFlowsTurbulence}

\subsection{Self-propelled rod (SPR) model}

\begin{figure}
\includegraphics[width=0.3\linewidth]{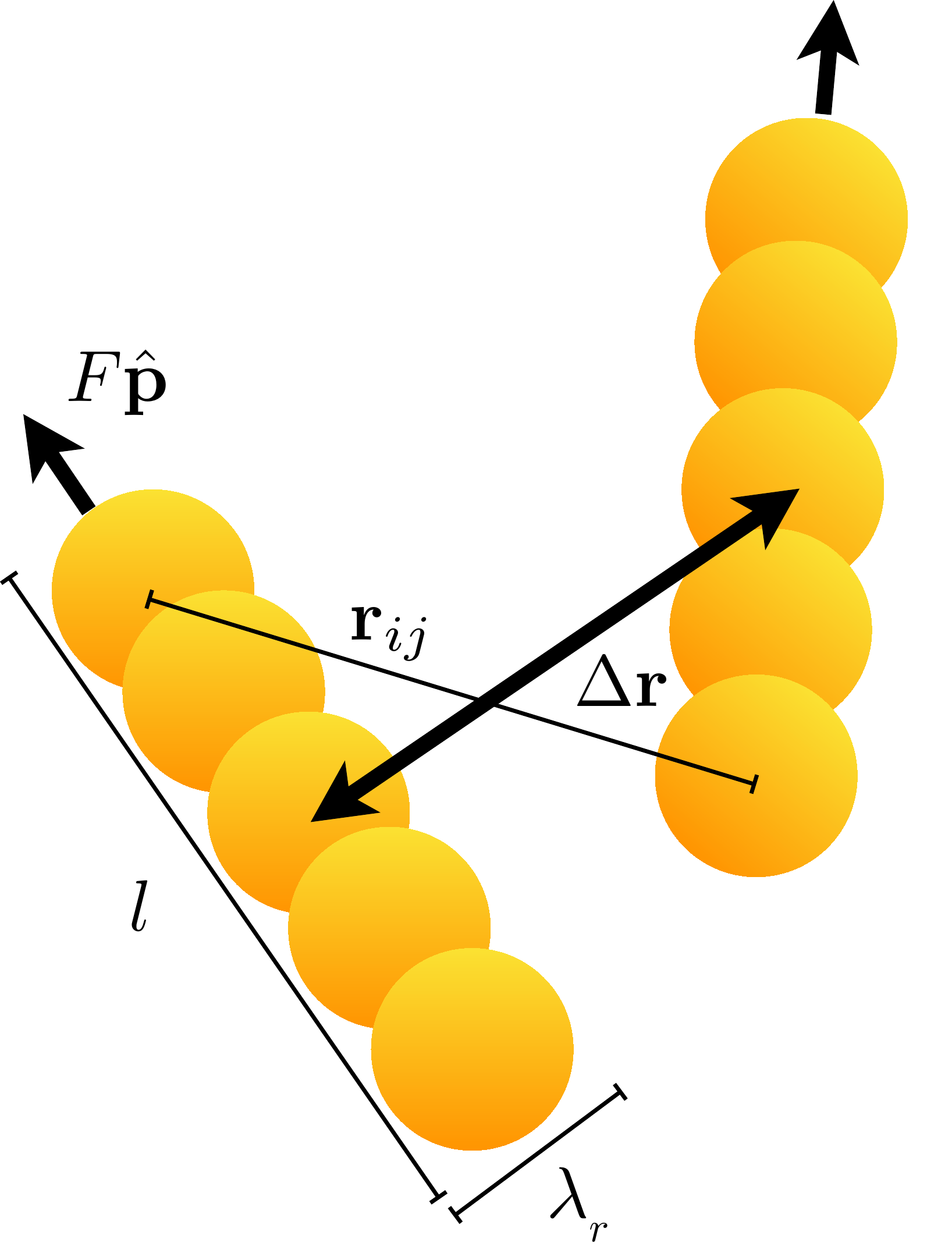}
\caption{\label{SIFig:SPR} 
Diagram of a pair-wise interaction in the self-propelled rod (SPR) model.
Rods of aspect ratio $\Gamma=\ell/\lambda_r$ are composed of $n=5$ repulsive Yukawa segments.
Self-propulsion arises from a constant force $F$ acting along the rod axis, indicated by the unit vector $\vec{\hat{p}}$.
The overall pair-wise interaction is obtained by summing the Yukawa potentials over all segment pairs with separation $\vec{r}_{ij}$, which decays rapidly with the centre-of-mass separation $\vec{\Delta r}$.
}
\end{figure}

We model the bacterial bath in two spatial dimensions by $N$ rod-like self-propelled units [Fig.~\ref{SIFig:SPR}].
For a detailed description, also see Ref.~\cite{wensink2012emergent}. 
Each rod has an aspect ratio $\Gamma= \ell / \lambda_r = 5$ which is chosen in order to model {\it Bacillus subtillis} suspensions, as considered in experiments dealing with bacterial turbulence.
Rods of length $\ell$ and width $\lambda_r$ are discretised into $n=6$ spherical segments equidistantly positioned, with a displacement $s=0.85\lambda_r$, along the main rod axis $\bhu = (\cos \varphi, \sin \varphi)$.
Between the segments of different rods a repulsive Yukawa potential is imposed 
 \cite{kirchhoff1996dynamical}. The resulting pair potential of a rod pair $\alpha$, $\beta$ is given by
\begin{eqnarray}
U_{\alpha \beta} = \sum_{i=1}^{n} \sum_{j=1}^{n} U_i U_j \exp [-r_{ij}^{\alpha \beta} / \lambda_r] /  r_{ij}^{\alpha \beta}, 
\end{eqnarray}
where $\lambda_r$ is the
screening length and $ r_{ij}^{\alpha \beta} = |{\bf r}_{i}^{\alpha} - {\bf r}_{j}^{\beta}|$ the distance between segment
$i$ of rod $\alpha$ and segment $j$ of rod $\beta$ $(\alpha \neq \beta)$.
Any overlap of particles is prohibited by choosing a large interaction strength $U_j^2=2.5F_0 \ell$. Here $F_0$ is an effective self-propulsion force directed along the main rod axis and leading to a constant propulsion velocity $v_0$. We do not resolve details of the actual propulsion mechanism or hydrodynamics interactions.

Micro-swimmers move in the low Reynolds number regime. The corresponding over-damped equations of motion for the positions 
$\br_{\alpha}$ and orientations $\bhu_\alpha$ are
\begin{eqnarray}
{\bf f }_{\cal T} \cdot \partial_{t} \br_{\alpha}(t) &=&  -\nabla_{\br_{\alpha}}
 U(t) +  F_{0} \bhu_{\alpha}(t), \\
{\bf f}_{\cal{R}} \cdot \partial_{t} \bhu_{\alpha}(t) &=&
-\nabla_{\bhu_{\alpha}} U(t),
\label{eom}
\end{eqnarray}
in terms of the total potential energy  
$U=(1/2)\sum_{\alpha, \beta (\alpha \neq \beta)} U_{\alpha \beta} + \sum_{\alpha, \gamma} U_{\alpha \gamma}$ 
with $U_{\alpha \gamma}$ the potential energy of rod $\alpha$ with the carrier $\gamma$. 
The one-body translational and rotational friction tensors for the rods ${\bf f}_{\cal T}$ 
and ${\bf f}_{\cal R}$ can be decomposed into parallel $f_\parallel$, perpendicular $f_\perp$
and rotational $f_{\cal R}$ contributions which depend solely on the aspect ratio $\Gamma = \ell/\lambda_r $
\cite{tirado1984comparison},
\begin{eqnarray}
\frac{2\pi}{f_{||}} &=& \ln p - 0.207 + 0.980p^{-1} - 0.133p^{-2},
\\
\frac{4\pi}{f_{\perp}}&=&\ln p+0.839 + 0.185p^{-1} + 0.233p^{-2},
\\
\frac{\pi a^2}{3f_{\mathcal{R}}} &=& \ln p - 0.662 + 0.917p^{-1} - 0.050p^{-2}.
\end{eqnarray} 
Accordingly, the propulsion velocity is given by $v_0 = F_0 / f_{||}$ and sets the characteristic
time unit $\tau = \ell / v_0$.

The total number of rods is $N_0 =10,000$ and we use a quadratic simulation domain of size $L=200\mummy$ with periodic boundary conditions, $(x_0,y_0) \in [-L/2,L/2] \mummy$, to establish a uniform swimmer density of $n=0.25\text{bact.}/\mummy^2$.
The dimensionless packing fraction $\Phi= \lambda_r \ell  N_0/ L^2$ is fixed to $\Phi=0.7$ to achieve a turbulent bacterial bath \cite{wensink2012meso}.
The initial swimmer configuration is a smectic lattice of rods, where the rods are randomly orientated up- and downwards.
We then simulate $301$ time steps of $\delta t = 0.01$ seconds, so the dimensional simulation times are $t = [0, \delta t, 2 \delta t, \dots, 3\text s]$.
Movie S2 shows the resulting dynamics, with some swimmers coloured so they can be identified throughout the turbulent motion.

\subsection{Movies of flow due to bacterial turbulence}

\begin{figure}
\includegraphics[width=1\linewidth]{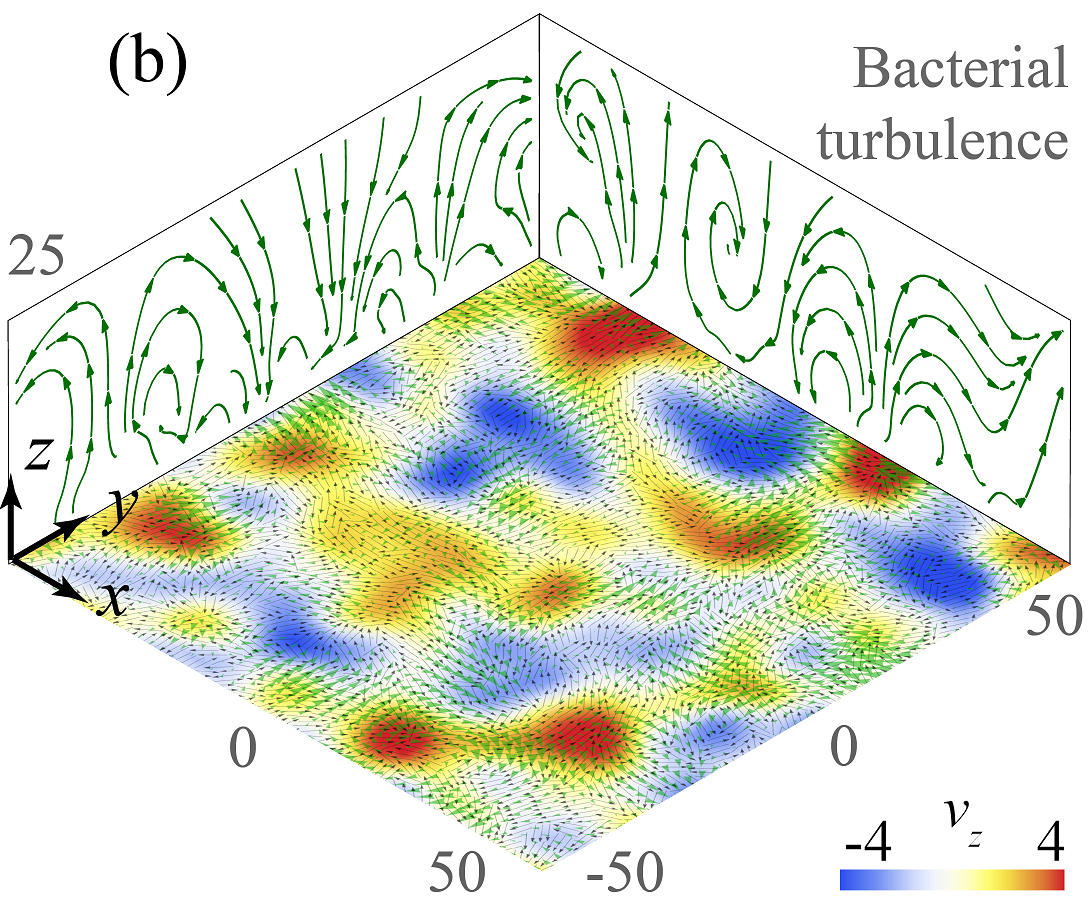}
\caption{\label{SIFig:Turbulence} 
Enlargement of Fig.~4b of the Main Text. 
Flows above a carpet of bacterial turbulence, simulated with the SPR model with aspect ratio $\Gamma=5$, packing fraction $\Phi  = 0.7$ and swimmer density $n=0.25 /\mummy^2$.
Colours indicate vertical flows in $\mummys$, simulated for $z=10 \mummy$, and side views at $x=- 50 \mummy$ and  $y=50 \mummy$. Green arrows are stream lines.
Black arrows show the individual swimmer positions and orientations.
}
\end{figure}

To compute the long-ranged flows, and to avoid edge effects, we enlarge the carpet by duplicating the periodic swimmer positions, 
\begin{align}
x_s = x_0 + 200 \mummy \times i, \quad \quad \quad
y_s = y_0 + 200 \mummy \times j, \quad \quad \quad i,j = [-2, -1, 0, 1, 2],
\end{align}
so that the total number of swimmers is $N = 5 \times 5 \times N_0 = 250,000$, in the domain $(x_s,y_s) \in [-500,500] \mummy$.
The flows are then computed using Eq.~\ref{SIEq:SimulatedTotalFlow}, as shown enlarged in SI Fig.~\ref{SIFig:Turbulence}.

Movie S3 shows the flows due to bacterial turbulence in the plane $z=10\mummy$, for the local area $(x,y) \in [-50,50] \mummy$.
Colours indicate vertical flows, $v_z \in [-4,4] \mummys$. Green arrows are stream lines of the lateral flows, and black arrows show the individual swimmer positions and orientations.

Movie S4 shows these flows in the plane $z=25\mummy$, again for the local area $(x,y) \in [-50,50] \mummy$.
Colours indicate vertical flows, $v_z \in [-1,1] \mummys$. Green arrows are stream lines of the lateral flows, and black arrows show the individual swimmer positions and orientations.

Movie S5 shows a side view of these flows, for the cross section $y=0$, with lateral position $x \in  [-50,50] \mummy$ and heights $z \in [2,25] \mummy$.
Colours indicate vertical flows, $v_z \in [-3,3] \mummys$. Green arrows are stream lines in the plane.

\subsection{Temporal correlation functions}

\begin{figure}
\includegraphics[width=1\linewidth]{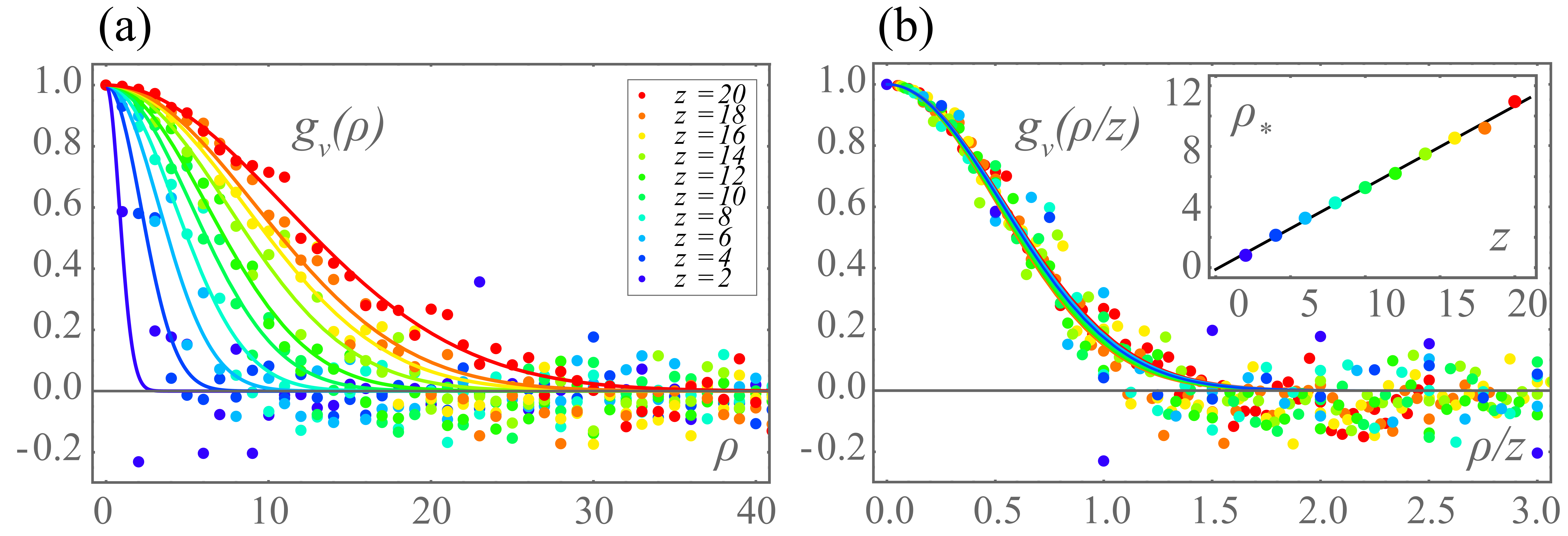}
\caption{\label{SIFig:CorrelationCollapse} 
(a) Spatial correlation functions of vertical flows generated by a carpet of bacterial turbulence, $g_{v_z}(\rho)$, for heights $z\in [2,20]~\mummy$ (blue-red). Same as Fig.~4d of the Main Text.
(b) Collapse of these spatial correlation functions onto one curve, when rescaling the lateral distance with the distance from the wall, $g_{v_z}(\rho/z)$.
}
\end{figure}

The bacterial turbulence flows are computed as in the previous section, with $N = 250,000$ swimmers in the domain $(x_s,y_s) \in [-500,500] \mummy$.
Next, the temporal correlation function of the vertical flows is defined as
\begin{align}
c_{v_z}(t) = \frac{\langle v_z (t_1) v_z(t_2) \rangle }{ \sqrt{\langle v_z^2 (t_1)  \rangle \langle v_z^2 (t_2)  \rangle}}; 
\quad 
t = |t_1 - t_2|,
\end{align}
where the average is over lateral space. 
This is implemented numerically by sampling the flow with Eq.~\ref{SIEq:SimulatedTotalFlow} at $N_q=200$ points with positions uniformly distributed over the carpet, $(x_q,y_q) \in [-100,100] \mummy$, and fixed height $z$ for each correlation function. 
The average is then taken over all points but only for the last 251 time steps, $t \in [0.5, 3]$s, in order to exclude the initial phase where the swimmers still develop turbulent motion after initiation.

Equivalently, the temporal correlation function of the swimmer orientations is defined as
\begin{align}
c_{\vec{p}}(t) = \langle \vec{p} (t_1) \cdot \vec{p} (t_2) \rangle; \quad t = |t_1 - t_2|,
\end{align}
where the average is performed over all $N_0$ swimmers, and again only for times $t \in [0.5, 3]$.

To determine the typical correlation time, $t_*(z)$, the resulting temporal correlation functions are fitted to exponentials,
\begin{align}
c(t) \to \exp \left( - \frac{t}{t_*} \right),
\end{align}
which is the most elementary function with one parameter that well describes the data. We tried other functions (Gaussian, $1/t$ decay) and these give similar results.
Hence, the fitted correlation times are plotted against $z$ in Fig.~4c of the Main Text.

\subsection{Spatial correlation functions}

The equal-time spatial correlation function of the swimmer orientations is defined as
\begin{align}
g_{\vec{p}}(\rho) = \langle \vec{p} (\vec{r}_1) \cdot \vec{p} (\vec{r}_2) \rangle; \quad |\rho - \sqrt{(x_1 - x_2)^2 + (y_1 - y_2)^2}|< \epsilon,
\end{align}
where the average is performed over all $N_{ps}$ \textit{pairs of swimmers} between the $N_0$ that are separated a distance $\rho$ in the simulation with a margin of $\epsilon = 0.5 \mummy$.
The number of pairs grows with $\rho$, and is still fairly large; $N_{ps} \sim 100$ for $\rho = 10 \mummy$.
A larger value of $\epsilon$ increases the number of sample pairs but reduces the resolution of the correlation function.

The equal-time spatial correlation of the vertical flows is defined as
\begin{align}
g_{v_z}(\rho) = \frac{\langle v_z (\vec{r}_1) v_z(\vec{r}_2) \rangle }{ \sqrt{\langle v_z^2 (\vec{r}_1)  \rangle \langle v_z^2 (\vec{r}_2)  \rangle}} ; 
\quad 
\rho = \sqrt{(x_1 - x_2)^2 + (y_1 - y_2)^2}.
\end{align}
This is implemented numerically by sampling the flow with Eq.~\ref{SIEq:SimulatedTotalFlow} for $N_q=200$ \textit{pairs of points}.
These are sampled by selecting a midpoint $\vec{r}_m$ with a position uniformly distributed over the carpet, $(x_m,y_m) \in [-100,100] \mummy$, and fixed height $z$. Next, a random orientation $\theta_m \in [-\pi, \pi]$ is sampled, so that the pair is given by $\vec{r}_{1,2} = \vec{r}_m \pm \frac \rho 2 (\cos \theta_m, \sin \theta_m, 0) $.
Hence, the correlation functions are found by averaging over all the pairs, with the fixed time $t = 1$s.

To determine the typical correlation length, $\rho_*(z)$, the resulting spatial correlation functions are fitted to Gaussians,
\begin{align}
g(\rho) \to \exp \left( - \frac{t^2}{2 t_*^2} \right),
\end{align}
which is the most elementary function with one parameter that well describes the data. 
We tried other functions (exponential, $1/t$ decay) but these do not fit as well.
Thus, perhaps surprisingly, the spatial and temporal correlation functions are best fit to different functions, a Gaussian or exponential respectively,  to give the best representation (least overall $R^2$ fit)  of both the correlation length and correlation time.  Still, the results would not change qualitatively if another definition were chosen.

Figure~\ref{SIFig:CorrelationCollapse}(a) shows these spatial flow correlations, $g_{v_z}(\rho)$, the same as in Fig.~4d of the main text.
Figure~\ref{SIFig:CorrelationCollapse}(b) shows a collapse of the correlation functions when rescaling with respect to the distance from the surface, $g_{v_z}(\rho/z)$.

\section{List of Movies}

The following Supplementary Movies are available online:

\begin{itemize}
\item Movie S1. Downward attraction of tracer particles above a dynamic bacterial cluster.
\item Movie S2. SPR model: Turbulent dynamics of $10,000$ bacteria, where twenty cells are labelled in colour to trace their motion.
\item Movie S3. Top view of flows generated by bacterial turbulence, at the horizontal plane $z=10\mummy$.
\item Movie S4. Top view of flows generated by bacterial turbulence, at the horizontal plane $z=25\mummy$.
\item Movie S5. Side view of flows generated by bacterial turbulence, at the vertical plane $y=0$.
\end{itemize}

\end{document}